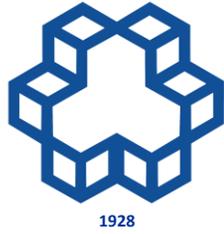

**Faculty of Electrical Engineering**
K.N. Toosi University of Technology

1928

# *An Intelligent Assistive System Based on Augmented Reality and Internet of Things for Patients with Alzheimer's Disease*

Fatemeh Ghorbani Lohesara

Supervisor: Dr. Mehdi Delrobaei
Advisor: Dr. Quazi Rahman

A Thesis Presented in Partial Fulfillment of the Requirements for the Degree of
Master of Science in Electrical Engineering (Mechatronics Engineering)

Academic Year 2019-2020

# Abstract

Independent life of the individuals suffering from Alzheimer's disease (AD) is compromised due to their memory loss. As a result, they depend on others to help them lead their daily life. In this situation, either the family members or the caregivers offer their help; they attach notes on every single object or take out the contents of a drawer to make those visible when they leave the patient alone. The aim of this thesis is to provide multi-level support and some helping means for AD patients and their family members through the integration of existing science and methods.

This study reports results on an intelligent assistive (IA) system, achieved through the integration of Internet of Things (IoT), augmented reality (AR), and adaptive fuzzy decision-making methods. The proposed system has four main components; (1) a location and heading data stored in the local fog layer, (2) an AR device to make interactions with the AD patient, (3) a supervisory decision-maker to handle the direct and environmental interactions with the patient, (4) and a user interface for family or caregivers to monitor the patient's real-time situation and send reminders once required. The system operates in different modes, including automated and semi-automated. The first one helps the user complete the activities in their daily life by showing AR messages or making automatic changes. The second one allows manual changes after the real-time assessment of the user's cognitive state based on the AR game score.

We provide further evidence that the accuracy, reliability and response time of the IA system are appropriate to be implemented in AD patients' homes. Moreover, the system response in the semi-automated mode causes less data loss than the automated mode, as the number of active devices decreases. We have also found that playing an audio message instead of displaying an image message has better performance and less battery consumption.

# Table of Contents

# List of Figures

# List of Tables

# Acknowledgments

The work in this thesis would not have been possible without the help and support I received along the way.

Firstly, I need to thank my ever supportive supervisor and advisor, Dr. Mehdi Delrobaei and Dr. Quazi Rahman, for giving me the opportunity to work on this topic for my thesis. I thank you both for the help and support I have had throughout this process. I hope we get the opportunity to work together again in the future. Thank you Dr. Delrobaei, for providing not just the research space and an engaging lab to be part of, but also for the continual confidence you had in my abilities to complete this research. I have also learned from you how to develop, evaluate, express, and defend my ideas.

I want to thank everyone in the Mechatronics Lab, including my friends and teammates, for making my time enjoyable and memorable. Thank you all for making the lab such a fun place to be; it was a brilliant place to work.

Finally, I would like to express my most profound gratefulness to my family for their continuous and unparalleled love, help and support. I am forever indebted to my beloved parents, Ashraf and Mousa, for giving me the opportunities and experiences that have made me who I am. You make me strive to make things I am proud of and give me the confidence to pursue my dreams; I love you.

This has been one of the best experiences of my life, and once more, to everyone who was there with me on the journey, whether big or small, thank you.

# Dedication

I would like to dedicate this thesis to my grandmother, Hourie. Despite her experiences with Alzheimer's disease, which affected us all, my grandmother remained a unique individual, dearly loved and respected by everyone who knew and cared for her.

# Chapter 1 - Introduction

In the following sections, we present some of the critical concepts in intelligent assistive tools, before providing an overview of the proposed model and thesis structure.

## 1.1 Alzheimer's Disease

Alzheimer's disease (AD) is one of the main common causes of dementia, and nearly 5.8 million patients are currently suffering from AD in the United States alone. This number may reach 13.8 million in 2050 [1]. People with AD or other dementias make up a large proportion of all older adults who receive adult day services and nursing home care. Caring for our elderly population living with dementia and AD raises issues over resources in terms of financial aid and time.

Loss of memory and insufficient capacity to make decisions are two main difficulties experienced by patients with AD. They generally have problems in remembering recent information and completing everyday tasks. Therefore, they should always be reminded of the required tasks that improve their confidence and quality of life.

As the disease progresses, neurons in different parts of the AD patient's brain are damaged or destroyed [1]. Activities that the patient used to do independently and memories related to the individual's identity, such as planning family events or participating in sports or groups, may no longer be possible or remembered.

Since AD affects the person in various ways, each person may present with symptoms differently [1]. The disease process results in memory impairments, personality conflicts, and reasoning problems. However, this is inimitable to each person with dementia [2]. Different studies have also indicated dementia as being distinct to the individual [2].

Professor Kitwood identifies dementia as a disability and discusses that it has a complex interaction of five defining characteristics that have consequences for the person with dementia [3]. These characteristics are the key components that lead to a person-centered theory for dementia care [3]. They are recognized as personality, biography, neurological impairment, social psychology, and physical health condition [2]. Each of these is considered to intermingle with the others and affect the person with dementia in different manners.

Everyone has a distinctive personality, and the primary focus should be on supporting this fact. Dementia might affect the person and change their personality, but their distinctiveness endures [2]. Besides, the person's personal life history can affect how they interact with other people, the words or phrases they may use in their daily life, and even their behavior in their workplace. Therefore, knowledge of the individual's biography allows experts to receive some insight into the person based on their background, which is distinctive to them [2].

On the other hand, people with dementia not only have to cope with the usual worsening in their cognitive abilities but also the physical disabilities in performing their daily life tasks, which can lead to experiencing illnesses and disorders that can significantly have a destructive impact on their health and quality of life. Neurological impairment refers to the kind of dementia the person suffers and the impact this can reflect on them.

Many experts argue that very little is recognized about the diagnosis of dementia an individual might have [3]. Some experts state that establishing and receiving a diagnosis can result in stigmatization and labeling of the sufferer. However, knowledge and identification of how dementia changes the needs of the person are fundamental because there is no two people with dementia have the same diagnoses [2].

Moreover, individuals with dementia are influenced by their surrounded people and activities over time. This includes the home or healthcare environment in which they are taken care of that might or might not meet their needs [2]. Dementia is classified as a gradual progression through which the individual shows particular traits and symptoms which begin from stage 1 (early diagnosis) and continue across the individual life (stage 4). It involves a gradual progression and changes in the individual's needs, which can continue over a long period [4].

Because stages can overlap with each other, it might be challenging to identify a person with AD in an explicit stage. As abilities and skills change through the progression of the disease and time slowly, it is essential to focus more on what activities a person can still perform rather than what they cannot do anymore; thereby, the person can live as well as possible with dementia and his cognitive disabilities.

The stages refer to the various aspects of the person at different stages of their personal life, and the impact dementia can create on their daily life. Thus, considering the person's needs and abilities is effectively achieved within a much higher background of the sufferers and their involvement of dementia [3]. To achieve this goal, person-centered methods are developed for each stage to support the individual with dementia and their caregivers [4].

More in detail, Alzheimer's disease typically progresses gradually in three stages, generally: mild, moderate, and severe [1]. The three stages make a general idea of how abilities alter once symptoms display and must only be provided as an accurate guide for experts and physicians. The following information provides more explanation about the stages and their diagnoses.

### *1.1.1 Early-stage*

In the early stage of AD, individuals might perform tasks independently. They may still drive, work, and take part in social activities, although the person with AD may experience having memory impairments, such as forgetting familiar words or the location of everyday items [1]. The person is more aware of mistakes and forgetting or misplacing things and may become frightened as they think they may be losing their mind.

Symptoms may not be generally apparent at this stage, but family members and close friends might perceive, and a physician would be able to recognize symptoms using particular diagnostic instruments [3]. The person may seem orientated, even though attention level might be decreased.

Loss of independence and control may be noticeable, and the sufferer may face some challenges in finding the right words that may be self-protective, tense, and nervous [4]. In short, common difficulties may include [1]:

- Coming up with the correct expression or name.
- Remembering names when met different people.
- Having difficulty performing activities in a social or work environment.
- Forgetting information that was just read or told.
- Losing or misplacing a valuable object.
- Experiencing enhanced trouble with planning or arranging.

During the early stage, people with dementia can live independently on their own by taking control of their health and well-being with some aids provided by assistive tools. The main aim of this thesis is to provide an intelligence assistive tool for the patient with dementia in the early stage, which is explained above.

### *1.1.2 Middle-stage*

Middle-stage AD is generally the longest stage and can last for many years [1]. Therefore, the disease progresses, and the person living with AD requires a higher level of care and support. During this stage, the person may make a mistake in choosing correct words, get frustrated or anxious, and behave in unexpected manners, such as refusing to bathe or changing clothes.

Destruction to nerve cells in the brain can also make it worsen for the person to share thoughts and perform usual activities without assistance [1]. The individual might be disorientated, and there is proof of a loss of time, day, and season priority or their personal information. They may also refer to essential memories relating to parents or mention them as being alive or return to more joyful memories of their life.

In some cases, the person may start to become more incontinent in their behavior and say what they feel directly [3]. Symptoms, which differ from person to person, may include [1]:

- Being forgetful of special events or personal skills and history.
- Feeling moody or abandoned, particularly in socially or cognitively challenging circumstances.
- Being incapable of recalling information about themselves, such as their address or telephone number, and the high school or university they joined.
- Experiencing misperception about where they are or what date it is.
- Requiring assistance choosing appropriate clothing for the season or the event.
- Having problem with controlling their bladder and bowels.
- Experiencing disorders in sleeping and patterns, such as sleeping during the day and turning to be restless at night.
- Showing an enhanced leaning to wander and become lost.
- Representing personality and behavioral variations, including dishonesty and misunderstandings or compulsive, recurring behavior such as hand-wringing or tissue shredding.

In the middle stage, the individual living with AD can still take part in daily activities with help and assistance. It is essential to find out how the person can still do or discover ways to simplify activities. As the necessity for more intensive care grows, caregivers may need to consider respite care or an adult day center; thereby, they can have a momentary break from caregiving while the person living with AD remains to receive proper care in a secure environment [1]. The IA system proposed in this thesis could also make assistance for people with AD in the middle stage based on the disease progression.

### *1.1.3 Late-stage*

In the final stage of AD, individuals who have dementia lose the capability of responding to their environment, to carry on a dialog, and eventually to control movement. They may still express words or phrases, but conversing pain gets worse. As memory and cognitive skills further weaken, personality variations may happen, and individuals need extensive care.

The experiences for the individual in this stage affect abilities being restricted, and the person may still communicate through short phrases [3]. This may consist of certain words such as mum, dad, names of children, or call out the word 'help' or 'nurse' continually. Other behaviors may include sufferer singing and whistling, and there might be minimal attempts to make communication verbally. Besides, evident be poor visual and sensory perception.

The person may also have repeated behaviors such as patting, tapping, rocking, and organizing clothing [3]. Often the person might distinguish the faces of family members but have struggle with recalling their names or who they are [4]. At this stage, individuals may [1]:

- Require around-the-clock assistance with daily personal care.
- Lose consciousness of prior experiences as well as of their surroundings.
- Experience variations in physical abilities, including walking, sitting, and swallowing ultimately.
- Have trouble with communicating.
- Become vulnerable to infections, specifically pneumonia.

Unfortunately, the last stage is vegetation in which the person loses most of their abilities and not be able to recognize their loved ones anymore [3]. The individual living with AD may not be capable of carrying out engagement as much during the late stage, but he can still benefit from making interaction in an appropriate way, such as listening to peaceful music or receiving support from their one-beloved.

During this stage, caregivers might want to take advantage of support services, such as hospital care services or nurses, which focus on giving comfort and self-respect at the end of life [1]. Hospital and healthcare services can be of considerable benefit to people in the final stages of AD and their families.

## 1.2 Caregivers Difficulties

Eighty-three percent of the help and assistance given to older adults in the United States comes from family members, friends, or other informal or unpaid caregivers [5]. Almost half of all caregivers (48 percent) offer help to older adults with AD or another type of dementia [6]. In 2018, caregivers of people with AD or other dementias lent an approximated 18.5 billion hours of informal and unpaid assistance, a contribution to the nation estimated at $233.9 billion [1].

The three main explanations caregivers offer care and assistance to an individual with AD, or another dementia are (1) the need to keep a family member or friend at the house (65 percent), (2) proximity to the individual with dementia (48 percent) and (3) the caregiver's perceived responsibility to the individual with dementia (38 percent) [1].

In fact, caregivers usually must assist in performing most of the tasks such as driving, cooking, bathing, and much more to keep the person with AD or other dementias active. The highest priority should be given to maximize the person's abilities and help them to perform tasks independently through choosing the best architecture and designing approaches for the in-home environment.

More in details, we can explain the critical dementia caregiving responsibilities as following [1]:

- Helping with instrumental tasks of daily life, such as household routines, shopping, making snacks and meals, providing transportation means, scheduling for doctor's appointments, handling finances, and legal affairs, and responding phone calls.
- Assisting the person with taking medications appropriately, either through cues or direct management of medications schedule.
- Helping the person accept treatment recommendations for dementia or other medical circumstances.

- Helping with personal tasks of daily life, such as bathing, dressing, cleaning, and feeding and assisting the person walk, move from bed to chair, go to the toilet and manage fecal incontinence.
- Handling behavioral signs of the disease such as aggressive manners, wandering, depressive temper, agitation, anxiety, repetitive activity, and nighttime turbulences.
- Finding and offering encouraging services such as support groups and adult day service programs.
- Making arrangements for paid in-home, nursing center, or assisted living care centers.
- Hiring and supervising others who can provide care for the person with dementia.
- Supposing additional concerns that are not essentially particular activities, such as:
    - Performing general management of getting through the day.
    - Tackling family challenges linked to caring for a relative with AD disease, including making communication with other family members about care strategies, decision-making, and agreements for respite for the primary caregiver.
    - Taking control of other health situations, such as arthritis, diabetes, or cancer.
    - Lending emotional support and feeling of security.

As a result, caregivers can quickly become overwhelmed by the daily challenges facing them. However, despite these difficulties experienced by the person with AD or other dementias and their caregivers, there is much that can be achieved through technology to help make the lives of them better and more fulfilling.

Intelligent assistive tools can be used to improve the quality of life of people with AD to help them feel safe, secure, and valuable and, in turn, create a tremendous meaningful and manageable experience for their families or caregivers. The IA system, which we propose in this thesis, can facilitate the daily life of AD patients in the early stages, alleviate caregivers' worries and reduce the health services cost. We introduce our primary motivation and the main parts of the system in section 1.6.

## 1.3 IoT in Healthcare

Internet of Things (IoT) is a conception indicating a connected array of anything, anyone, anytime, and anywhere. IoT is a big trend in next-generation and future technologies that can influence the whole business field and can be considered as the interconnection of distinctively recognizable smart items and devices within today's internet infrastructure with prolonged benefits [7].

Benefits typically consist of the advanced connectivity of these objects, systems, and services that surpass machine to machine (M2M) scenarios [8]. Thus, presenting automation is feasible in almost every field. IoT introduces appropriate solutions for a wide range of applications, including smart cities, retails, traffic congestion, emergency services, structural physical and mental health, logistics, waste management, safety and security, industrial control systems, and health care.

Medical care and health care signify one of the most appealing application fields for IoT [9]. IoT can give rise to many medical applications such as remote fitness programs, chronic and cognitive diseases, health condition monitoring, and elderly care. Fulfillment with treatment and medication at home and by healthcare providers is another important potential application of IoT in healthcare. Thereby, different medical devices, sensors and actuators, and diagnostic and imaging devices can be considered as smart objects or devices, creating a focal part of the IoT structure.

IoT-based healthcare services are supposed to decrease costs, improve the quality of life, and enhance the user's experience [7]. From the healthcare providers' point of view, the IoT has the potential to lessen device interruption through the remote establishment.

Additionally, the IoT can properly detect optimum times for refilling supplies for different devices for their efficient and continuous operation. Further, the IoT provides for the efficient planning of restricted resources by confirming their best usage and service of a more significant number of patients. Fig. 1.1 demonstrates recent healthcare tendencies [10]. Ease of cost-effective interactions through smooth and safe connectivity across individual patients, healthcare centers and clinics, and healthcare organizations is a significant trend.

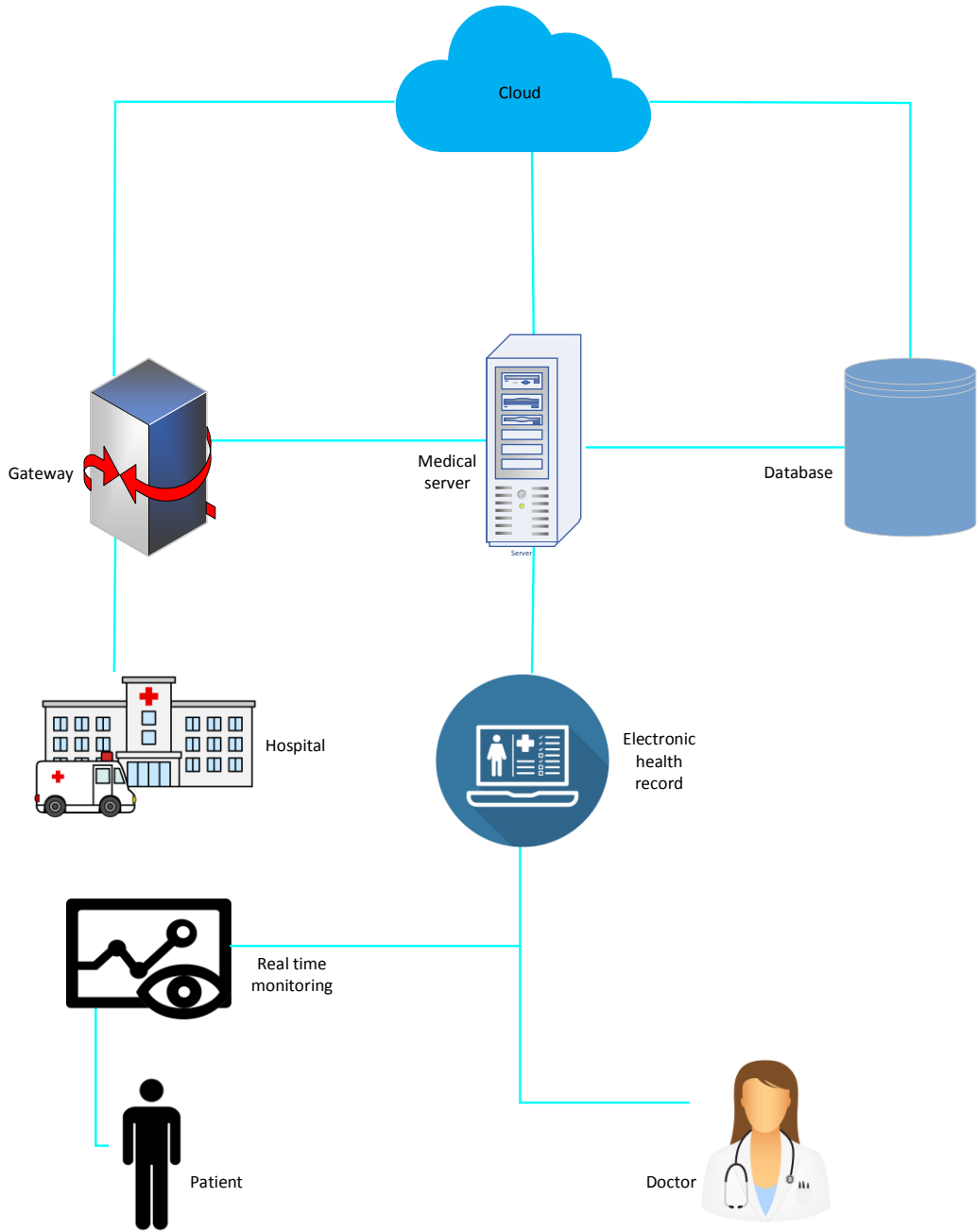

Fig. 1.1 Smart healthcare system.

The IoT can improve several features of quality of life concerning the health condition of individuals. Logistics processes, as well as care-taking in hospitals, can be expanded with regard to identifying everything from blood to joint hips, tools, as well as people, in particular new-born children.

By implementing IoT solutions, various devices can develop more and more integrated effectively within the human body. It is supposed that body area networks are capable of being formed, communicating with emergency services, treating physicians, and caregivers of older adults [11]. This advancement is not new. A current example can already be noticed in the Cardioverter-Defibrillator, which is built into the human heart and is able to decide when to give the patient shocks to defibrillate autonomously and is completely networked such that a physician can follow up on their patient.

Implantable wireless recognizable devices can be used to store health condition records that are expected to protect the patient's life in emergency circumstances. Being capable of having access to the data and health information on these situations, hospitals would instantly realize how to treat the incoming patient. This possibility is incredibly helpful for people with cognitive impairments, diabetes, coronary heart disease, different types of strokes and cancers, chronic obstructive pulmonary disease, seizure disorders, as well as people with complex medical device implants, such as stents, pacemakers, joint replacements, and organ transplants [12].

Moreover, information about patients and health status permits for hospitals to check before administering medication to a patient, whether it is the right drug, at the correct time, and in the correct dosage based on the weight, age, and height of the patient. Drug compatibility can be confirmed by attaching codes to drug packages, thus, constructing a drug knowledge base that alerts people of allergies or interferences with other medication [13].

The IoT as a tool to prevent counterfeiting, and it also improves the health of individuals due to the fact that it can ensure that people are not taking wrong or counterfeited and possibly lethal medications [13]. Improvements and developments of life through the IoT solutions are also achievable with regard to forecasting the weather conditions and to the avoidance of other accidents.

Besides, human life can be spared through early alerts, and people saved from accidents if they are notified of dangerous circumstances. This leads to offering significant benefits to insurances, which can lessen the risks and dangerous conditions. However, determining the precise amount of financial contribution to the development of IoT solutions is the challenging part [7].

On the other hand, the more excellent benefits are the savings stemming from the aids of ambient assisted living systems that allow people to stay at home for a much more extended period of time. Particularly with the predicted aging of society and the associated growth of health care fees, the IoT can become extremely beneficial.

Healthcare systems based on IoT can be applied to a varied array of fields, such as taking care of pediatric and elderly patients, the management of chronic diseases, and the supervision of private health condition and fitness, among others. Based on [7], IoT-based applications are further divided into two categories: single-condition and clustered-condition applications. A single-condition application indicates a particular disease or infirmity, while a clustered-condition application refers to several diseases or circumstances together as a whole. Fig. 1.2 demonstrates this categorization.

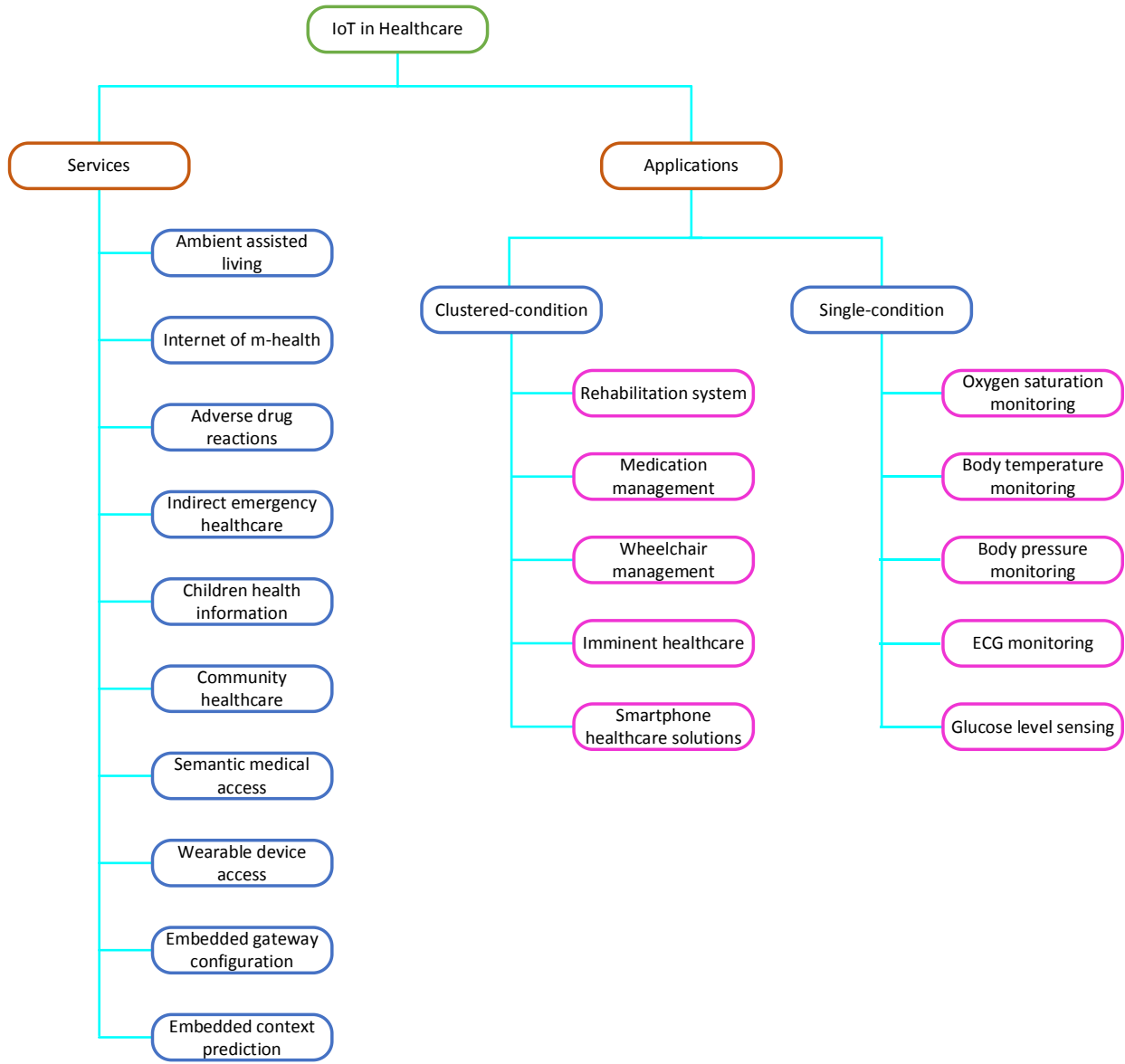

Fig. 1.2 IoT healthcare services and applications [7].

There is also a diverse set of services which is created based on IoT infrastructures, such as ambient assisted living systems. Generally, smart homes and a typical IoT-based medical service can also offer specialized services to elderly individuals or people with cognitive impairments.

Ambient assisted living (AAL) is an IoT platform created based on artificial intelligence that can address the healthcare difficulties of aging and debilitated individuals. The main goal of AAL is to enlarge the independent life of elderly people in their homes and improve their quality of life. Solutions recommended by AAL services can boost the self-confidence of elderly individuals by ensuring more considerable autonomy and offering them more help in completing everyday tasks and activities. Many studies have discussed AAL based on the IoT concept. We represent some examples of this type of IoT service in the next Chapter.

## 1.4 Assistive AR

Augmented Reality (AR) is a useful tool for assisting people in a wide range of concepts such as developing learning approaches, assessing the cognitive state of users, training people to complete complicated tasks, improving rehabilitation, and helping people at their workplace. To define the AR, we can take advantage of the concept that it creates a distinctive experience of the virtual environment, which is known as virtual reality (VR). AR and VR share similar fundamental features, such as employing virtual contexts in the field of view of the user.

However, VR signifies an entirely immersive space, which is created on an entirely synthetic world, the AR enables the user's contact with the real environment, completing it by augmenting the virtual data on the physical world in real-time [14]. One of the most significant aspects that concerns VR is the trouble with replicating a synthetic environment through which it reproduces aspects of the physical world tangibly. Using the fundamental descriptions and definitions of AR potentials can be highlighted three features of AR process:

- AR integrates real and virtual data.
- AR is communicating in real-time.
- AR works and is operated based on the three-dimensional environment [15].

Moreover, AR is a capable tool for many circumstances in which assistance is required, as it provides instructions and feedback for the end-user. While research and development in this field have been mostly driven by industry, AR could also have a considerable impact on those who need assistance and help the most: individuals of all ages with cognitive impairments. In recent years some vital research on applying AR for task completing assistance and learning in the context of this target population has been proposed [14].

Based on the literature, AR can be used to build multimodal sensations that facilitate learning through visualization, animation, and contextualization in learning. Many concepts for learning have been introduced before with diverse types of media and tangible components. While compared to the last most AR solutions have divested haptic feedback, AR recommends an improvement that software-based solutions scale better to large groups and are hence less expensive while supporting visual and acoustic reliability [14]. Some of the most appealing advantages of AR can be seen in different cases, for instance, in the support for animations and storytelling that develop AR solutions feasible for periods of self-learning.

Moreover, AR can offer more help to cognitively impaired people as action assistance, for example, by suggesting precise support where otherwise, the assistance of families or caregivers would be needed. Aid tools for cognitively impaired people or individuals with AD are usually given on paper as a note, for example, in the form of text and informative images. AR technology enables more adaptable instruction designs created based on three-dimensional visualizations or audio.

In this thesis, we mostly focus on finding a new approach that makes use of AR techniques to create solutions for people with mild AD or mild cognitive impairments. To the best of our knowledge, this research area has only been sparsely investigated by research. This might be due to the fact that smart devices useful enough to support high-quality AR are only recently accessible at affordable prices, and enabling technologies, including ARKit and ARCore, have been released as free software development kits (SDKs) to software developers only beginning in 2017 [14]. We go through some of the previous studies which have been carried out in designing AR-based assistive tools in the next Chapter.

## 1.5 Human-Computer Interaction

Research into human-computer interface (HCI) almost regularly includes the participation of human subjects. It is explained as a discipline that has been focused on design, evaluation, and implementation of human-computer interaction [16]. Thus, it is profoundly connected to the communication procedure amongst humans and computers. Because it is related to the process of designing interfaces, websites, or social platforms, sometimes it is called HCI. However, the fundamental of it is a system development of the software that can be efficiently used by humans, for instance, computer users [17]. Therefore, it must enable computer systems to be safe, efficient, easy to use, as well as effective.

In this approach it is closely linked to the elements of usability. It means that the software should be:
- Easy to use,
- Effective to use, and
- Want to be used.

The functionality is determined as a set of software services that are available to the users. Still, the importance of the functionality is visible when it is linked to the functional efficiency to the user. The efficiency is measured by the task completion time and learning time. Hence, these two characteristics are closely related to another [18].

Moreover, the system can be clarified as communication between two subjects. One side is a human side or an expert user, while the other is technology, for example, computers or smart devices. Each of them has its input and output parts. The input of the computer denotes the output of users. The output of the computer represents the input of users. This circumstance is demonstrated in Fig. 1.3.

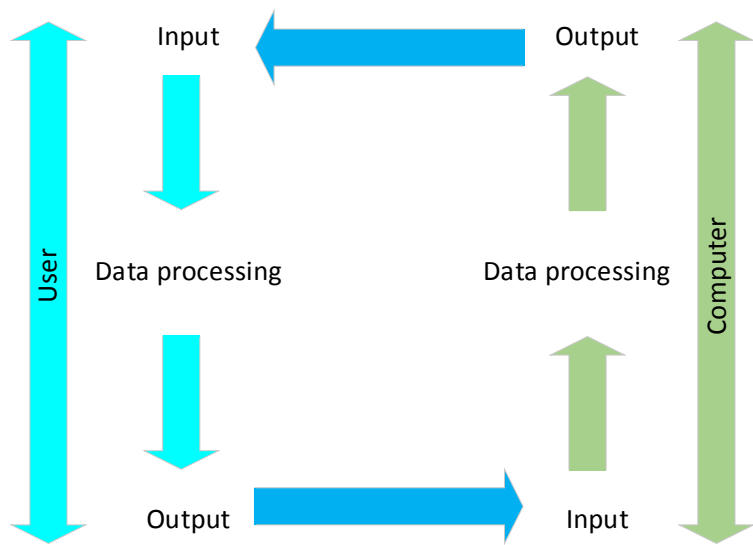

Fig. 1.3 Human-computer interaction data processing flow [17].

For making an effective communication, the system should include the following elements:
- Human-computer interaction,
- Communication based on the agreements on terms defined in the dialogue, and
- Communication created on the agreement on the circumstance of the communication.

If all three of the given criteria are not met, then the communication cannot be effectual [19]. As previously mentioned, the end-user knowledge and cognitive behavior are two main factors in designing assistive tools. Generally, the user significant independent variables are [19]:

- Knowledge,
- Motivation,
- Discretization.

Knowledge refers to the user's knowledge and experience working with computers and smart devices. Motivation is a crucial element. If the user has a high motivation, then he/she inlays more energy to overcome given obstacles such as problems with using the system. In using assistive tools by individuals suffering from mild to moderate AD, they intend to live more independently from their families. Hence, user cooperation is expected to be high. Discretization indicates that the user can satisfactorily decide a part of a system to use for its intentions, which fulfill their requirements [17].

Accordingly, the human-computer interaction should be considered in the process of development as well as the process of designing our IA system and user interfaces. The target users should be included in the process of development and implementation of the system, and the user interface evaluation, which includes the cognitive and behavioral aspects of the target population.

Using an uncomplicated language it must involve careful researching in communication between humans and computers and expanding technological input/output techniques to improve efficiency, effectiveness, and faultless use of their interaction [17]. The primary purpose of human-computer interaction is to implement a computer system that can meet the users' requirements and special needs. In designing an IA system, this goal can be determined according to the final users' cognitive state. To effectively conduct this process, the following is needed to be considered in our IA system:

- Involvement of users in the designing process,
- Integration of diverse scientific disciplines such as computer science, and cognitive science
- Iteration of the process to be improved as much as possible.

To make our IA system user-friendly, many features that influence the HCI should be considered [19]:

- Organizational factors,
- Environmental factors,
- Health and safety factors,
- User,
- Comfort factors,
- User interface,
- Task factors,
- Constraints,
- System functionality, and
- Productivity factors.

We take all of these features into account to create the optimized interaction and design a more profitable IA system for the end-user.

It is important to appropriately choose the best device as an assistive tool for making interaction with the end-user. The comparison between visual devices and contact devices is shown in Table 1.1 [9]. From Table 1.1, it is evident that contact devices require user cooperation more. On the other hand, the vision-based devices do not require user cooperation. However, they are complex to be configured for designing assistive tools and dementia patients.

Table 1.1 Comparison between visual and touch interaction devices

| Criteria | Contact devices | Visual devices |
|---|---|---|
| **User cooperation** | Yes | No |
| **User intrusive** | Yes | No |
| **Precise** | Yes/no | No/yes |
| **Flexible to configure** | Yes | No |
| **Flexibility to use** | No | Yes |
| **Occlusion problem** | No (yes) | Yes |
| **Health issues** | Yes (no) | No |

In our work, we have introduced two wearable devices that the user should be worn. The IA system is accessible by the user while he or she wears a watch and smart glasses (or using smartphones) for the position estimation and making interaction with different objects, respectively.

### 1.5.1 HCI in AR

Some of the augmented reality applications are overlapped with human and computer interaction concepts. For this reason, a little more user-centric approach can give them an advantage of not only being more robust with respect to usability but also be popular in the markets for the same purposes [20]. There exist many HCI guidelines for user interfaces. However, they cannot wholly and quickly be applied to every AR project.

HCI builds standards for product development, emphasizing the focus on unique and robust designs for hardware and software components to facilitate human interaction with the system in terms of gestures, behaviors, and commands. Thereby indicating, implementing HCI can result in more efficient and natural ways of user interaction of any product [21].

Today, in a technical aspect, the significant subfields under the spectrum of HCI are machine learning (to predict and aid for user requirements), recognition of gesture, voice, behavior, and mood/emotions. Its implementation ranges from large-scale laboratories to a small wristwatch and in various fields. Compares to HCI, there are still some challenges in the designing AR-based systems, which are as following [20]:

   a. The wearable is expected to be small, light-weight, consumes less power, and produces less heat giving scalability issues. As fascinating as AR may seem, some individuals may not prefer wearing a headgear all day in public. It's bizarre and socially unbalanced, so that it may give a social rejection. To overcome this challenge, we have suggested Google Glass as an assistive tool because of its minimal heads-up display and light-weight, so it could be used by each mild to moderate AD patient simply.
   b. Since in any AR application, the demand for synchronization in real-time is high for it to give the user, the correct information limited bandwidth gives latency. It also deals with capturing, tracking, and processing, that needs to be accurate, to filter the required and correct data to the user. To keep virtual and real-world in sync, the device needs the quickest data communication and high bandwidth. According to the evaluation result, which is presented in Chapter 5, the network latency in our IA system is satisfactory, and data loss does not occur after performing a series of data transfers.
   c. In some cases, AR can sabotage the privacy of the user and start storing personal preferences and information that facilitate tracking, leading to lawful issues. In the proposed system, collected data can only be shared with physicians to monitor disease progression and treatment response. Furthermore, individuals with AD or memory impairments can control personal data based on their cognitive stage.

### *1.5.2 Technology design issues*

One of the most significant applications of intelligent assistive technology is as a tool to help individuals with dementia through tasks of daily living due to the fact that support for these activities is usually provided by informal, unpaid caregivers or their families. IATs can assist people with AD in one or more functional impairments associated with the disease. For example, they could help individuals to make daily life decisions through memory aids.

Accordingly, the development of AT for recognizing and supporting the patients has become the main research area to maintain the person's ability to independently complete daily tasks as well as reduce the burden experienced by their caregiver [22]. One of the main challenges in IA systems is the acceptance of the end-user. This leads to requiring caregivers or family members to be able to utilize any application to support the patient in its use of the assistive tool.

Heerink *et al.*, in the Almere Model [23], introduce an assessment of the acceptance of assistive technology by older adults (see Fig. 1.4). Based on the data of the study conducted, it results in the conditioning criteria such as attitudes, gender, age, and technological knowledge [24].

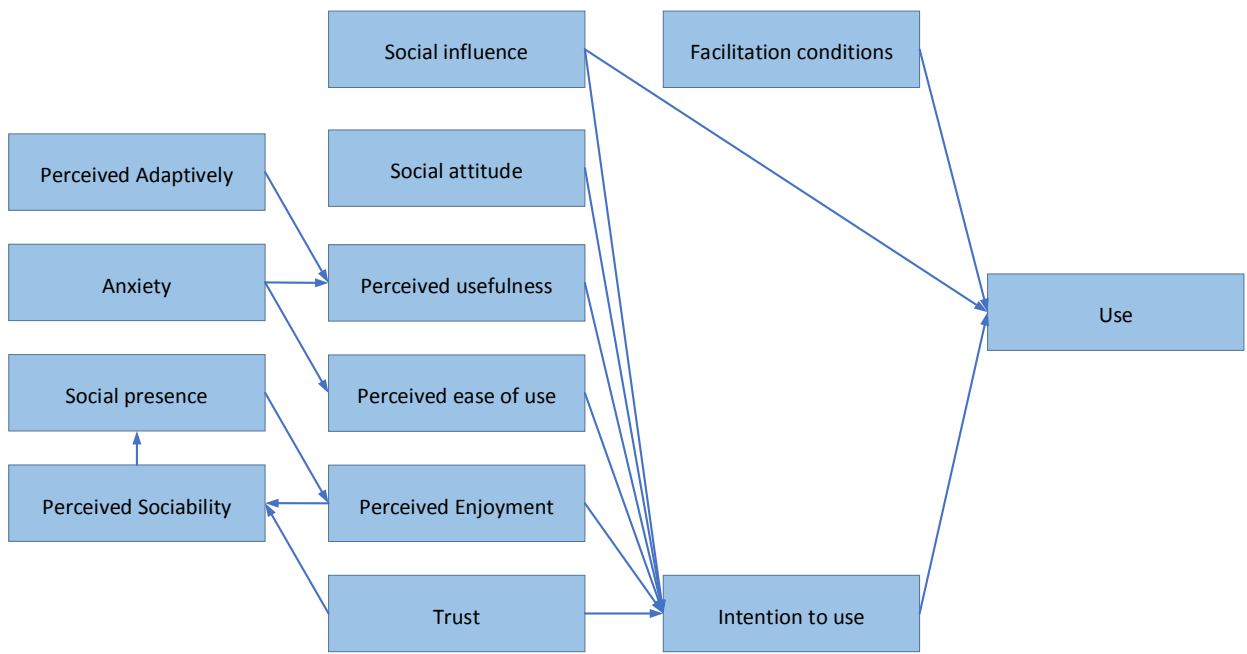

Fig. 1.4 Evaluation model proposed by Heerink and others [23].

These results indicate the fact that IATs can have great acceptance between the end-users. In the designing process of our system, we should also consider a user-centered approach with our design because individuals with dementia needs may change over time, depending on how their dementia is influencing them. For taking this factor into account, we take advantage of the patient's cognitive serious game score several times a day, which we explain in detail in Chapter 4.

In another research, Robillard *et al.* propose the model of "ethical adoption": The deep integration of ethical principles in the design, development, deployment, and use of technology.

Ethical adoption is based on five main elements, supported by experimental evidence: (1) Inclusive participatory design; (2) emotional alignment; (3) adoption models; (4) evaluation of ethical standards; and (5) education and training. To reduce the gap between adoption study, ethics, and practice, a set of 18 practical suggestions created on these ethical adoption pillars is introduced [25] (see Table 1.2).

We also consider the definition of stimuli and their impact on the engagement of the end-user. Interventions that involve objects or activities with meaning specific to the user with dementia are more expected to engage with that person [26].

Table 1.2 Ethical adoption in the context of dementia

| Number | Pillar | Keywords |
| --- | --- | --- |
| 1 | Inclusive participatory design | User engagement |
|  |  | Usability |
|  |  | Culture |
|  |  | Benefit |
|  |  | Customization |
| 2 | Emotional alignment | Emotion |
|  |  | Implicit bias |
| 3 | Adoption modeling | Barriers and facilitators data |
|  |  | Data |
| 4 | Ethics | Consent |
|  |  | Privacy and confidentiality |
|  |  | Conflict of interest |
|  |  | Accuracy |
|  |  | Evidence |
|  |  | Responsible use |
| 5 | Training and education | Intuition |
|  |  | Training courses |
|  |  | Social support |

According to [24], four challenges are also categorized to expedite the evaluation of ethical models in IAT design: (i) Rational method, (ii) uncertainty evaluation process, (iii) evaluation results, and aggregation process, (iv) software tool. For designing intelligent assistive technologies that should operate in the homes of people with dementia, we have to also investigate the required factors and functions for the users and their caregivers.

Caregivers must be able to alter the changing capabilities of an assistive tool for the person with dementia autonomously. Besides, customized settings of the system may be appropriate for the particular patient with dementia but not necessarily another. To achieve these goals, our IA is capable of scaling up to more sensors, actuators, and rules to improve the AD patient's independency according to their daily cognitive state. This can also result in reducing the burden associated with being a caregiver.

Caregivers also want any in-home assistive tools to be recognizable and inconspicuous, indicating an exciting challenge design for intelligent assistive tools designers; obtain sufficient sensor information and offer helpful and instinctive user interfaces without compromising the user's home environment [27].

As we previously mentioned, one of the most critical issues in employing an IA system is users' privacy and security. One of the challenges is how we ensure that the patient with dementia maintain a level of privacy in their home environment and indeed from any collected data shown their health condition. In the proposed IA system, aggregated data can only be shared with physicians to monitor disease progression and treatment response. Furthermore, individuals with AD or memory impairments can control personal data based on their cognitive stage.

However, the experts note that AD patients may try to take off any additional object from their bodies. In our work, the IA system is accessible by the user while he or she wears a watch and smart glasses (or using smartphones) for position estimation and interaction with different objects, respectively. Therefore, the included sensors are naturally carried by the patient from one room to another, so they do not cause them any annoyance. These wearable devices are hands-free, entirely understandable, and light-weight enough to provide non-invasive human-computer interaction.

Moreover, there is no sign of disease, fear, or anxiety, and user-friendly interaction can be achieved easily. To improve users' acceptance of the IA system, AR messages such as voice reminders can be entered by caregivers. This may result in increasing patients' tendency to stay home longer and more independently. People providing care to individuals with dementia are at greater risk than others. IA systems can facilitate the delivery of care and enrich the quality of life for the patients and their families.

The main goal of this section is to define the functionality and general design of our IA system with the purpose of helping mild AD patients and their caregivers at home. In order to achieve this goal and based on the discussion above, there is an essential process of requirement extraction. In general, this process includes the following steps (Fig. 1.5) [28]:

- First, it is compulsory to carry out an analysis of the needs of both AD patients and their caregivers.
- Second, it has to be formed how an IA system can meet these needs: in which scenarios it can be employed and which are the typical features it has to be achieved.
- Third, identifying the useful cases and non-functional characteristics, a study on the possibility of each of them must follow. This study results in finding short-term and long-term necessities.

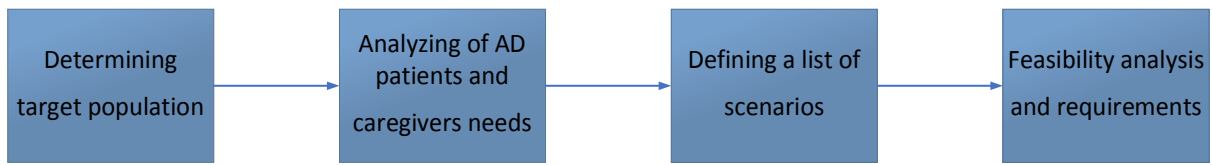

Fig. 1.5 Diagram of the typical steps for requirements extraction.

These steps can be performed in different means. In our case, they are carried out according to the design approach previously mentioned, which is a well-known technique in HCI field. This approach focuses on cooperating with the end-users from the beginning, determining the target population, and designing based on their needs.

Based on Mayer and Zach's Lessons Learned from Participatory Design with and for People with Dementia [29], realistic prototypes facilitate the process of extracting the target user's needs and permit design evaluations with people with dementia. Our proposed system is designed to help patients with mild memory problems or having difficulties in completing everyday tasks based on their needs, and we have tried to create a realistic framework.

Most of the AD patients in the early stage suffer from a disability to sequence tasks correctly. In this stage, individuals with self-awareness of their disease can understand the technology they are using and its expected assistance and risks. The proposed IA system can also be used by people with mild memory impairments. Conversely, individuals diagnosed with AD who are unable to use smartphones and interact with the user interface cannot take advantage of the assistive system.

One of the primary goals of the proposed IA system is to facilitate recall of past experiences of the user, thereby promoting intrapersonal and interpersonal functioning and so improve the user's wellbeing and quality of life. Therefore, the proposed system can remind momentous memories of the individual with dementia and result in a reduction in their anxieties and stress level.

## 1.6 Proposed Model

As we mentioned before, caregivers try to create the best home design to make a simple environment for the patient when he or she is alone at home. For instance, they may use clear covers to make objects more noticeable or leave reminder annotations when the patient is alone at home. Unfortunately, such ways have not been entirely effective [30]. This is the primary motivation for our work.

In response to the increasing number of individuals with dementia and lacking treatment that slows or stops its progression, emerging technologies seem to be a significant factor. Recent developments in Artificial Intelligence (AI), Pervasive and Ubiquitous Computing (PUC), robotics, and IoT combined with new progress in Wireless Sensor Networks (WSNs) and HCI open up new prospects for dementia care with intelligent technologies [31].

Moreover, the pervasive deployment of IATs for dementia could have a disruptive effect on dementia care and reducing caregiver stress. IATs can (i) reduce the burden on public finances throughout the postponement or removal of institutional care, (ii) lessen the psychological burden on formal caregivers and families, (iii) compensate for the severe lack of human caregivers while improving and optimizing the quality of care, and (iv) empower older adults with dementia and thereby enhancing their confidence [31]. The number of researches in this area has been significantly expanding recently [32]; however, few researchers have addressed the development of IATs with possible application into dementia care.

The aim of this study is to provide an IA system for mild AD patients and improve their ability to complete everyday tasks on their own without compromising patients' privacy [33]. To develop the final IA system, we have first designed a task prompting system based on the AR messages without any adaptive decision-making engine to evaluate the general cognitive condition of the user or any positioning data.

The preliminary system has two main parts: the first one is the smartphone or windows application that permits caregivers or family members to monitor a patient's status at home and be notified if the patient is at risk. The second part allows the patient to use a smartphone to identify QR codes in the home environment and receive information related to the tags in the form of audio, text, or three-dimensional image.

In the next step, we have taken advantage of the user's indoor positioning data. The upgraded system has four main components; a location and heading measurement system in the local fog layer, an AR device to make interactions with the AD patient, a supervisory decision-maker to handle the direct and environmental interactions with the patient, and a user interface for family or caregivers to monitor the patient's real-time situation and send reminders once required. The location information is stored in a cloud database. The collected data is then transmitted via the Internet to a fuzzy decision-making engine.

Finally, to increase the intelligence of the system and so adaptation, in this work, we take advantage of an AR-based serious game assessment tool. The collected data of the user's game result is then transmitted via the Internet to a fuzzy decision-making engine. This adaptive decision engine analyzes and compares data to make an appropriate decision based on fuzzy rules set for making interaction with the user.

Our proposed IA system can facilitate the daily life of AD patients in the early stages, alleviate caregivers' worries, and the health services cost. In our IA system, two main rule-bases are essential to creating an adaptive decision-making process and multi-level support.

The first one helps the user to complete the activities in their daily life by showing AR messages or making automatic changes such as actuators activation. The second one allows manual changes after real-time assessment of the user's cognitive state according to the AR game score. In this situation, some of the smart home sensors and reminders can be turned off or disabled according to the fuzzy rules. This feature can improve the patient's self-management and self-care, and it would also slow the progression of the disease.

The system is also capable of scaling up to more sensors, actuators, and rules to enhance the AD patient's independency without compromising privacy.

To the best of our knowledge, despite significant interest, there has not been much work done on the integration of AR, IoT, cloud-based adaptive fuzzy decision-making engine, and indoor positioning technologies to develop an effective IA system. This framework is capable of scaling up to more sensors, actuators, and rules to improve the AD patient's independency.

## 1.7 Approach

More in detail, based upon the above discussion, this thesis presents the following features:

- To implement smart home features, Message Queuing Telemetry Transport (MQTT) protocol is applied. All of the data values, such as indoor positioning data, embedded sensors, and cognitive game scores, can be published to the server in the system's automatic mode. Family or caregivers can also subscribe to each particular topic and override the message for the patients to send reminders if needed.
- In the automatic mode, once an event happens in a specific indoor location, all the relevant fuzzy rules are checked on the server. Thus, following the decision-making algorithm, data values are updated. Both the patient and/or the caregiver receive appropriate notifications.
- To monitor the user's real-time location by Unity game engine and provide their interaction with the objects, the user wears an indoor localization tag. The localization tag sends the positioning information to the monitor module, and the monitor module receives all the published data. The real-time position and orientation data values also form part of the decision-making engine's inputs. Moreover, patterns of the patient's activity can be generated and stored on the server for further analysis.
- The AR messages are sent to the user's smart device while they interact with selected objects, or an event is detected from the sensor data. These messages are the output of the decision-making engine.
- By real-time assessment of the user's cognitive state according to the AR serious game score, the performing mode of the IA system changes to the manual mode. This leads to saving energy of the devices, enhancing accuracy and performance, and providing cognitive enhancement for people with mild AD.

- Simulations are carried out with respect to various performance evaluation parameters such as AR messages response-time and the accuracy and reliability of the decision-making engine to confirm the effectiveness of the proposed system.

## 1.8 Research Scope

In order to provide a focused and precise model as described above, the scope of the research is limited as follows:

- The performance of the system requires more evaluation in AD patients' real life; however, we believe this framework paves the way to develop novel IA systems for further studies. Therefore, this thesis is not concerned with "Ethical Standards" of the system.
- The target population and primary end-users are individuals with mild AD and older adults experiencing memory impairments, who can understand the technology they are using and its expected assistance and risk.
- In this work, we recommend AR messages to users to complete their daily life activities. The main focus of this research does not consider if users are finished or if they are appropriately performed or not a task.
- Although we present adaptive approaches to assist users of the IA system to live more independently, we have not directly studied whether our IA system and AR messages help to reduce cognitive impairments in users, instead leaving this future work.

This work is part of a project that has been supported by the Cognitive Sciences and Technologies Council of Iran. The main goal of the project is to develop a framework that can assist AD patients to complete their everyday tasks, provide a real-time platform for monitoring daily activities through indoor positioning techniques, and evaluate the patient's cognitive state by designing an AR-based serious game. In this thesis, we have taken advantage of indoor positioning data and the AR game score to assess our IA system performance.

## 1.9 Organization

The rest of this thesis is organized as follows; the related work is introduced in Chapter 2. The system design, implementation, and evaluation are presented in Chapter 3, 4, and 5, respectively. After that, the potential improvement of the system and conclusion are discussed in Chapter 6.

## 1.10 Summary

In this Chapter, we have presented an overview of our research topic and the main concepts of the proposed model. In the next Chapters, we go into more detail as we first describe related work to help illustrate the research gap, before detailing the study and evaluation of our IA system.

# Chapter 2 - Background and Related Work

In this Chapter, we first review the state-of-art of Augmented Reality (AR) and Internet of Things (IoT) rehabilitation approaches and then discuss the architectures of intelligent assistive (IA) systems and their mechanisms for dementia care.

## 2.1 Wearable AR

Visual lifelogging [34] captures real-life images through a camera embedded in glass-like devices to create a personal photo or video-based memory prosthetics. The goal of lifelogging is to support people's self-awareness and self-management for memory recall [35]. SenseCam [36] is one of the first feasible lifelogging cameras. It comprises a sensor-enhanced image capture device worn on the neck to record imaged passively and context information.

Several studies have shown SenseCam's benefits for supporting the recall of episodic memories [37],[38], both as remembering and knowing about the past. For instance, Lee and Dey [39] have shown that SenseCam photos capture four distinct types of cues: people, objects, places, and actions. Kalnikaite *et al*. [40] have found that visual cues provide better memory recall than other cues. Furthermore, other types of cues have been investigated to assist memory recall. Examples include geo-locational cues [40] and audio cues [41], such as ambient sound recording.

A considerable amount of work has focused on the design of wearable augmented reality applications [42]. For example, iShadow [43] designs a real-time eye-tracking system for Google Glass. This system improves the execution efficiency by moving redundant pixels in the eye images but is still limited by the poor hardware performance. Other Google Glass applications, such as Chroma [44] and Expression [45], are respectively developed to help color blind and visually impaired people. These applications usually use libraries, such as Optical Character Recognition (OCR), Motion Classifier, and Object Recognition, where most of the functions are computational exhaustive [46].

Recently researchers have considered that AR can be used as a cognitive aid tool to help individuals suffering from AD, specifically in the early stages [22]. The adoption of virtual reality (VR) technologies could provide a cost-effective, flexible, and comprehensive solution for realizing complex cognitive training environments [47].

On the other hand, AR can remove some of the boundaries compared to VR in which the virtual layer can be available even if the user is making moves. This also presents opportunities for development within AR-based assistive tools, and there has been a range of interventions with people living with dementia (PwD) in this area [22].

In the early stages, they may experience memory failures, such as forgetting how to deal with daily tasks or the location of mostly used objects. In cognitive psychology, memory is the process by which we encode, store, and retrieve information [48].

The ARCoach system has been designed by researchers at Chung Yuan Christian University, which was based on using a personal computer with an external web camera to add information to a real object [49]. They studied the possibility of training three people with cognitive impairments using an AR-based task prompting system through which it provided picture cues, identified incorrect task steps on the fly, and helped users make corrections. Based on a multiple baseline design, the data showed that the participants considerably increased their target response, which improved their vocational job skills during the intervention phases and enabled them to maintain the acquired job skills after the intervention [49].

Furthermore, other types of AR-based systems have been investigated to assist memory recall. The Ambient aNnotation System (ANS) [50], was included two main sections; the first one permitted the caregiver to create new AR tags by choosing objects or locations and then annotating them with new data that could assist the person with AD. The second one operated on a smartphone, was used by a person with AD and warned the persistence of tags in the environment.

Bolestsis and McCallum [51] have developed an AR-based gaming experience utilizing Cognitive Augmented Reality Cubes (CogARC). The physical cubes were watched through a tablet fixed in a physical angle bracket. Users could then notice different games that encouraged pattern and color matching tasks with the cubes. The design repetitions of the application experienced some interaction concerns such as cube recognition and an interval between the real movement of the physical item and the screen display. Some of these problems were overcome by exploiting Unity Game Engine and an extension known as Vuforia Unity Extension.

An impressive integration of AR and Bluetooth beacons have been developed to make a system known as a Memory Palace [52]. This system inspires caregivers and PwD to attach beacons to frequently used objects in the house environment. They then make use of a mobile application to attach media to the object of interest, including pictures, audio, and video. The PwD can then employ the smartphone and application to walk around the home with memory replays as they go around a tagged item. The implementation of the system required caregivers to be involved in the construction of the object memories, and the PwD found the application more beneficial as an aid to speak about a specific memory or event.

An AR-based system, Kognit, proposes a future platform that operates on bespoke depth-sensing AR glasses to focus on body sensor interpretation, activity recognition and as a proactive episodic memory rehabilitation, including assistance and monitoring of daily activities such as the taking of medication and drinking water [53]. This particular system extends AR with Artificial Intelligence in order to provide task prompts at appropriate points near the house and outdoors.

The above future platform and feature support are extended in a novel design study entitled "The adventures of Gladys in (an augmented) wonderland" [22]. This research introduces a set of design standards to address the challenges that individuals with dementia experience within public places. The proposal is to promote the development of an AR device that also has depth-sensing technologies to provide context-sensitive data when a PwD is walking in a public area [22]. The aim of this study is to provide some helping means for AD patients and their family members through the AR-based interaction and other methods. Details are presented in the sections of system design and implementation.

### *2.1.1 Levels of AR*

In the implementation process of AR, tracking is the first and primary step. It identifies the position and orientation of the end-user, including their point of view, expediently represented by a virtual camera. This step can be based on various types of methods of tracking, and it depends on the technology, hardware, and software utilized in this process. The input data can be gathered from different types of sensors (magnetic, infrared, inertial, and GPS) or taken from the camera. Within the subject of Augmented Reality applications, there is a classification of levels depending on the complication of the tracking technology employed [15]:

- Level 0: systems that suggest hyperlinks in the physical space. This level would remain out of the description of Augmented Reality from the key features recommended by Azuma, lacking to fulfill the provision of tridimensional content. It is created based on the technology of QR codes and bar codes lecture, recognized by the smartphone camera that automatically retrieves a web page through the predefined smartphone web browser.

- Level 1: is based on the utilization of QR/AR markers and natural feature tracking. It relies on systems of identification of two-dimensional markers, with the capability of displaying three-dimensional objects. This level lets the insertion of any kind of virtual content in the physical environment.
- Level 2: focuses on tracking by mixing GPS data with compass information or other types of sensors, such as inclinometers or accelerometers.
- Level 3: indicates applications that use three-dimensional tracking focused on head-mounted display devices. According to Rice, it permits us to move away from the conception of a monitor or screen, converting Augmented Reality into what it describes the Augmented Vision and provides global experience instantaneously becomes more applicable, appropriate, and personal.
- Level 4: is defined based on the use of retinal stimulation devices or other devices, with the ability to link to the nervous system of the user's brain, producing the combining virtual contents with the physical space observed by the user. This level has not been achieved today, but we can express that is much related to virtual reality than to augmented reality.

## 2.2 Ambient Assisted Living

Recently, the development of IoT services has drawn considerable interest in the scientific association. However, few researchers have addressed the development of a model designed for seniors with chronic diseases and particular demands, such as dementia and AD. Ambient Assisted Living (AAL) is an IoT-based service that supports the care of elderly or debilitated patients [54].

AAL not only contributes a safer environment but also provides independency and encourages the user to be more physically and mentally active. According to [55] AAL has achieved the first rank among various IoT applications in health care and has had the potential to be a reliable opportunity in the near future.

A variety of researches has been carried out on AAL. In a research work called Autominder, researchers designed a system established on a mobile robot [56]. This system assists older people to adapt to their cognitive impairment and improve the performance of their daily activities, thereby potentially enabling them to remain in their own homes longer. The system achieved this goal by providing adaptive, personalized reminders of (basic, instrumental, and extended) activities of daily living.

The smart system proposed in [57], has addressed the needs of people suffering from dementia to recognize their families and have better interaction and quality of life, without the need for training courses. The system suggests a ubiquitous recognition system, using smart devices such as smartphones or smart wristbands. When a familiar person is recognized in the home environment, then a sound is replayed on the smart speaker to stimulate the patient's memory.

The primary step in designing an AAL system is to determine types of help. A study was carried out in Portugal on "Adherence to Medication Regimen in Elderly" [58] implied that the majority of the elderly people want to be externally helped for managing medication. The study was very detailed, with a population of elderly people. It stated that 60.5% of total patients gave forgetfulness as the reason for non-adherence to medication. While 24.4% stated, they did not have medication with them at the time of intake. It was also emphasized that interventions, such as providing counseling on pharmaceutical drugs and their control) are effective in increasing adherence to medication.

From Table 2.1, we can deduce the majority of elders need help in medication control [59]. This table states one significant aspect where AAL systems, automated or semi-automated, are crucial for elderly patients. Extensive work has been carried out for AAL systems. To make AAL a reality, technologies such as smart homes, e-textile, assistive robotics, wearable, and mobile sensors (Table 2.1) are heavily explored.

Firstly, mobile and wearable sensors play an essential role in our daily lives somehow. Multiple types of sensors are being used to monitor our environment. For monitoring mobility and any kind of activity, sensors such as gyroscopes and accelerometer are available in our smartphones and smartwatches.

Table 2.1 Necessities of elderly patients

| Type of Help | Total Number of Elders | Percentage |
|---|---|---|
| Manage medication | 119 | 36.1% |
| Get info on medication | 63 | 19.2% |
| Explain medication | 44 | 13.3% |
| Interpret medication regime | 26 | 7.9% |
| Monitor medication regime | 20 | 6.1% |
| Remind dosage time | 19 | 5.8% |
| Refilling medicine dispenser | 10 | 3.0% |
| Label reading | 8 | 2.4% |
| Monetary help | 7 | 2.1% |
| Getting drugs out of the box | 4 | 1.2% |

The extensive efforts of researchers have given fruitful results in epidermal electronics. To monitor health signals, largely non-invasive sensors, such as a pulse oximeter, glucometer, GVR, ECG (see Table 2.2) [59] can be put in use. E-textiles or smart clothes are bidding to provide us completely non-invasive real-time health monitoring.

In smart clothes, sensors can be embedded over the garment, or it can be woven in the fabric itself. Fabric-based sensing is the ultimate goal to achieve efficient and accurate Body Sensor Network functionalities [60].

In this thesis, we have taken advantage of localization data by representing the data from the tag worn by the user and MQTT (Message Queuing Telemetry Transport) messaging protocol. Moreover, to ensure that the device is worn, Heart Pulse Sensor (SEN-11574) is used.

Table 2.2 Various wearable sensors available for vitals measurement

| Functionality | Sensor |
|---|---|
| Orientation | GVR(Galvanic Skin Response) |
| Acceleration | Accelerometer |
| Blood Glucose | Glucometer |
| Body Temperature | Thermal |
| SPO2 (Blood Oxygen) | Pulse Oximeter |
| Cardiac Activity | ECG (Electrocardiography) |
| Perspiration | GVR(Galvanic Skin Response) |

## 2.3 Fuzzy-Based Healthcare

Decision support in healthcare applications is typically provided by means of declarative logical frameworks able to represent clinical guidelines, to provide smart and case-specific advice to individuals and doctors. The broader set of medical and health care decision-making have several kinds of variables that are related to specific decisions, including environmental, social, physical, organizational social, and professional criteria [61].

Healthcare and medical industries play a significant role in improving living standards and human life quality. In many situations, making-decisions in the healthcare sector is difficult due to its complexity and critical effects on the life quality of individuals. Such decisions are mainly aimed to maximize the health benefits, minimize the health risks, increase patient choice, maximize physical involvement, satisfy the resources and other limitations, and be fair and equitable.

Several decision-making techniques are supporting health care and medical decision-making processes. Among these declarative logical frameworks, fuzzy logic has been profitably used for modeling uncertainty and vagueness of clinical guidelines in the form of if-then rules, aimed at reproducing the reasoning process followed by doctors when health data are evaluated [62-65].

The usage of fuzzy logic to build a clinical decision support solution foresees that, on the one hand, the main concepts characterizing the medical domain are represented through a fuzzy linguistic model composed of linguistic variables, linguistic terms, and membership functions whose shapes encode uncertainty and vagueness [66]. Moreover, researchers have applied fuzzy method to develop a decision support system for monitoring patients who live alone at their homes.

In [67], the authors introduced a fuzzy logic based home healthcare system for chronic heart disease patients, who are in stable conditions, for out-of-hospital follow-up and monitoring purposes. The proposed system can provide an innovative, timely resource and a supplement for the existing healthcare systems helping experts to treat efficiently to cardiac patients who lived alone at their homes.

Recently, applying fuzzy decision-making models through IoT implementation has been introduced for monitoring elderly people [68]. A fuzzy probability transformation based system has been proposed to make sure that the elderly are secure in their houses. However, this study was limited to learn the bathroom usage model of an older person. If an unusual pattern is recognized, the system could notify the caregivers instantly. The proposed system utilized the idea of the IoT to collect, transmit, and present data to the stakeholders. We discuss the corresponding adaptive fuzzy model principles in Chapter 3 and the details of implementation in Chapter 4.

## 2.4 Intelligent Assistive Technologies

Active assistive technologies can provide advanced and proactive support to an individual while he or she performs activities. Such systems typically include modules based on artificial intelligence and can be denoted intelligent assistive systems. Since personalization is the key to providing optimal support, such technology needs to include the following functions: (i) assessing the individuals current ability to perform and complete an activity in a way that is satisfactory to the individual in a particular situation, and (ii) deciding on the extent and nature of the assistance to be provided in a specific situation [69].

A significant trend is the combination of synergistic efforts between information science, design science, and behavioral science to design and develop AI systems. In response to the increasing number of people with dementia and lacking treatment that slows or stops its progression, emerging technologies seem to be a significant factor. Moreover, the pervasive deployment of intelligent assistive technologies for dementia could have a disruptive effect on dementia care and reducing caregiver stress. The number of researches in this area has been significantly expanding recently [32]; however, few researchers have addressed the development of IATs with possible application into dementia care.

IATs are currently being implemented into dementia care for different purposes (Fig. 2.1). The results of [31] revealed that the most common application of IATs in dementia care is supporting older adults with dementia in the completion of ADLs (n = 148) such as eating, bathing, dressing, toileting, and continence. These results reflect the oft-stated wish of elders with dementia to enhance their independent living and the need for healthcare systems to delay or obviate institutional care, hence age-in-place [71].

With 100 systems, monitoring is the second most common application. Monitoring is a crucial function for enhancing a person's safety as it allows identifying patterns of abnormal behavior, prompting responses from caregivers in case of danger, and collecting data for other connected applications. Physical (n = 88) and cognitive assistance (n = 85) also compose a substantial proportion of the general applications of IAT to dementia care.

Cognitive assistants are intelligent devices capable of supporting or augmenting cognitive functions in cognitively impaired individuals, functioning as external cognitive processors. These include memory aids and other cognitive orthotics [70].

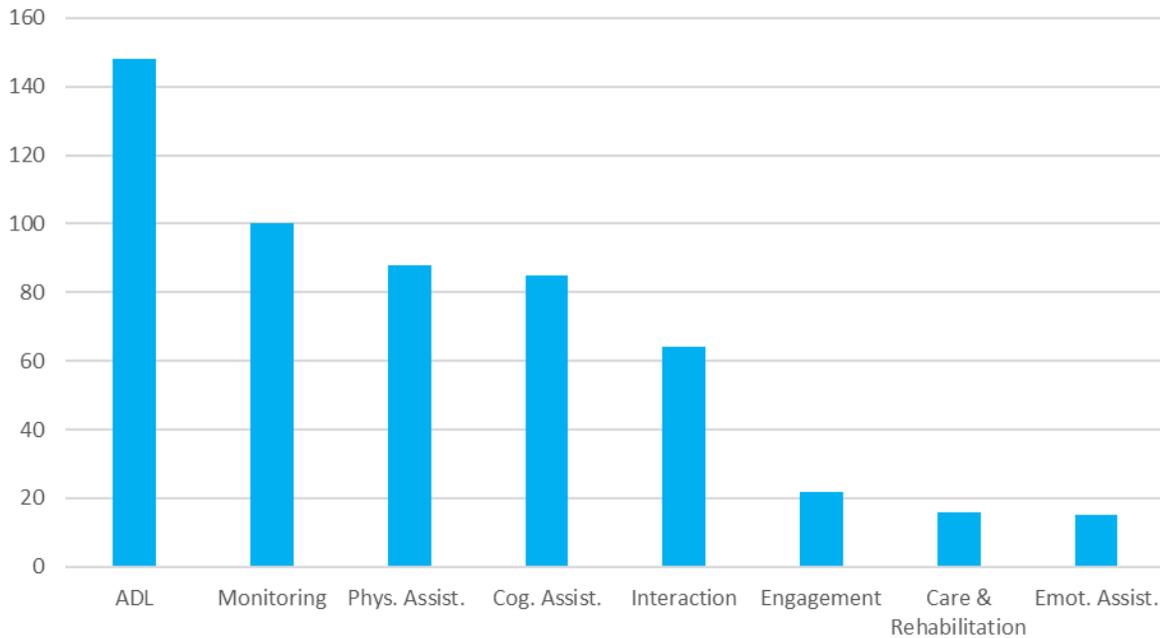

Fig. 2.1 Most common IAT applications.

An example is the COGKNOWDay Navigator, a digital prosthetics to support persons with mild dementia in their daily lives, with memory, social contacts, daily activities, and safety [72].

In contrast, physical assistants compensate for motor and locomotive deficits associated with dementia-related disability. An example is a MOBOT, an intelligent physical assistant that aims to support elderly patients with mobility disabilities during gait and sit-to-stand transfer [73]. Emotional support and assistance represent a smaller (n = 15) but rapidly developing portion of the IAT application. Finally, promoting interaction (n = 64) and engagement (n = 22), as well as facilitating care and rehabilitation, complete the picture of possible applications enabled by current IATs for dementia.

As different forms of dementia and different stages of the disease progression may present inherently specific symptoms and care requirements [31]. While all proposed IATs could assist or compensate for one or more functional impairments associated with AD or other dementias, most of them were not exclusively designed for people with AD or other types of dementia but also the general elderly population with neurocognitive disability [31].

A significantly smaller portion of intelligent systems is more selectively designed to primarily target people with dementia, or the specific cognitive, physical, and behavioral symptoms of people with AD. The MED-AR system [69] presents to present a research methodology towards developing intelligent assistive systems that utilize augmented reality for tracking and distributing prescribed medicines for older adults in a home health care scenario. Their multi-faceted methodology brings together first-order logical proofs, activity theory, and targeted user studies.

More in detail, IATs can (i) reduce the burden on public finances throughout the postponement or removal of institutional care, (ii) lessen the psychological burden on formal caregivers and families, (iii) compensate for the severe lack of human caregivers while improving and optimizing the quality of care, and (iv) empower older adults with dementia and thereby enhancing their confidence [31]. In this thesis, we aim to provide an intelligent assistive system for mild AD patients according to their cognitive state during daily life and improve their ability to complete everyday tasks on their own.

# Chapter 3 - System Design

In this Chapter, we first describe the motivating scenario and system requirements that must be considered in order to implement different scenarios. After that, we introduce the modules included in our system and give an overview of the system's architecture.

## 3.1 Scenarios of Use

In this section, we first describe the motivating scenarios and system requirements that must be satisfied in order to implement the final scenarios.

### 3.1.1 Experiences of a person with dementia

As the disease progresses, the person with AD might become withdrawn and appear to lack interest and motivation, which can affect their relationships, mood and create problems when interacting with other people or family members. Concentration may also be a severe challenge, as the person might have problems focusing on activities and have difficulties with problem-solving tasks.

Challenges the individuals with AD face in expressing language may increase. A person with dementia might appear to do the same thing over and over again repetitively, such as folding and unfolding items of clothing, or frequently touching objects. The individual may no longer know their parents, husband or wife, close friends as the personality might reshape because of the deterioration appearing in the moderate to severe.

In the middle to severe stages, they might show behaviors that family might find challenging to deal with, such as what would be known as inappropriate use of the shower, language disorders that the family may find offensive and unsuitable, or maybe eating or swallowing items that are not considered as food items [4,][30].

The episodic memory seems to be the most involved in AD and dementia. This leads to memory impairment, and a person with dementia may have difficulties in remembering sequential tasks, such as the detailed process involved in cooking various foods. As the person with dementia loses the ability to perform a rapid movement, they are unable to perform tasks rapidly and might become aggravated if rushed by family members.

Consequently, the family should let people with dementia time achieve their goals and complete the task correctly, thereby reducing stress level and anxiety. The person with dementia is not responsible for such actions because these behaviors are a consequence of brain destruction and their cognitive impairments.

People with dementia often face challenges in finding their direction around new or even some familiar areas, a term known as spatial orientation and navigation disorder. They might identify the voice of their family, but not their face or their personal information. The person may also have problems with attending to one task at a specific time, and completing complicated daily activities or tasks can be frustrating for a person with dementia. They might also have difficulties in naming items (anomia), locating words when writing (agraphia), reading (alexia), drawing things, distinguishing from left to right, mathematics (dyscalculia), being aware of particular parts of the body or environment, visual concentration problems, and eye and hand coordination.

There are other reported problems by caregivers, such as identifying colors, seeing objects correctly, recognizing and drawing items, the inability to distinguish the movement of objects, and difficulties with reading and writing skills.

The person might have problems with understanding the use of objects, such as not deciding what to do with clothing, how to use the toilet or shower. For example, the family might take a person with dementia to the toilet and want them to use it; the person may know it as a toilet but does not identify what it is used for, and thereby they face problems in performing the task. In these cases, the main focus should be on expanding what the person with dementia is still able to do and not try to get them to do parts of their care that they may no longer be able to reach.

### 3.1.2 Proposed scenarios

To define a list of usage scenarios for our IA system, we have taken advantage of the experts' discussion about AD patients and their caregivers' concerns in [28] as following:

- Safety:

One of the main challenges of the caregivers is to keep the AD patient monitored. This may result in AD patients becoming disoriented and can enter dangerous areas (for example, the balcony), may fall, or even leave their homes. Caregivers cannot be continuously watching the patients, so that an assistive tool can be beneficial in this case.

- Personal Assistance:

Psychologists indicate that the engagement of patients with AD to routine activities is essential, and an assistive tool which have some functionalities of a personal assistant could be beneficial, mainly because AD patients have problems with short term memory and recent events.

- Entertainment:

This feature is mentioned mostly by caregivers, due to the fact that AD patients intend to require much time from them. If an assistive tool can provide some entertainment for the patient, the caregiver can have more free time to spend on their own needs, which is generally what they want.

- Stimulation:

Finally, the therapists state that some physical and mental motivation is helpful to perform to slow down the progression of the disease. However, they also mention that these activities must be carefully designed for each patient based on their general mental and physical condition and monitored by an expert.

We are considering the possibility of using our IA system in the daily life of an AD patient. Here, we have described some distributed mechatronics scenarios, evaluation of which are presented in the next Chapter. First, the scenarios are defined based on the QR codes and without using real-time indoor positioning data of the patient. The following sample scenarios describe how the system would be used as a memory aid:

- When a relay actuator is activated by the smartphone application, the drawer lock is opened, and the user can notice an image of pills on the screen. Besides, the user receives an audio message as a reminder for their medication schedule.
- When the QR tag attached on the bedroom door is recognized by the camera, the user notices their family picture and get some personal information about them as an audio message.
- If the user enters the kitchen and the PIR detects this event, then they can see the picture of the dishes. Furthermore, an audio alarm is played about the locations.

- If the temperature sensor detects that the indoor weather is cold, then a text message is shown to the user on the screen. If this message is confirmed by the user, then the heater can be turned on.
- If the flame sensor detects a flame or fire in the oven, then the relay actuator is activated, and the oven can be turned off.

In the next step of the designing IA, we have used indoor localization system real-time data instead of the QR codes. According to AD patients' needs and daily life activities:

- If (rain status is yes) and (distance from object1 is near) and (heading angle is small), then (image message is picture3) and (voice message is audio3).
- If (distance from object2 is near) and (heading angle is small), then (image message is picture2) and (voice message is audio2).
- If (plants humidity is very dry), then (text message number is text 1).
- If (distance from object4 is near) and (heading angle is small), then (image message is picture 6) and (voice message is audio6).
- If (time is late afternoon) and (distance from object5 is near) and (heading angle is small), then (image message is picture4) and (voice message is audio4).
- If (time is evening), then (image message is picture 5) and (voice message is audio 5).
- If (flame status is yes), then (relay status is yes) and (voice message is audio7).
- If (gas status is yes), then (relay status is yes) and (image message is picture8).
- If (temperature is hot), then (text message is text3).
- If (temperature is cold), then (text message is text4).
- If (distance from object3 is near), then (voice message is audio8).
- If (distance from object1 is near) and (heading angle is small), then (image message is picture9) and (voice message is audio9).

In our experimental tests, each of these messages has been defined based on the AD patients' real-life scenarios. We describe some of them in the next Chapter. Finally, we have used AR-based serious game results to make the system more intelligent. This part presents a description of the usage scenarios categorized according to the final and completed model of the proposed IA system. All of the scenarios are indicated in Table 3.1. These scenarios are precisely defined to choose the most proper ones for the first experiment with AD patients [19, 74].

Table 3.1 Scenarios of use

| Areas | User's location | Command type |
|---|---|---|
| Entertainment/cognitive assessment | Not specific | Launching AR game |
| Medication schedule/personal assistance | Not specific | Reminder |
| Leaving the stove on/safety | Not specific | Alert |
| Trouble with cooking/ personal assistance | Kitchen | Reminder |
| Putting on clothes/ personal assistance | Bedroom | Reminder |
| Mealtime | Not specific | Reminder |
| Lack of movement/stimulation | Not specific | Alert |

According to [30], AD patients often lose their ability to sequence in activities such as putting on clothes or cooking. When the user walks in different rooms and tries to interact with other objects, the IA system tries to predict the patient's target based on the distance between the user and predefined objects and other inputs.

Lack of movement is also an important issue that should be considered in the patients' daily life. The system measures the user's movement to check this factor by localization tag worn by the patient. If he moves between 3.5 to 4.5 hours per day, the IA system sends a reminder to the patient and their caregiver. In another scenario, once the patient is near the drawer of a wardrobe or closet in the bedroom, the IA system can suggest to them what to wear. This approach can help the patient make everyday decisions before they get nervous. To customize the IA system, family or caregivers can determine some image messages, including the patient's family pictures, first house, and childhood events to be sent to the patient when they are alone at home.

## 3.2 Hardware requirements

### 3.2.1 Wemos D1 board

As shown in Fig. 3.1, the Wemos D1 is a mini Wi-Fi IoT module based on ESP-8266EX microcontroller and provides 4Mb flash memory. Its nine GPIO pins make this board suitable for a large IoT target audience. It is an excellent MCU that can be programmed with both Arduino IDE or Nodemcu. It has a micro USB for auto programming, and it can also be programmed using Over-the-Air (OTA). One side of the board features the ESP-8266 module, and the other side has CH340 serial to USB chip, and a reset button, and a PCB antenna. It is even compatible with Android and iPhone. It can use an external Antenna and a CP2104 USB to UART IC.

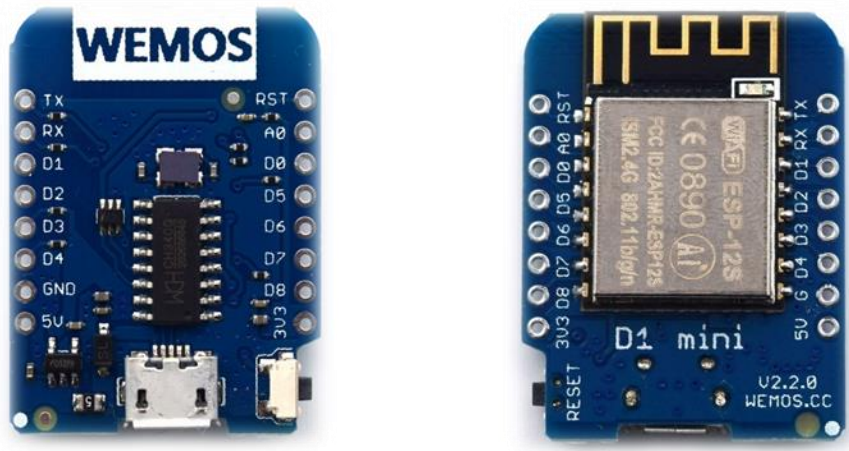

Fig. 3.1 Wemos WiFi board.

ESP-8266 boards provide communication abilities to the sensors through a built-in Wi-Fi interface. Thus, through a Wi-Fi home Access Point (AP), sensors are allowed to connect directly to the broker and publish messages whenever an event is detected via MQTT protocol.

In our case, sensors are connected to ESP8266 in order to publish messages on topics such as temperature and flame data values, and family from a smartphone or a computer could read these values by subscribing to this topic. When the smartphone is offline, all the notifications can be stored by the server and then sent again when the smartphone is connected to the Internet.

The technical specification of this module is shown in below Table 3.2. In this proposed system, the Wi-Fi module of Wemos board is connected to the station having a SSID and password, which access the cloud services and retrieve the data. This system is cost-effective as it is not using any kind of sensor for retrieving data.

Table 3.2 Technical specification of Wemos

| Microcontroller | ESP-8266-FX |
|---|---|
| Operating Voltage | 3.3 Volts |
| Digital I/O Pins | 11 |
| Analog Inputs Pins | 1 |
| Clock Speed | 80MHz/160MHz |
| Flash memory | 4Mb |

We use several Wemos D1 mini with a DHT shield to send the data values on topics such as humidity and motions to a MQTT topic and apply HiveMQ as a MQTT messages broker. We describe the implementation process more in detail in the next Chapter.

### 3.2.2 IoT objects

IoT devices are divided into two parts: objects and Google Glass or smartphones. Each object can operate as one of the following groups:
- sensors
- actuators

The objects equipped with a sensor communicate and then exchange data according to the type of sensor communication. These sensors record the patient's environmental information and wait for a request from Google Glass or smartphone. Similarly, any object may work as an actuator, such as activating a relay to turn a device on and off or controlling the speed of an engine. Table 3.3 shows the sensors and actuators designed for Wemos boards in our work.

Table 3.3 Sensors and actuators

| PIR sensor | Relay actuator | Temperature and humidity sensor |
|---|---|---|
| 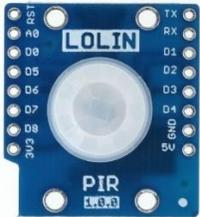 | 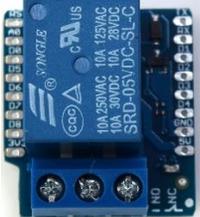 | 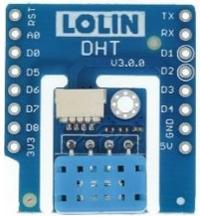 |
| **Buzzer sensor** | **LCD module** | **Rain sensor** |
| 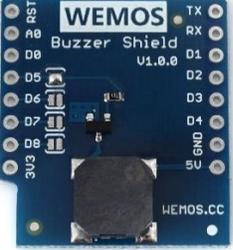 | 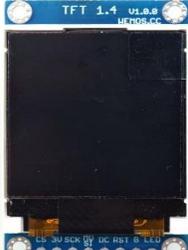 | 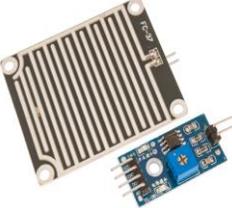 |
| **Pressure sensor** | **Gas sensor** | **Flame sensor** |
| 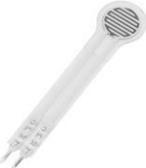 | 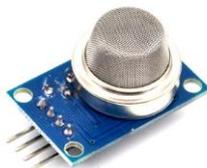 | 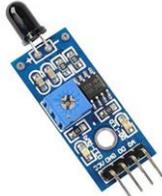 |

## *3.2.3 AR Glasses*

We have chosen Google Glass for AR implementation because of its minimal heads-up display and lightweight, so it could be used by each mild to moderate AD patient thoroughly after short-term training that does not require any special skills. Google Glass has a straightforward utility in the clinical setting, for example, surgery, assistive device for people with Parkinson's, and patient monitoring. It also has applications in robotics and remote control of a mobile robot.

In comparing to other AR glasses, it gives patients access to information in the simplest possible way that does not distract their daily life, and it makes enough essential virtual information while the patient is interacting with the physical world. Moreover, if the patient is a glass-wearer, real glasses can have Google Glass screwed onto them. For example, Microsoft HoloLens is too bulky and distracting for the patient, so it is not appropriate for this purpose. We are aware that our application could run on Google Glass because the glass has the same operating system as the Android smartphone. Fig. 3.2 shows Google Glass and its features.

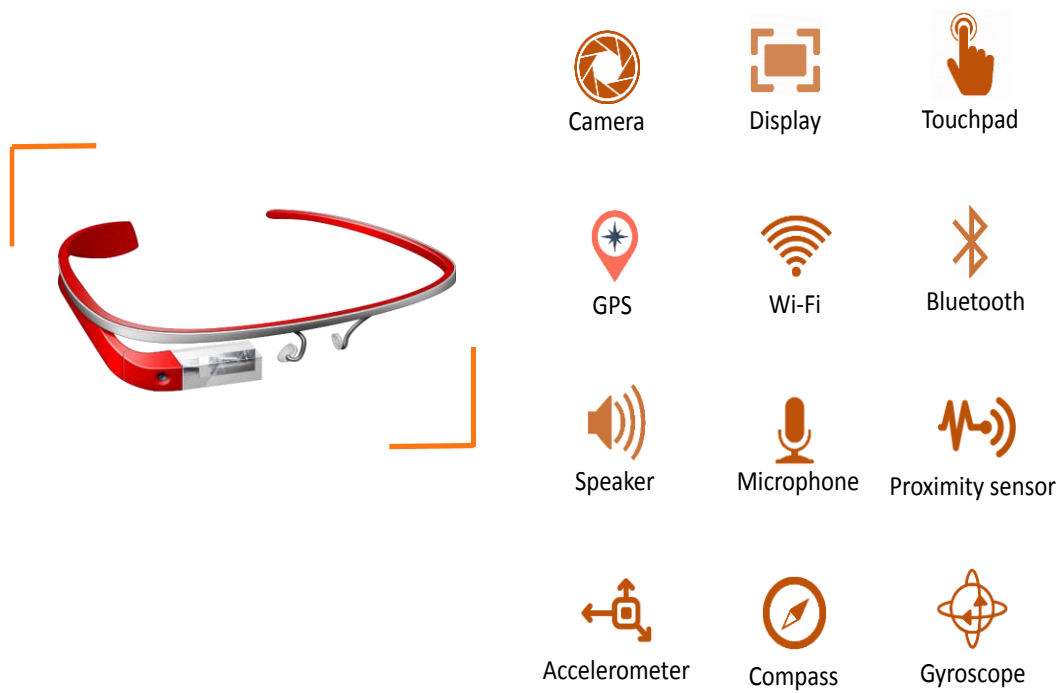

Fig. 3.2 Google Glass.

## 3.3 System Requirements

The system described in the above scenario is expected to be running on wearable devices such as Google Glasses and to be used on a daily basis. Hence, the system must be easy to use and must efficiently provide accurate and useful information. We refine the requirements and summarize them as follows.

### 3.3.1 Low Latency

As an assistance tool for real-time usage, the system is expected to recognize objects and return relevant information with a short response delay. A long delay would primarily affect the user experience.

### 3.3.2 Convenience

Convenience is twofold in terms of hardware and functional aspects. As a system for daily use, the hardware devices on which the system runs should not be bulky nor weighty. Meanwhile, the operation of the system should be simple enough for users to quickly and easily invoke functions. Too many functions require additional manipulations and distract the user from the system intentions. Also, functions should be invoked within a few steps so that users can access them easily and quickly.

### 3.3.3 Energy Efficiency

Wearable devices are powered by batteries. Hence, the system running on wearable devices must be energy-efficient. Heavy computations on wearable devices should be avoided as far as possible. Most executions are expected to be rendering related functions.

### 3.3.4 Adaptivity

In some situations, such as completing daily tasks, people occasionally require assistance for memory recall. A fire-and-forget usage model is suitable so that users can accept service quickly and continue with their main activities. In other situations, such as a big social event, they need continuous assistance, so an always-on model is preferable.

Our IA system is designed to implement the functionalities described in the motivating scenario while satisfying the listed requirements. We take the modular design approach and follow the design principles explained in the following section.

## 3.4 Decision-Making Algorithm

Table 3.4 compares various decision-making models [75]. As shown in the table, the fuzzy logic method provides minimum computation time with the least complexity and highest accuracy, compared to artificial neural networks, Game Theory, and Bayesian decision-making methods.

The structure of artificial neural networks is complex; it may give a non-deterministic outcome, so it is less accurate. On the other hand, Bayesian decision-making models have an intricate design; it results in either occurrence or non-occurrence of an event.

Game theory decision model cannot be analyzed for all competitive problems. It is considered an unrealistic/impractical approach since one player has knowledge of another player's pay-off matrix. Compared to these decision models, fuzzy logic is highly accurate as it yields to an outcome in between yes (1) or no (0).

Table 3.4 Comparison of decision-making methods

| Decision Making Models | Computation time | Complexity | Accuracy in outcome |
|---|---|---|---|
| **Artificial Neural Networks** | High | Complex | Least Accurate |
| **Bayesian Decision Making** | High | Highly Complex | Less Accurate |
| **Game Theory** | High | Complex Pay-Off Matrix | Accurate |
| **Fuzzy Logic** | Low | Simple | Highly Accurate |

## 3.5 IA System Architecture

In an attempt to send all the data to the cloud, we considered MQTT protocol as the messaging protocol. We have utilized HiveMQ server as a MQTT message broker and Wemos D1 Mini (gateway/sensor/actuator processor: ESP8266) modules for each sensor, actuator, and data coordinator in the positioning system to collect user's indoor positioning data. Fig. 3.3 shows the general architecture of the proposed IA system in which the IA system works in its automatic mode.

The application for making interaction with the patient and assessing his cognitive functions is developed on an Android operating system, implemented both on a smartphone and the AR glasses. We have also designed a service-based application (based on the C# cloud service) for caregivers to present the data related to the events detected by the sensors and to control the actuators embedded in the smart home environment. The patient cognitive functions state and game results can also be monitored by the caregivers.

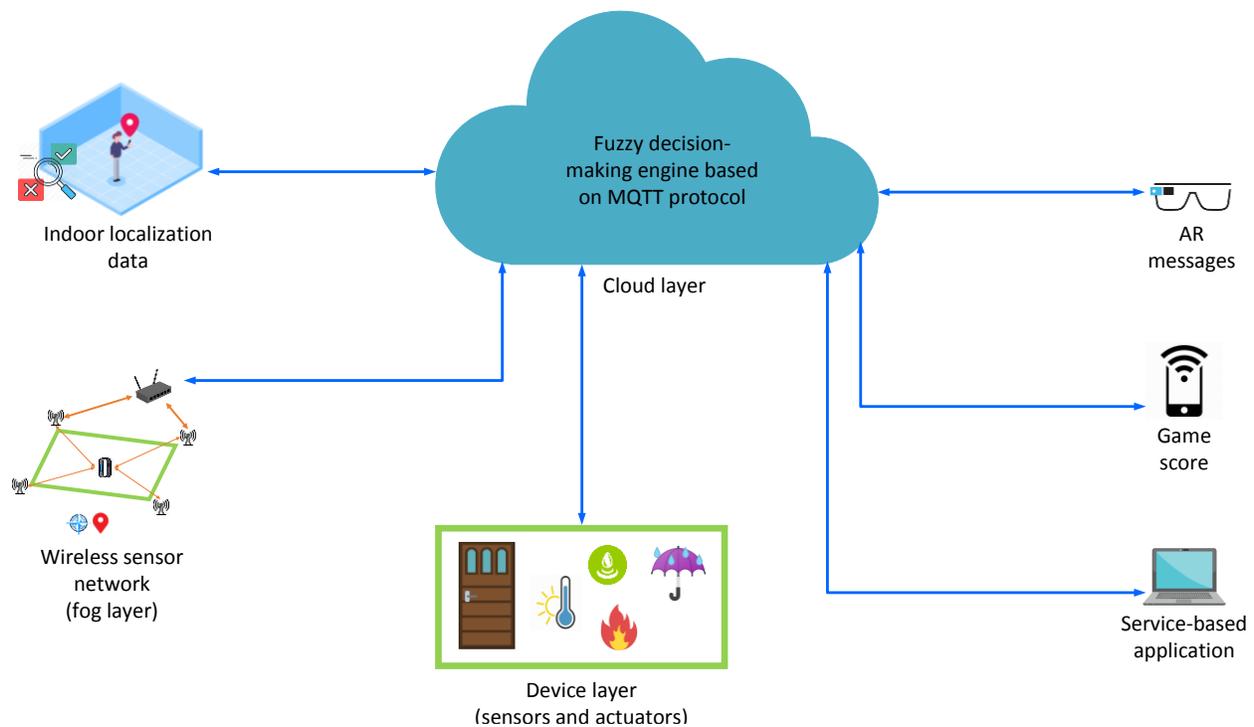

Fig. 3.3 Intelligent Assistive system architecture.

As the daily routine of an AD patient is contingent on their surrounding objects in the environment, the data about the topics such as temperature, humidity, and $CO_2$ level, is published on the cloud layer by placing different sensors and actuators in several locations such as kitchen, bedroom, and washing room.

Furthermore, WSNs have been deployed in the local fog layer to collect data about the user's location. Three anchors are placed in the home environment to estimate the patient's real-time location. All the communication to our IA system is done over MQTT with data serialized in JavaScript Object Notation format (JSON).

To avoid unnecessary transmission of the data, the performing mode of the IA system can be changed to the manual based on the decision-making algorithm results. This leads to saving energy of the devices, enhancing accuracy and performance, and providing cognitive enhancement for people with mild AD. We propose our model based on the localization tag for positioning system and AR glasses for receiving AR messages and evaluating user's cognitive functions.

## 3.6 Summary

In this Chapter, we have presented the selection criteria and an overview of the main design methods we used throughout the thesis. We have also described our IA system's architecture. Each individual section in the following Chapter contains a more detailed description of how the design, development, and evaluation of each method was undertaken.

# Chapter 4 - System Implementation

In this Chapter, we present the implementing process of the proposed model, including scenarios of use, IoT architecture, IA system architecture, and AR application. All the libraries, protocols, sensors and actuators, which have been used in our system, are discussed.

## 4.1 Internet of Things Implementation

In this section, we explain IoT architecture and its different layers in our system.

### *4.1.1 IoT healthcare network*

The IoT healthcare network for health monitoring goal is one of the vital elements of the IoT in designing healthcare services. It enables access to the IoT backbone, facilitates the transferring and acquisition of the patient's mental and physical data, and supports the use of healthcare-tailored communications. As shown in Fig. 4.1, the current section presents the topology, architecture, and platform, which we have used for implementing the IA system presented in this thesis.

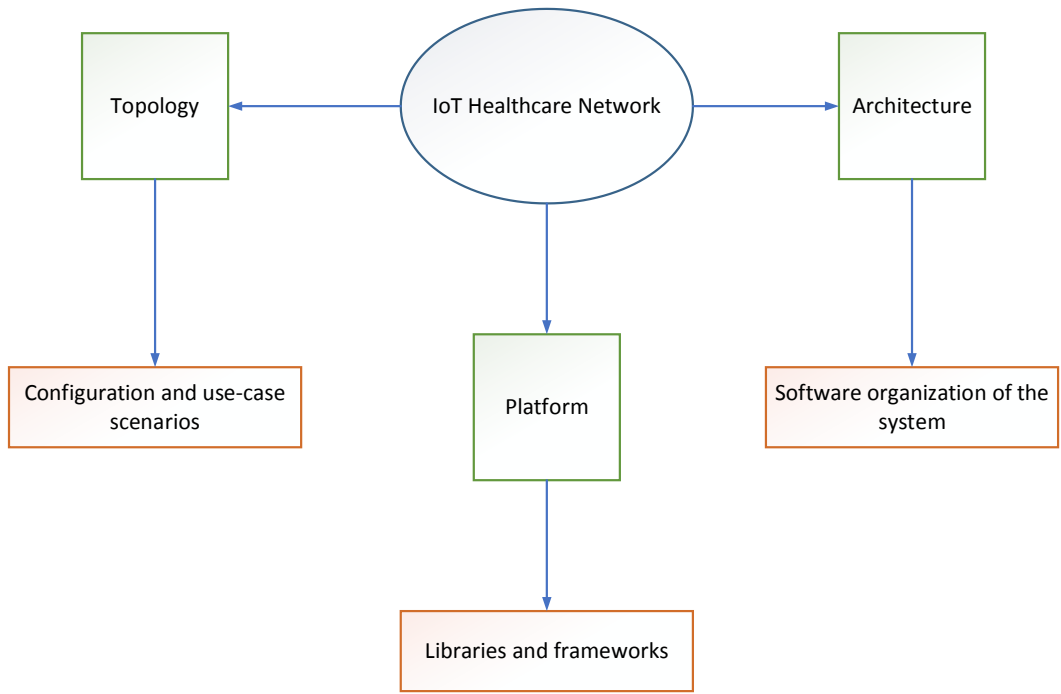

Fig. 4.1 IoT healthcare network features.

### *4.1.2 IoT architecture*

There is no unique model for explaining IoT architecture, which is generally used. Different architectures have been introduced by different researchers. In our study, we have implemented the primary system based on the architecture presented in Fig. 4.2, which consists of the IoT applications, applications model, resource management, infrastructure monitoring, fog devices, and IoT sensors and actuators layers. This model has We go through each layer and explain them further.

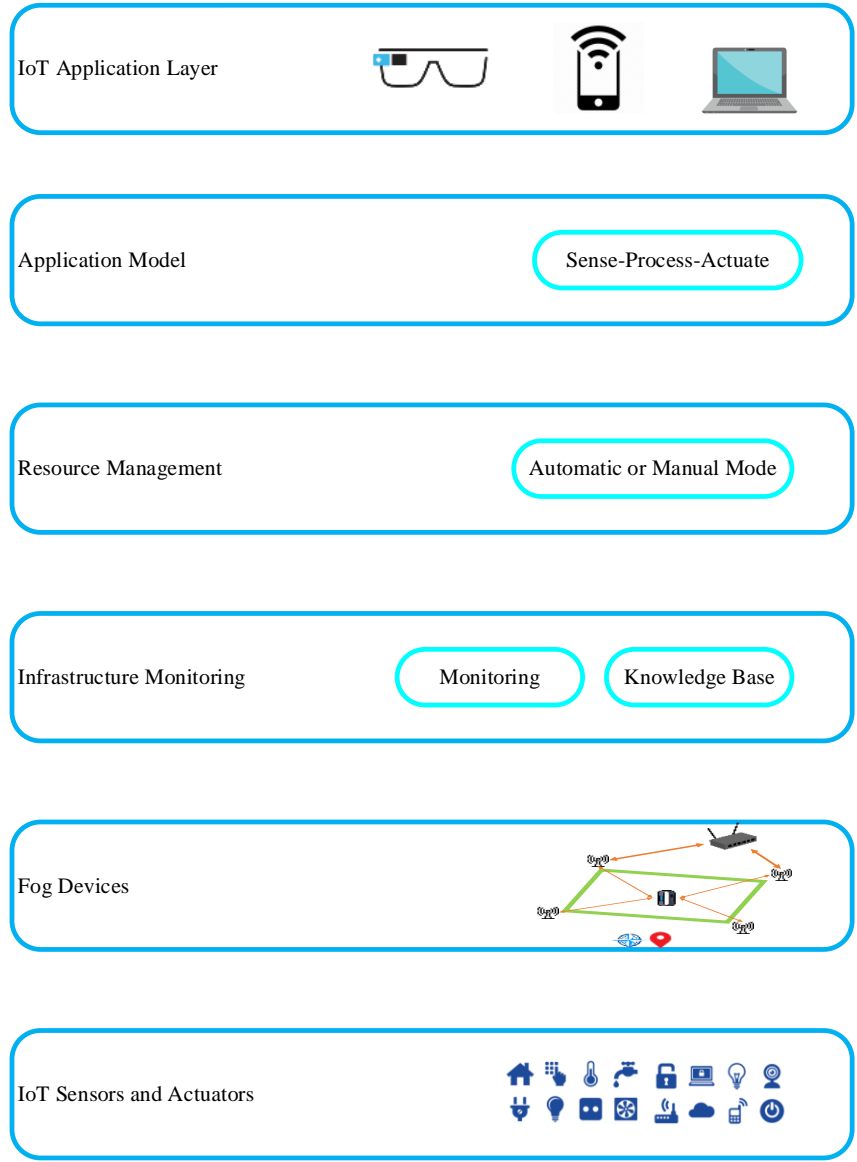

Fig. 4.2 IoT architecture.

### *4.1.3 MQTT protocol*

Many IoT standards are introduced to facilitate and simplify programming applications and designing program services. Different teams and group of experts have been founded to provide protocols in support of the IoT, such as efforts directed by the World Wide Web Consortium (W3C), Internet Engineering Task Force (IETF), EPCglobal, the European Telecommunications Standards Institute (ETSI) and Institute of Electrical and Electronics

Engineers (IEEE). Among the proposed protocols, MQTT intends to connect embedded sensors and actuators, networks with applications layer, and middleware layer.

MQTT is a commonly used application layer protocol to transmit data between the devices in IoT architecture, due to its simplicity and scalability. In this protocol, the client that sends messages is called a publisher, and the one that receives messages is called a subscriber. Numerous applications utilize the MQTT, including health care, monitoring, energy management systems. Hence, the MQTT protocol represents a perfect messaging protocol for the IoT and M2M communications and is able to offer routing for small, cheap, low power, and low memory devices in low and vulnerable bandwidth networks [76].

The connection operation applies a routing method (one-to-one, one-to-many, many-to-many) and enables MQTT as an optimum connection protocol for the IoT and M2M applications [76]. It also uses the publish/subscribe model to provide an implementation straightforward and flexible. Thus, it is appropriate for resource-constrained devices that utilize unreliable or low bandwidth links. MQTT only consists of three components, subscriber, publisher, and a broker. Fig. 4.3 illustrates the publish/subscribe process utilized by MQTT [77].

A target device would register as a subscriber for specific topics for it to be informed by the broker when publishers publish topics of interest [76]. The publisher performs as a generator of interesting data. After that, the publisher transmits the information to the interested subscribers through the broker. Furthermore, the broker achieves security by checking the authorization of the publishers and the subscribers [78].

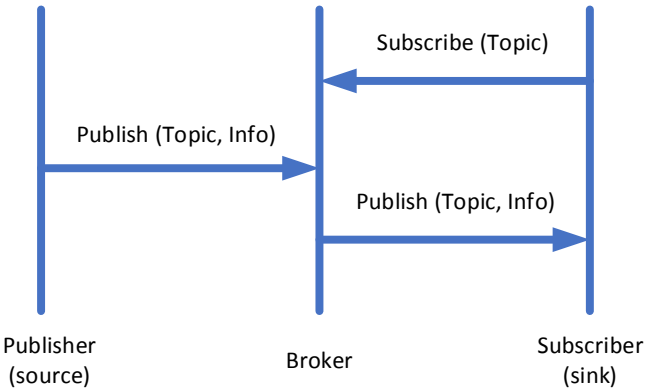

Fig. 4.3 Publish/subscribe process utilized by MQTT.

In this study, MQTT standard is applied because the message header requires only 2 bytes of a data packet, which makes it an extremely lightweight published-subscribed messaging protocol. To implement our AAL system, we considered the server as a MQTT broker and Wemos D1 Mini (ESP8266) modules for each sensor, actuator, or data coordinator positioning system. Fig. 4.4 shows the MQTT protocol operation in our study. The application for making interaction with the patient is developed on an Android operating system, implemented both on a smartphone and the AR glasses.

We have also developed a Windows application user interface for the patient's caregiver to show the notifications related to the actions detected by the sensors and to control the actuators embedded in the home environment. All the communication to our IA system is done over MQTT with data serialized in JSON format.

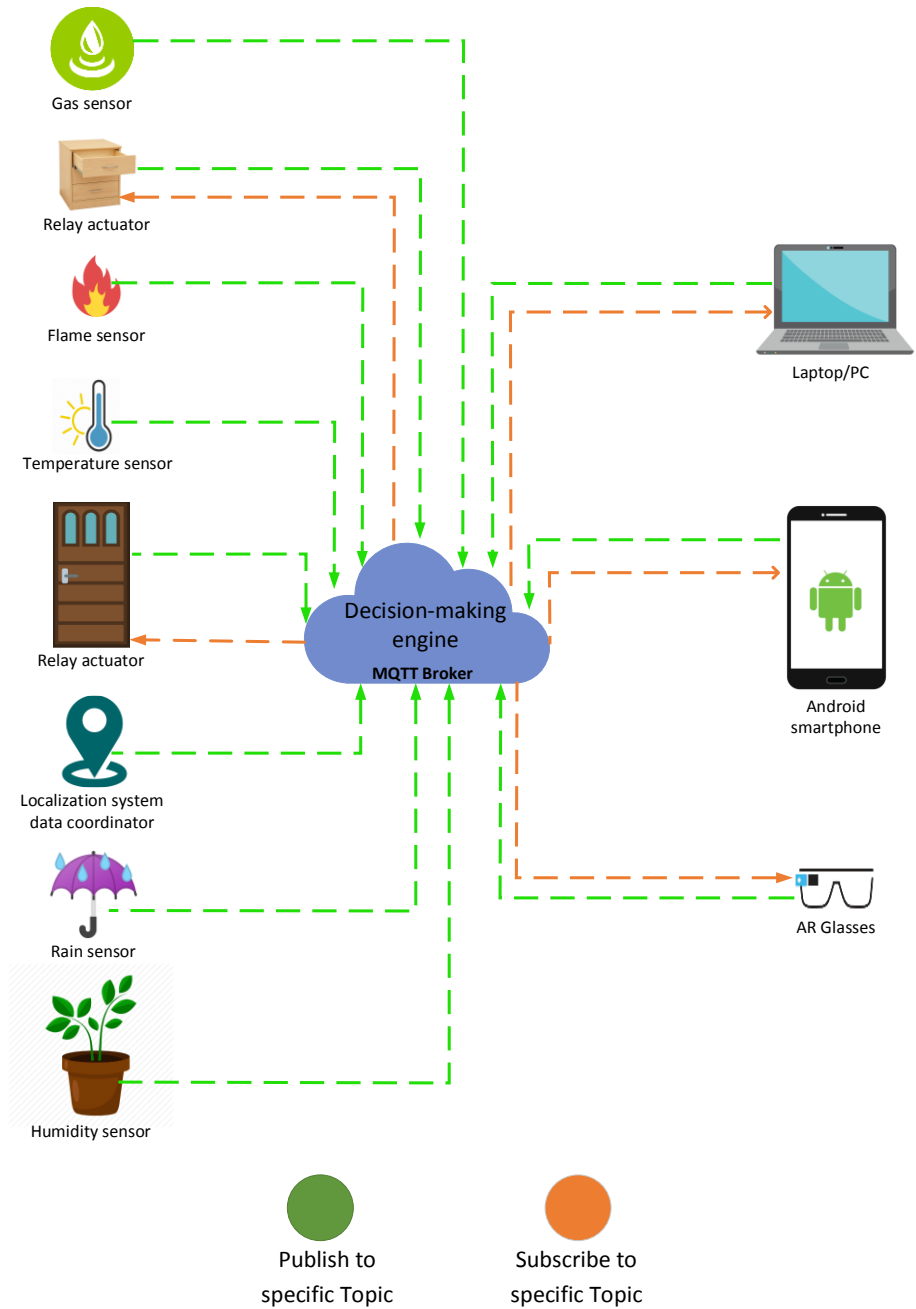

Fig. 4.4 MQTT protocol operation in the IA system.

### *4.1.4 Sensors and actuators*

As the daily routine of an AD patient is contingent on their surrounding objects in the environment, we deploy six types of sensors and actuators placed in different locations in their home, including: (1) rain sensor (in the terrace), (2) flame sensor (in the kitchen), (3) gas sensor (in the kitchen and bedroom) (4) temperature sensor in the TV room, (5) humidity sensors in the flower pots, and (6) relay actuators to open or close the drawers or the main entrance. The general architecture of the proposed IA system is shown in Fig. 4.5.

To estimate the patient's location, three anchors are also placed in the home environment to estimate the patient's real-time location. Our proposed model is created based on the localization tag for positioning system and AR glasses for receiving AR messages, worn by the patient. We have utilized a rain sensor in the terrace for rain detection, enabling the system to notify the user based on the event in the form of AR messages.

Furthermore, for reminding the weather condition changing, we utilized a temperature sensor in the TV room, as shown in Fig. 4.5. Relay actuators have been employed to open or close the drawers or cabinet doors based on specific events, which we describe in the scenarios section.

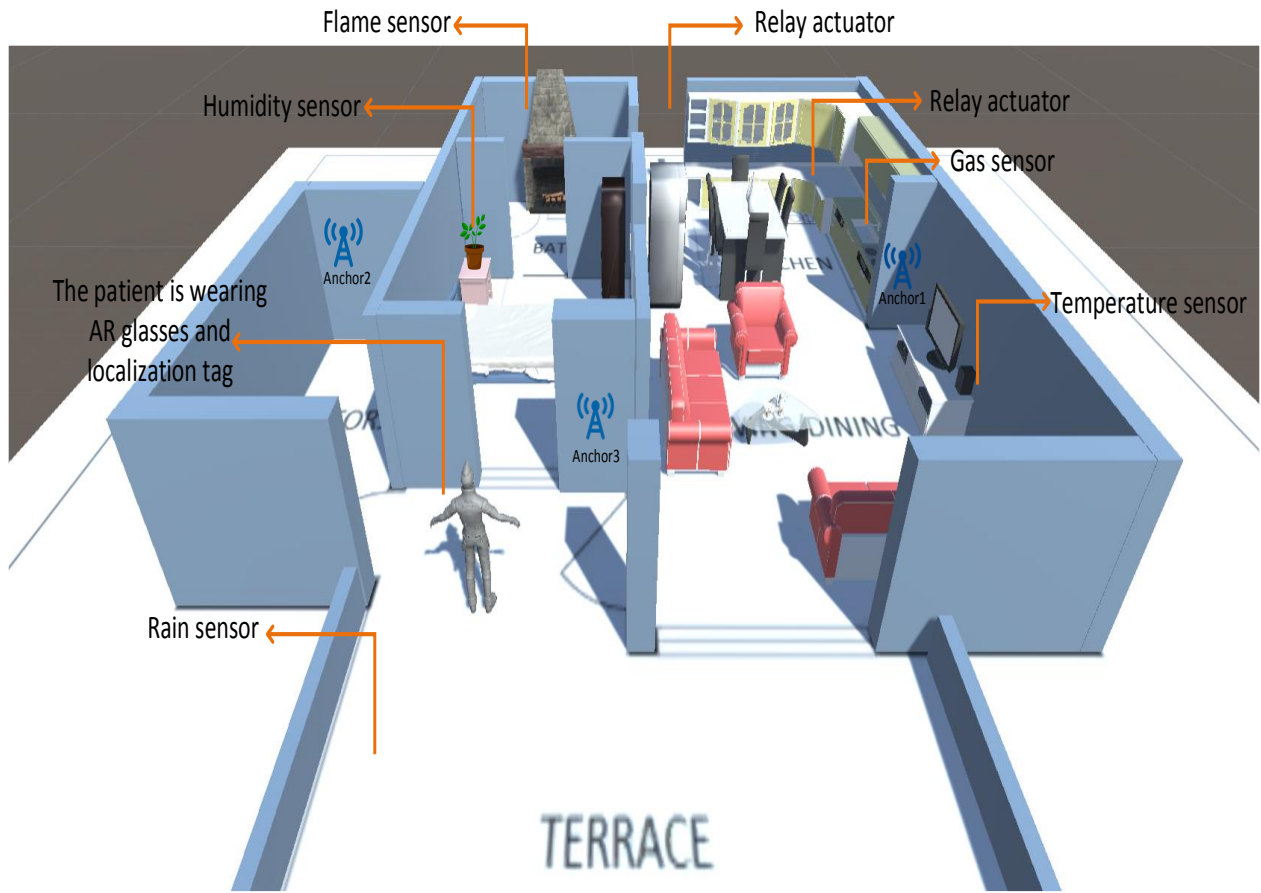

Fig. 4.5 General architecture of the IA system in AD patient's home environment illustrated by Unity game engine.

*4.1.5 Platform*

The ESP8266-12 [79] boards make communication abilities to the sensors through a built-in WiFi interface. Thus, by means of a WiFi home Access Point (AP), sensors are permitted to connect to the broker directly and publish messages whenever an event is recognized via MQTT protocol. Once a sensor activation is identified, the node that generally remains in a low power consumption mode starts the WiFi connection stage. When the connection is detected, the node can access the remote MQTT broker. This results in starting the MQTT connection, supposing the role of MQTT client. Once connected, it publishes a message on the established topic [80].

In our case, sensors were connected to ESP8266 to publish messages on topics such as gas, flame, and temperature data values, and family from a smartphone or a computer could read these values by subscribing to this topic. When the smartphone was offline, all the notifications could be stored by the server and then sent again when the smartphone connected to the Internet.

The family could also publish a value such as LED's or relay's states, and the ESP8266 required to subscribe to this value in order to read it. The published-subscribed messaging pattern needs a message broker. The broker handles and manages all these messages between all the subscribed clients. We applied several WEMOS D1 mini with a DHT shield to send the data values on topics such as humidity and motions to a MQTT topic and used HiveMQ as a MQTT messages broker.

## 4.2 AR Application

In this section, we describe the user interface and the AR application that we have developed to facilitate the user's interaction with different objects.

*4.2.1 AR process*

Different technological aspects make a significant impact on the creation of AR environments, which had been expanded and varied over the years and has found its way in several systems. Current solutions implement a more comprehensive range of software and hardware, as technical requirements may diversify, depending on the target results; Billinghurst and Thomas explain the process utilized in the operation of AR technology into the five main steps [81]:

- Build a virtual world with a synchronize system real world.
- Determine the position and orientation of the user.
- Pose the virtual camera accordingly to the position of the user.
- Render the picture of the physical space on the display of a device.
- Mix the real image of the environment with the virtual data [81].

As shown in Fig. 4.6, these five factors can be represented by three fundamental phases to carry out the process required to run AR applications [15]:

- Tracking: in this thesis, the position and orientation of the user are determined based on the indoor positioning data, including his point of view and his distance from the predefined objects. This phase can be operated by several types of methods of tracking, according to the technology, software, and hardware employed in the process. The input data are gathered from different types of sensors (including magnetic, infrared, inertial, and GPS) embedded in the user's home environment.
- Processing: creates a bridge between the coordinates of the physical space and the virtual world with respect to the virtual contents with the three-dimensional synchronizes and generating an image of the augmented environment. In our study, the virtual contents are retrieved online from the cloud and based on the knowledge base decision-making process. It can also be recovered offline from the internal repository. The efficiency of this phase is linked to the processing performance and battery usage of devices and the AR software exploited for this goal.
- Visualization: facilitates to the user the vision and perception of the augmented image and environment. In this phase, the quality of immersion is clarified based on the type of hardware employed to present the virtual content. In this study, we have taken advantage of Google Glass and an Android smartphone to display the AR messages. This phase indicates the operation of a different kind of hardware and software to create a realistic experience of the AR environment.

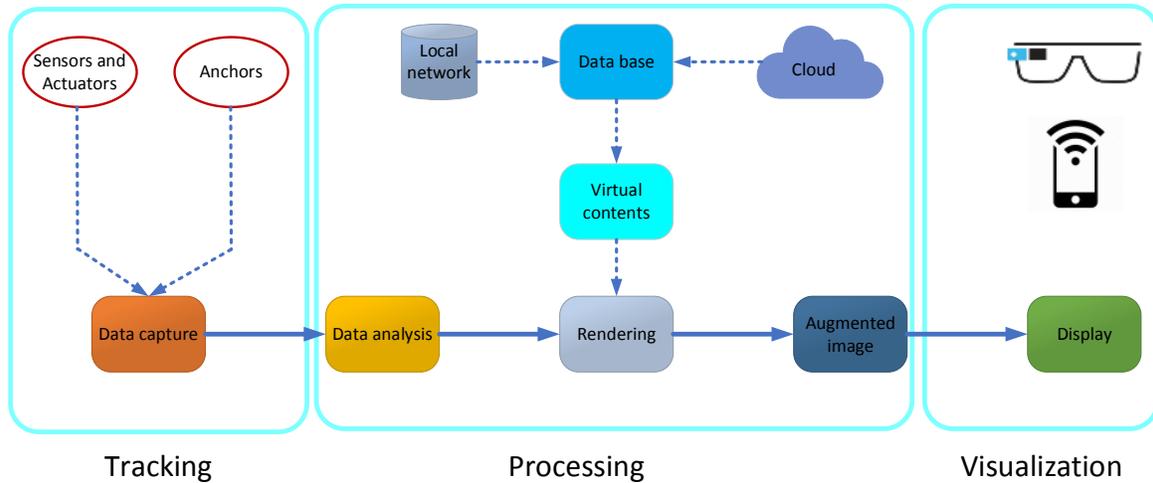

Fig. 4.6 AR process.

### *4.2.2 User-centered interaction*

There are two ways of interaction with the end-user; direct interaction (including switches and input devices) and indirect interaction (such as gesture, voice command, and automation). We recommend a balanced, mix user-side (direct) and automatic side (indirect) interaction according to the IA system operation mode and level of support, as shown in Fig. 4.7.

For example, the lighting and HVAC system are semi-automated with options for the users to override the system using manual switches. However, other safety factors, such as flame or fire detection, should be checked totally automated.

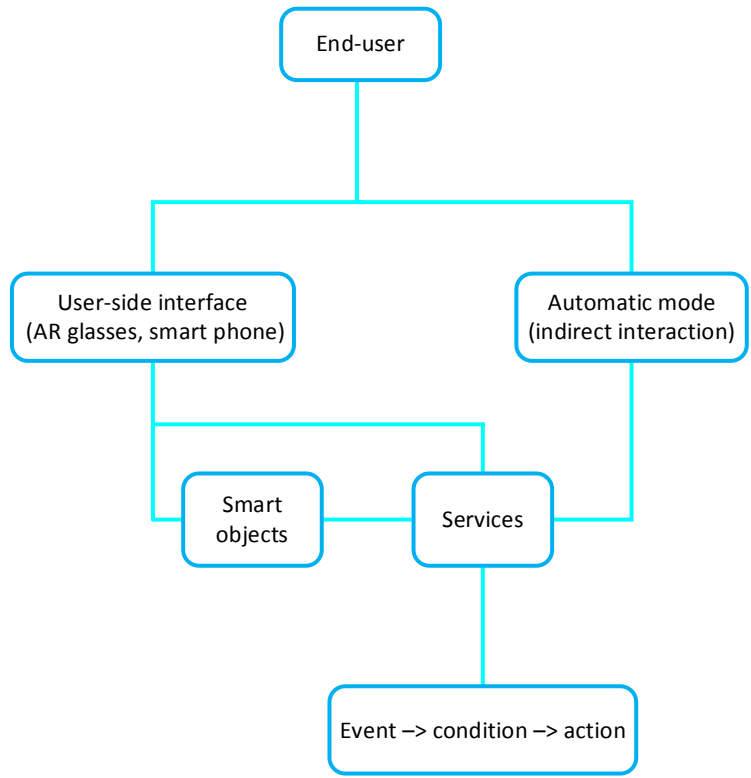

Fig. 4.7 User interaction structure.

To design our AR application for making interaction with the end-user, we have used Android ARCore platform, which is Google's open-source AR Software Development Kit (SDK), and it is capable of creating three-dimensional contents. In this study, we have developed our application based on an Android-based smartphone, and then we investigate the possibility of transferring this app on Google Glass as the ultimate assistive tool.

In this section, we first describe why we have chosen Google Glass as a user interface for AD patients; then, we consider the difference between AR and QR tags in AR applications. We have designed this platform based on both smartphone and AR glasses because a convenience design model can be more generally managed by the end-user.

Moreover, a person with dementia may not remember or indeed wish to wear a foreign object, then the project can be failed and no benefits experienced by the person with dementia. However, for example, wearable devices that measure heart rate can have an impact on any AR intervention. In our system for providing an assistive aid more helpful, we have taken advantage of the heart pulse sensor to ensure that the device is worn by the user.

*4.2.3 Google Glass*

We have selected Google Glass for AR implementation because of its lightweight and minimal heads-up display; thereby, it could be used by each mild to moderate AD patient thoroughly after short-term training that does not need any particular skills. Fig. 4.8 shows an example of the user's interaction with QR codes. In comparing to other AR glasses, it gives patients access to data and information in the easiest possible way that does not distract their daily life, and it makes enough essential virtual information while the patient is interacting with the physical world.

Moreover, if the patient is a glass-wearer, real glasses can have Google Glass screwed onto them. For instance, Microsoft HoloLens is too bulky and distracting for the patient, so it is not appropriate for this goal. We are aware that our application could operate on Google Glass due to the glass operating system, similar to Android smartphones.

Table 4.1 categorizes the specification of the Google Glass, which includes standard components found in smart devices, for example, central processing unit (CPU), camera, global positioning system (GPS), speaker, microphone, and display.

Table 4.1 Google Glass specification

| Specifications | |
|---|---|
| **Processor** | Dual-core 1.2-GHz Texas Instruments OMAP 4430 SoC with Power VR SGX540 GPU |
| **Connectivity** | Bluetooth 4.0 and WiFi 802.11 b/g |
| **Display** | 640*360 resolution |
| **Storage** | 16 GB memory with Google cloud storage |
| **Camera** | 5.0-megapixel camera, capable of 720p video recording with 30 frames per second rate |
| **Battery** | 570mAh lithium-polymer batt |
| **Charger** | Micro USB and charger (outlet or PC charging) |
| **Weigh** | 50 gr |
| **Compatibility** | Motion Process Library (MPL) Accelerometer, MPL Gyroscope, MPL Magnetic Field, MPL Orientation, MPL Rotation Vector, MPL Linear Acceleration, MPL Gravity |
| **Sensors** | LTR-506ALS Light Sensor<br>Rotation Vector Sensor<br>Gravity Sensor<br>Linear Acceleration Sensor<br>Orientation Sensor<br>Corrected Gyroscope Sensor |

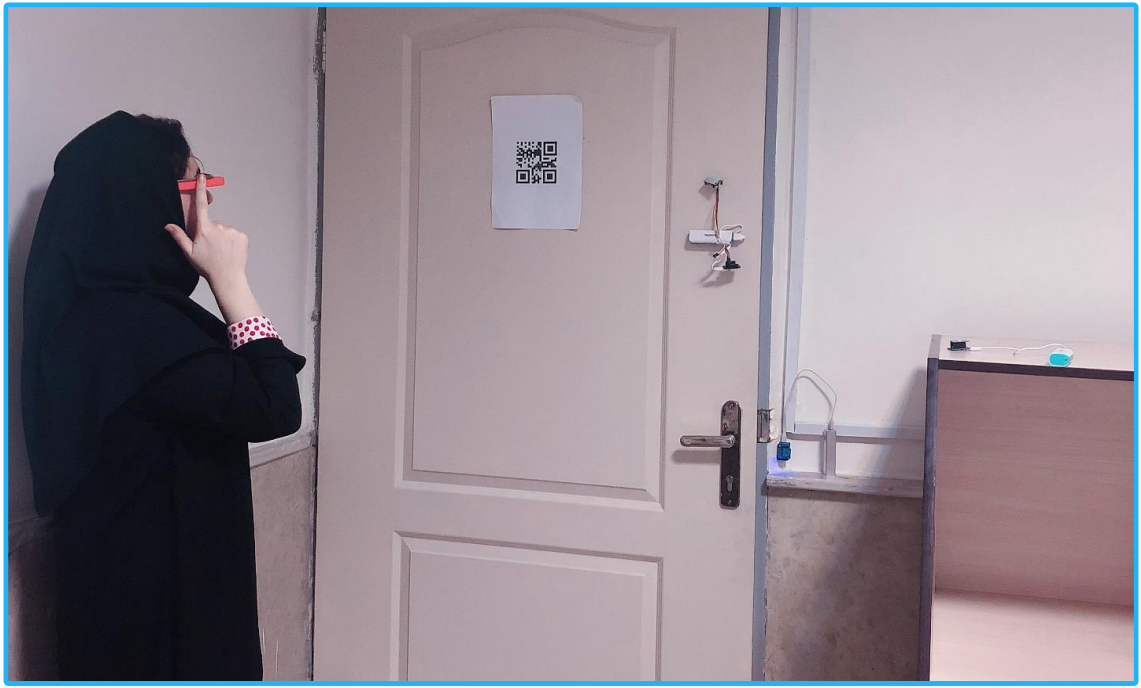
Fig. 4.8 User interaction through Google Glass.

Fig. 4.9 illustrates the structure of Google Glass AR application. It consists of two modules, a visual module, and an acoustic module. The visual module is responsible for detecting QR codes and identifying the information related to each tag. It takes advantage of the OpenCV engine to perform real-time image detection and recognition.

The acoustic module works in conjunction with the visual module to create multimedia contents and inserts them into the database. It utilizes the language processing engine to recognize speech and summarize conversations. The database stores a user's interaction with different objects. It is maintained locally on the smartphone side application and can be retrieved on the cloud layer.

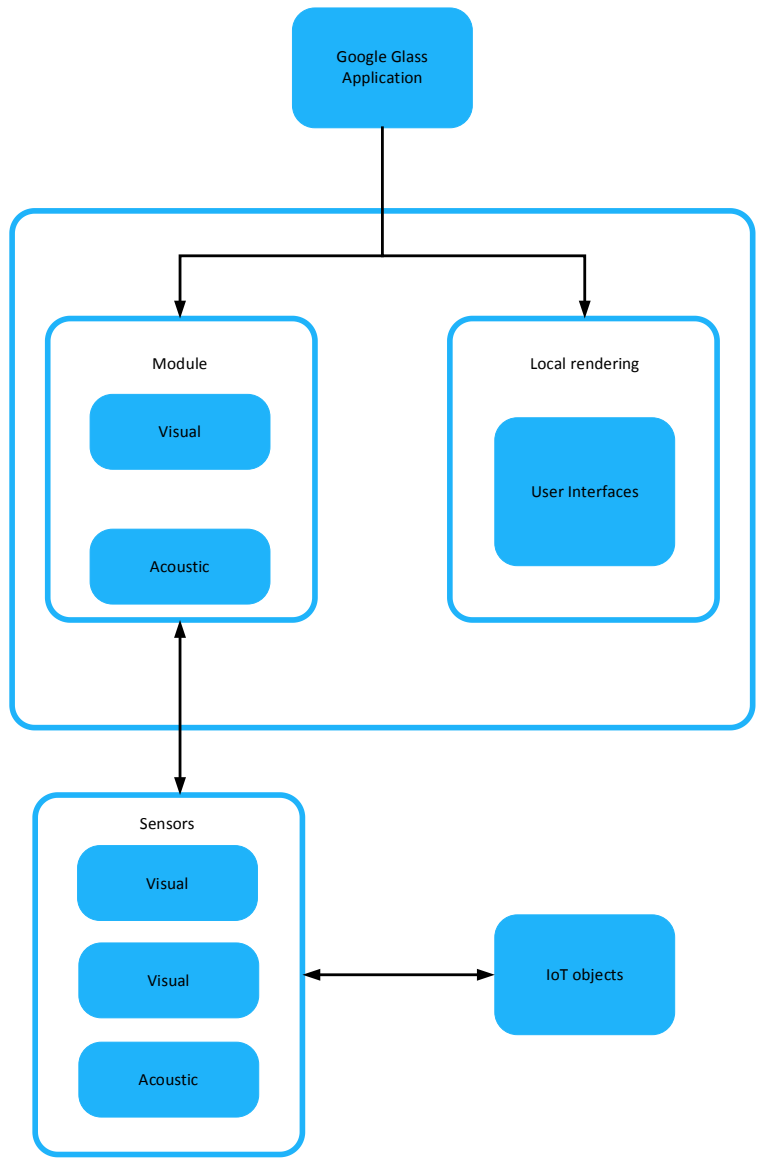

Fig. 4.9 The structure of Google Glass application.

### *4.2.4 QR code*

In this work, we have used six different QR codes because they can be simply processed using free open-source software. Furthermore, QR codes printed on paper are cost-effective. However, they can only be detected one at a time, which was not a limiting issue in this study because we needed simultaneous identifying of single symbols.

On the other hand, if we want to detect multiple symbols in any future study, we should use AR tags. Similar to QR codes, the AR tags can be printed on paper applying open-source software. Therefore, it is cost-effective too. In some scenarios, a 3second timeout happened in QR code recognition algorithm when we aimed to detect that the user was staring at the QR code. Obviously, if we use indoor positioning data, there is no need to attach QR codes in different places for making interaction with the end-user.

### *4.2.5 Visualization*

For simulation and evaluation of our IA system, we have used the Samsung Galaxy S7 smartphone, which has a quad-core Snapdragon 820 processor, 12MP rear camera, and 4GB RAM. Fig. 4.10 indicates some pictures based on scenarios we have defined our distributed mechatronics system. For instance, in Fig. 4.10(a) rain warning was recognized by the rain sensor, and at the same time user was staring at the main door. Thus, after this event, they could notice an umbrella and hear an audio reminder message associated with the event.

In Fig. 4.10(b), a flame sensor near the oven in the kitchen identified the existence of a flame, and so the relay actuator was activated that assisted the user received an image notification. The user could also see the current and previous picture states by scrolling through the received pictures. In Fig. 4.10(c) this state is illustrated on the LCD screen. The dishes picture for reminding the place of them in the kitchen indicates the current state, which is based on the user's field of view, and the right picture is the previous image that was sent to the user by the family to remind them of the medication time.

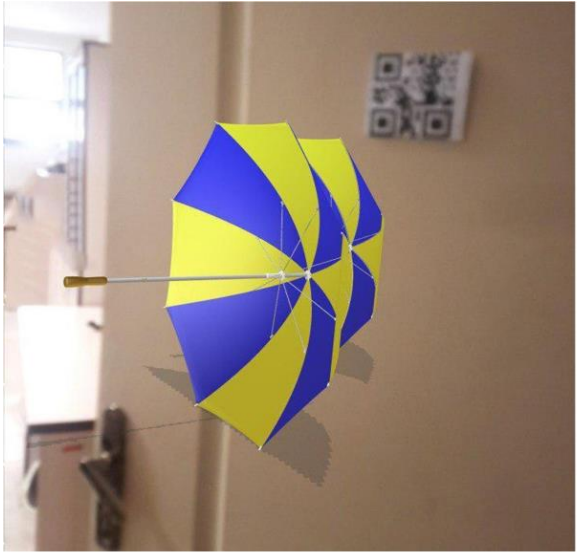

(a)

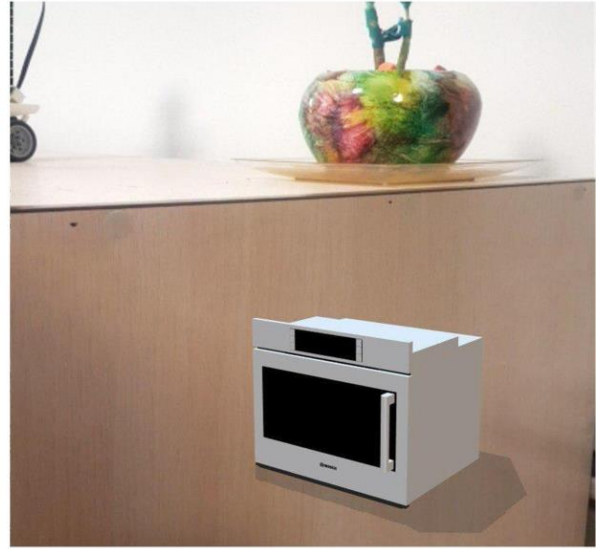

(b)

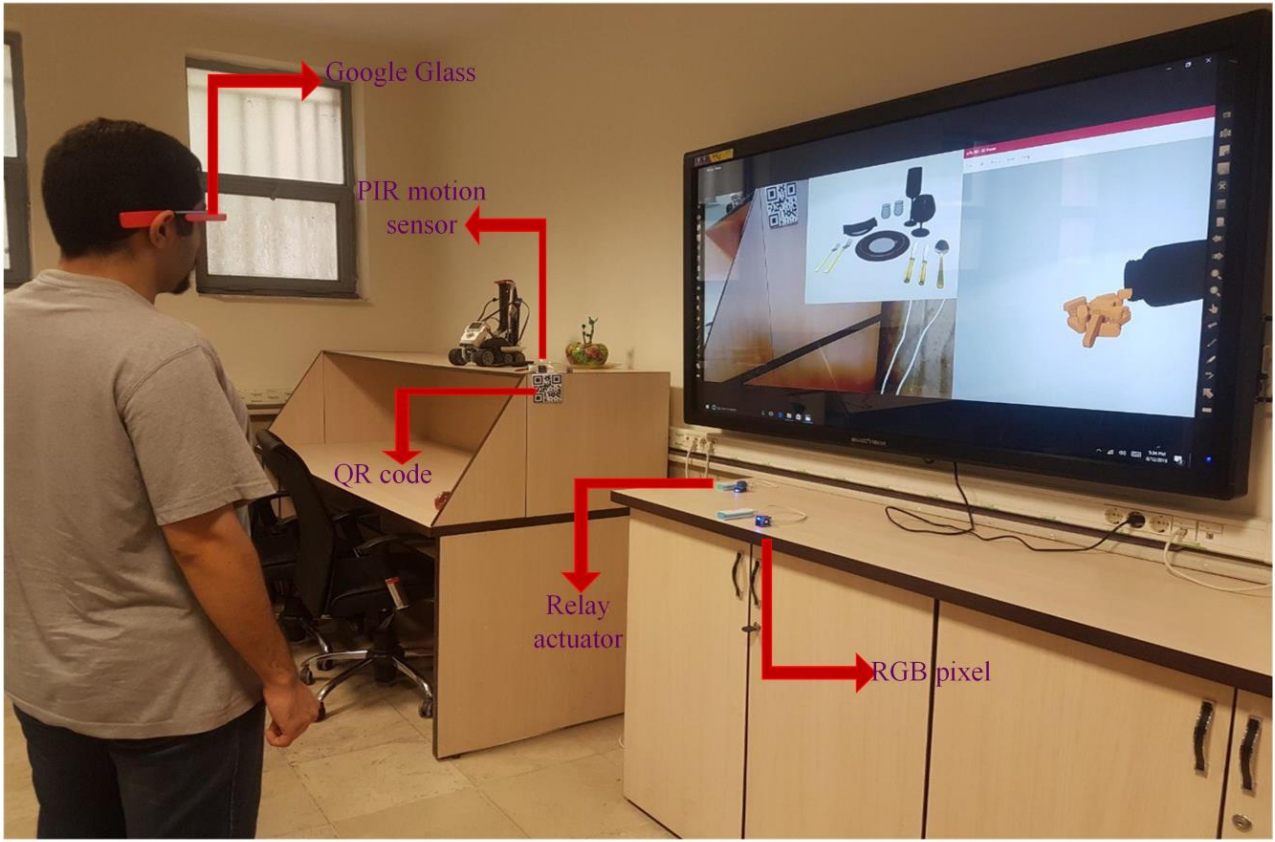

(c)

Fig. 4.10 Image messages received by the user

(a) The rain warning was detected by the rain sensor, and a picture of an umbrella is generated as a reminder, (b) the flame sensor in the kitchen near the oven detected the presence of a flame, so the relay actuator was activated, and the user received an image notification, (c) an illustration of the user's field of view shown on the LCD screen.

In this work, we defined three different types of messages for making interaction with the patient, including picture message, audio message, and text message. In some scenarios, the patient could receive two types of messages at the same time. The IA system uses two different user interfaces (UI) to interact with the patient and their family. Both user interfaces are developed by utilizing ARCore SDK (Software Development Kit) for Unity and Android features [9]. Unity is a cross-platform game engine that can be applied to design games and simulations for computers, consoles, and smartphones [82].

To display the user's localization information clearly, a 3D indoor positioning system based on the Unity 3D platform is created. This configurable user interface can represent complex indoor area geographic information and enhance the user experience [83]. Furthermore, to monitor and track the user's real-time location and provide their interaction with the objects, the user wears an indoor localization tag.

For presenting and transmitting positioning data in JSON format via MQTT protocol to the cloud server, MQTTnet and Json.NET libraries are also employed in the Unity application. The localization tag wearing by the user sends position information to the monitor module, and the monitor module receives all the positioning data.

To ensure that the device is worn, Heart Pulse Sensor (SEN-11574) was used. All the data related to the localization system and AR messages can be stored on the cloud for further studies. To improve the patient's safety, caregivers can add the location of their home's danger zone. Hence, the IA system can provide warnings if the patient is near to these locations. The second user interface is built on Android Studio and SDK tools to interact with the individual with mild AD as a personal assistant device [84]. Android Studio is the official integrated development environment (IDE) for the Android operating system created particularly for Android development. In this work, two different AR messages are defined to send reminders or alerts to the user, including visual and audio messages.

The AR messages show information using virtual data based on the predefined fuzzy rules. These multimedia contents are for taking medication, playing AR cognitive training games, reminding meal times, and alerting danger zones. For example, to recall medication schedule, a voice message, "Let's take our medicine," and an image message indicating the drug's picture are defined. These messages must not remind AD patients' disability. Initiating treatment at the early stage of AD would slow the progression of the situation and expand the patient's quality of life. In this study, an AR-based serious game provides entertainment and mental state examination simultaneously.

## 4.3 Decision-Making Process

To convert our system into an intelligent and context-aware system that promotes monitoring of mild AD patients and improving their ability to perform their daily tasks, we consider a fuzzy method. Fig. 4.11 shows the structure of our IA system decision-making process based on the fuzzy rules on the cloud layer.

The collected data of the sensors and actuators, the user's cognitive game result, and positioning information are then transmitted via the network to a fuzzy decision-making engine on the cloud. This adaptive decision engine analyzes and compares data to make an appropriate decision for making interaction with the end-user. In this section, we explain the decision-making process of the proposed IA system more in detail.

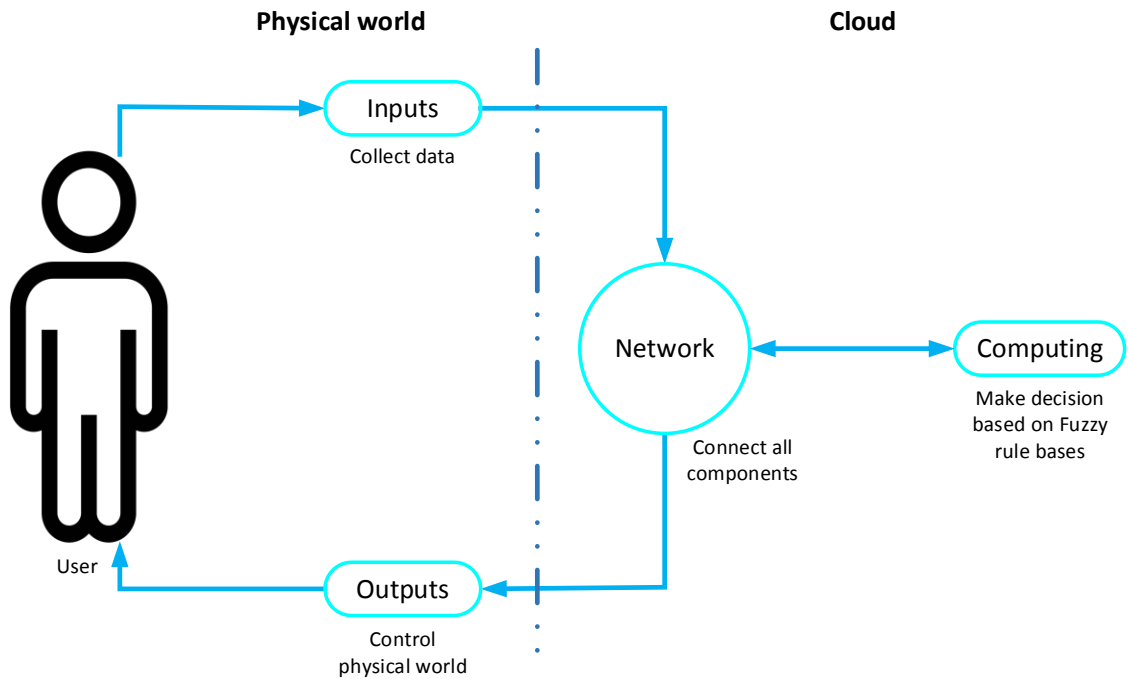

Fig. 4.11 Cloud base decision-making model.

### *4.3.1 Adaptive fuzzy control*

In order to update each variable on the cloud and making proper decisions based on the determined rules, we employ a fuzzy logic model. We consider the possibility of extending this method to convert our system into an expert system that promotes monitoring of AD patients and improving their ability to perform their daily tasks.

However, the design process for fuzzy controllers that is created based on the utilization of experimental data from human experts has been successful in many applications. Moreover, the approach to building fuzzy controllers via numerical input-output information is progressively finding use. Despite which method is implanted; however, several challenges are faced with practical control problems, including the following [85]:

- The design of fuzzy controllers is operated in an unexpected way so that it is often challenging to select at least some of the controller model parameters. For instance, it is usually difficult to recognize how to choose membership functions and rule-base to achieve a particular required level of performance.
- The fuzzy controller formed for the nominal plant might later perform ineffectively if considerable and unpredictable plant parameter deviations happen, or if there is noise or some other kind of disturbance or another type of environmental consequences. Thereby, it may be complicated to perform the preliminary synthesis of the fuzzy controller, and if the plant varies as the closed-loop system is performing, we might not be able to preserve acceptable performance levels.

On the other hand, a learning system demonstrates the capability to stage its performance over time by interacting with its surroundings. A learning control system is proposed so that its learning controller has the capacity to improve the performance of the closed-loop system by creating command inputs to the plant and exploiting feedback information directly from the plant. In this method, the adaptation mechanism detects the signals from the control system and adjusts the parameters of the controller to support performance even if there are variations in the plant.

Occasionally, the target performance is described with a reference model, and the controller then tries to make the closed-loop system work as though the reference model would even if the plant alters. This is called model reference adaptive control (MRAC), as demonstrated in Fig. 4.12 [85].

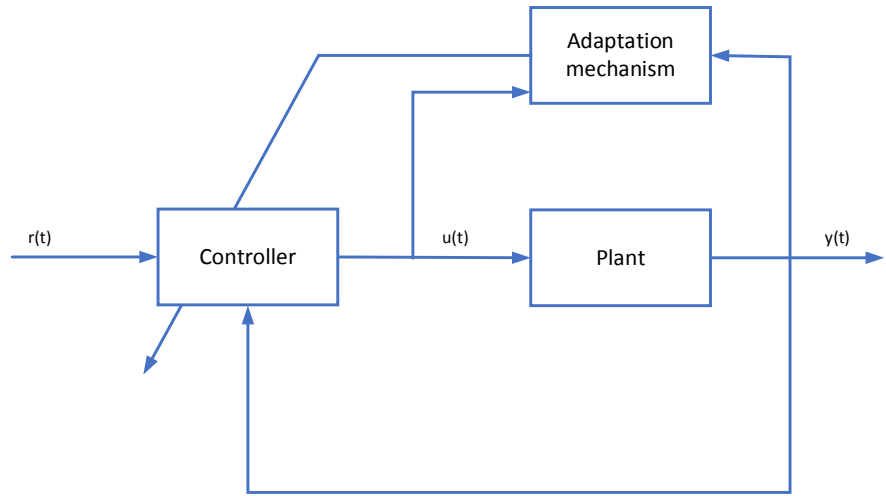

Fig. 4.12 Direct adaptive control.

In our work, the decision-making process should be adaptive and precise to avoid false notifications or false-positive AR messages and interact with the user correctly. Based on the above discussion and in order to update each variable of the IA system on the cloud and making the right decision, we have employed an adaptive fuzzy logic model that selects the most proper values. The model of fuzzy adaptive decision-making is shown in Fig. 4.13. According to [85], a fuzzy controller has four main components: rule base, inference system, fuzzifier interface, and defuzzifier interface.

- Rule-base: this includes a fuzzy logic quantification of the expert's linguistic explanation of how to obtain the best control. Two knowledge bases have been elicited from the user's cognitive state and the AR game score. These rule-bases are necessary for deciding when to provide aids and predict the user's target accurately.
- Inference system: this part emulates the decision of the expert to apply his knowledge about how good to control the plant. In this work, we applied Mamdani Fuzzy inference method.
- Fuzzifier interface: converts controller inputs into information so that the inference system can simply utilize to activate and apply rule-base.
- Defuzzifier interface: it converts the conclusions accessed by the inference system into crisp output values for controlling the process.

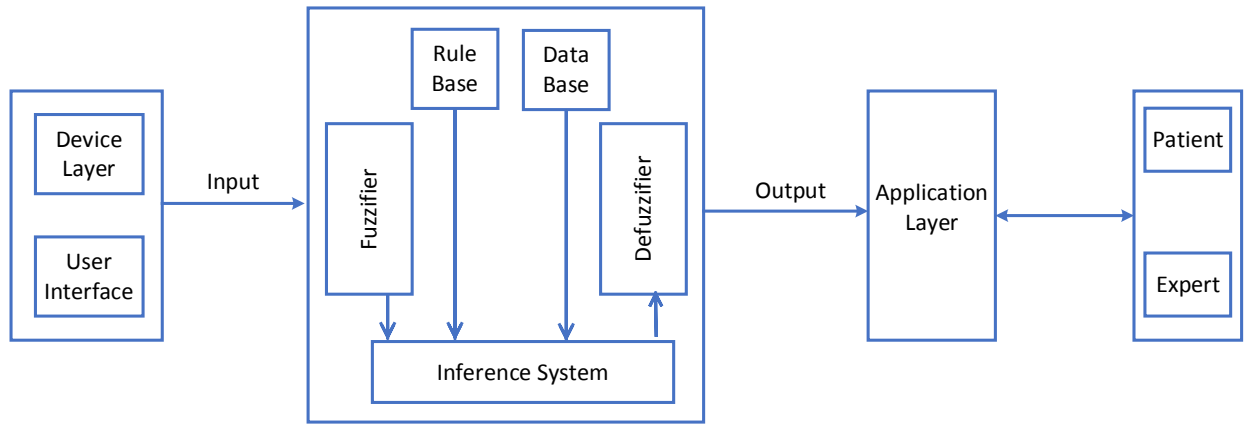

Fig. 4.13 Fuzzy adaptive decision-making model.

As mentioned before, the determination of input variables, fuzzy inference, and output variables are the basic concepts of a fuzzy logic implementation. For instance, in our model, the distance between the patient and a particular object is considered as an error signal.

Moreover, in our proposed system, the input variables are several types of data from embedded devices and sensors, the patient's real-time position and heading angle, and the output variables are different types of messages which are received by the patient, and the state of the actuators in the specified places. Table 4.2 indicates input and output parameters, data types, and their membership functions in our work. We can add other home objects and calculate the distance as a fuzzy input variable.

Table 4.2 Description of inputs and outputs of fuzzy system

| Parameter | Fuzzy membership functions | Data type | Direction |
|---|---|---|---|
| **Heading angle** | Small, medium, large | Linguistic | Input |
| **Game** | Start, Stop | Boolean | Input |
| **Time** | Early morning, morning, early afternoon, late afternoon, evening, night, midnight | Linguistic | Input |
| **Distance** | Near, far, very far | Linguistic | Input |
| **Game score** | Low, high | Linguistic | Output |
| **Reminder** | Yes, No | Boolean | Input |
| **Movement** | Low, medium, high | Linguistic | Output |
| **Humidity** | Very dry, dry, humid, very humid | Linguistic | Input |
| **Temperature** | Very cold, cold, cool, mild, warm, hot, very hot | Linguistic | Input |
| **Rain detection** | Yes, No | Boolean | Input |
| **Flame detection** | Yes, No | Boolean | Input |
| **Gas detection** | Yes, No | Boolean | Input |
| **Relay status** | Yes, No | Boolean | Output |
| **Voice message** | 1, 2, …, 20 | Integer | Output |
| **Image message** | 1, 2, …, 10 | Integer | Output |
| **Text message** | 1, 2, …, 10 | Integer | Output |

## 4.3.2 Input/output variables

We have considered each input and output variables and defined membership functions based on the data types and fuzzy rules. In our proposed model, the input variables are several types of data from embedded devices, the patient's real-time location, and their cognitive state. The output variables are different types of AR messages which are received by the patient or his caregivers.

We can add other home objects and calculate the distance as a fuzzy input variable. By considering each input and output variable, membership functions are defined based on the data types and fuzzy rules. The principal fuzzy membership function that is often used to represent vague linguistic terms is a Gaussian membership function for each language expression, which is shown in Fig. 4.14. With Formula 1, for each language expression, we obtained its membership as follows:

$$\mu_z(x(k), y(k)) = exp\left(-0.5 \cdot \frac{(x(k) - C(k))^2}{y(k)^2}\right)$$

$$C(k) = \frac{u(k) + U(k)}{2} \quad (1)$$

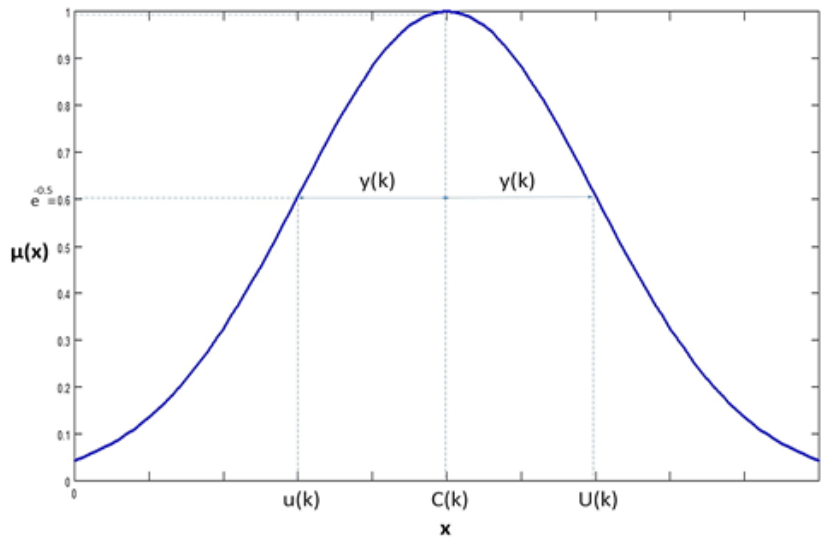

Fig. 4.14 Gaussian membership function.

Fig. 4.15 shows the Gaussian membership function for the distance between the patient and specific objects, as well as the patient's heading angle. Membership functions for time, plant humidity and temperature are shown in Fig. 4.16 as examples. Output variables are voice, image and/or text messages, defined as fuzzy singleton membership functions. All types of messages and their membership functions are shown in Fig. 4.17.

Each message has an identification number (ID) and can be activated by particular inputs. Moreover, the membership function for the patient's movement time, as well as the AR game score, are presented by the Gaussian membership functions, as depicted in Fig. 4.18. Variables with Boolean data type are described by triangular membership functions.

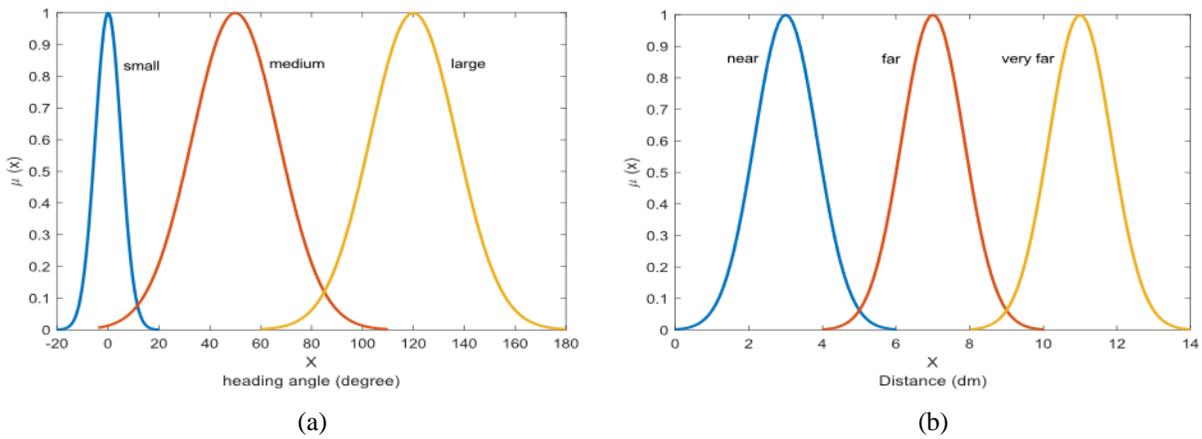

(a) (b)

Fig. 4.15 Membership function of (a) heading angle, and (b) distance.

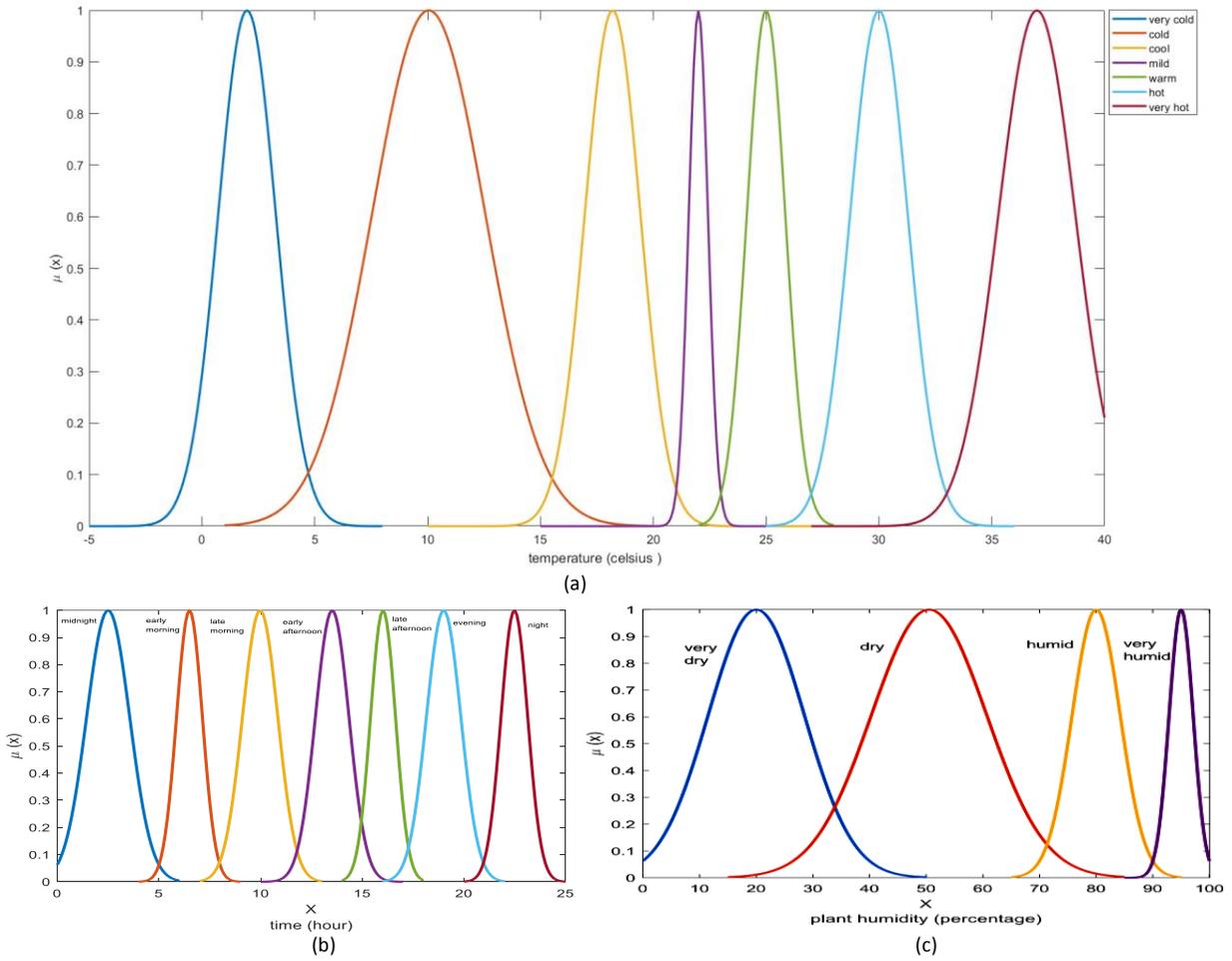

Fig. 4.16 Membership function of (a) temperature, (b) time, and (c) plant humidity.

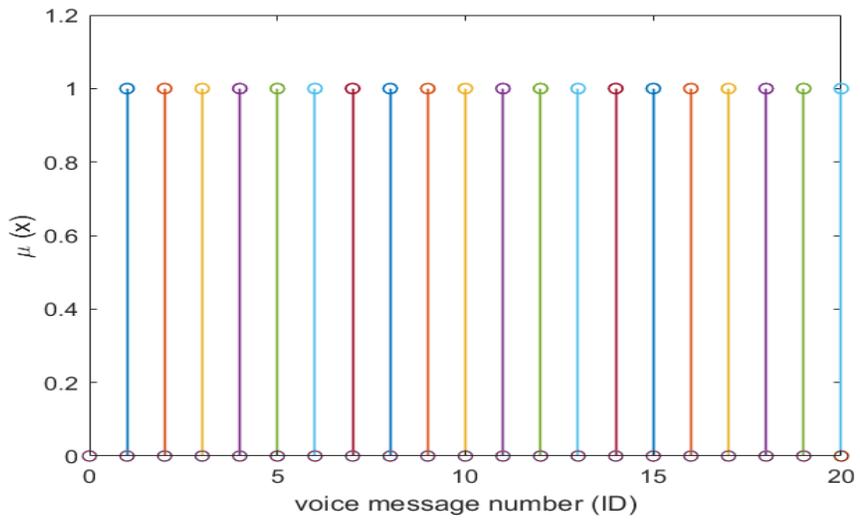

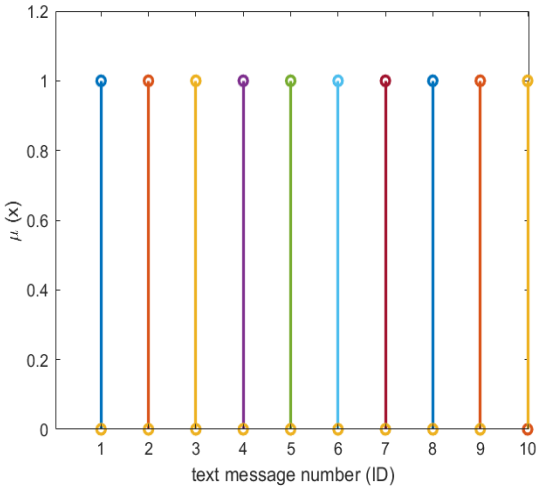
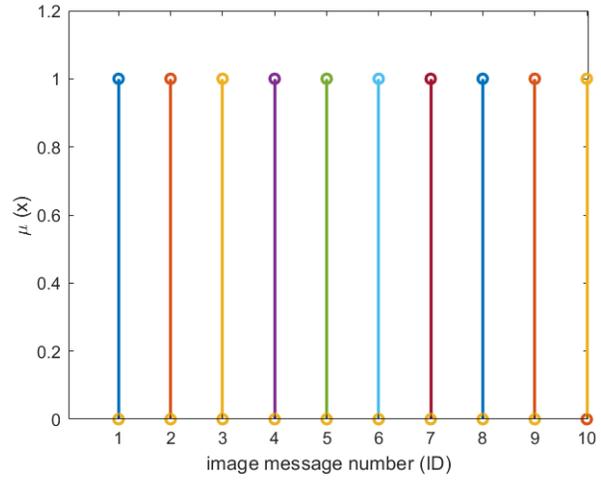

Fig. 4.17 Membership functions of (a) voice messages, (b) text messages, and (c) image messages.

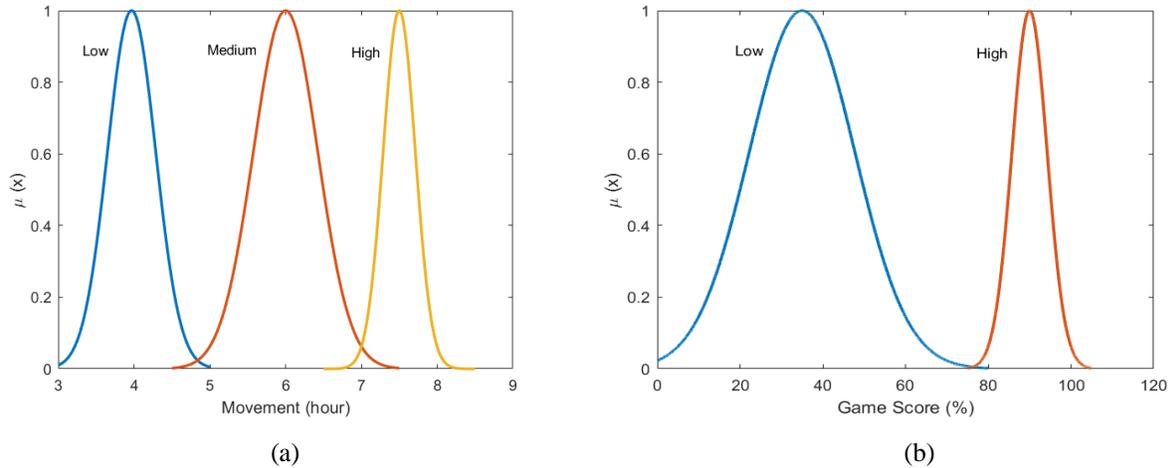

Fig. 4.18 Membership function of (a) movement, and (b) game score.

### *4.3.3 Rule-base initialization*

The input membership functions are defined to create the bases of the rules that define the different circumstances in which rules must be applied. The input membership functions are left invariable and are not modified by the fuzzy model reference learning controller (FMRLC).

The membership functions of the output are supposed to be unknown. Thus, the FMRLC should automatically tune or synthesize. To achieve this goal, the FMRLC aims to fill in what actions must be taken for the diverse condition that is created by the rule bases. The learning mechanism adjusts the rule-base of the direct fuzzy controller; thereby, the closed-loop system acts similar to the reference model.

These rule-base adjustments are made by observing and detecting data from the reference model, the controlled process, and the fuzzy controller. The learning mechanism includes two main components; a fuzzy inverse model and a knowledge-based modifier. The fuzzy inverse model presents the function of mapping the deviation from the desired behavior to modifications in the process inputs that are required to decrease the error signal to zero. The knowledge-based modifier presents the function of adapting the fuzzy controller's rule-base to make the necessary changes in the process inputs [85].

After defining each variable's membership function, we must build the rule-base, including all of the expert IF-THEN rules. In our IA system, two main rule-bases are essential to creating an adaptive decision-making process and multi-level support. The first one helps the user complete the activities independently in their daily life by showing AR messages or generating automatic changes such as actuators activation. The second one allows manual changes after real-time assessment of the user's cognitive state according to the AR-based serious game score. In this situation, some of the smart home sensors and reminders can be turned off or disabled.

Once an event occurs (for example, medication time alarm), input data values such as location, AR game score, and sensors data value are published on the cloud via MQTT protocol. Moreover, fuzzy rules are checked, and corresponding output variables are updated and published on their topics. Then, according to which rule-bases are activated, the user receives different types of messages, or the user's manual changes can be enabled. In our experimental tests, each of these rules has been defined based on the mild AD patients' real-life scenarios.

To improve system performance and decrease processing time, the refresh rate of the algorithm's control loop is set between 0.5 and 2 Hz. The adaptive fuzzy decision-making algorithm is capable of scaling up to more rules to improve the user's independence. The location of different objects and the most used ones can also be added to the database. Therefore, new rules can be added, or the other rules can be removed and updated by the caregivers based on the knowledge and dependent of each patient's lifestyle, and they can manually start a scenario or send simple reminders to the patient.

The fuzzy rules must be endowed with a base of knowledge of each patient provided by the caregiver, so all the activities are adapted to his preferences. Some parts of such scenarios and their fuzzy rules are presented in Table 4.3. These scenarios are precisely defined to choose the most proper ones for the first experiment with AD patients. According to [30], AD patients often lose their ability to sequence in activities such as putting on clothes or cooking. When the user walks in different rooms and tries to interact with other objects, the IA system tries to predict the patient's target based on the distance between the user and predefined objects and other inputs.

Table 4.3 Examples of fuzzy rules

| Event | User's location | Fuzzy rules If | Fuzzy rules Then | Command type |
|---|---|---|---|---|
| **Entertainment/cognitive assessment** | Not specific | *Time* is "*evening*" | *Game* is "*start*" | Launching AR game |
| **Medication schedule** | Not specific | *Time* is "*morning*" | *Voice message* is "*1*" and *Image message* is "*1*" | Reminder |
| **Leaving the stove on** | Not specific | *Gas detection* is "*Yes*" | *Relay status* is "Yes" and *Voice message* is "*2*" | Alert |
| **Trouble with cooking** | Kitchen | *Distance*\* is "*near*" | *Voice message* is "*3*" and *Image message* is "*2*" | Reminder |
| **Putting on clothes** | Bedroom | *Distance* is "*near*" | *Voice message* is "*4*" and *Image message* is "*3*" | Reminder |
| **Mealtime** | Not specific | *Time* is "*early afternoon*" | *Voice message* is "*5*" and *Image message* is "*4*" | Reminder |
| **Lack of movement** | Not specific | *Movement* is "*low*" | *Voice message* is "*6*" | Alert |
| **High game score** | Not specific | *Game score* is "*high*" | *Reminder* is "*No*" | Disable reminders |

*Distance between the user and particular object (such as refrigerator and drawer)

Moreover, the lack of movement is an important issue that should be considered in the patients' daily life. The system measures the user's movement to check this factor by localization tag worn by the patient. If he moves between 3.5 to 4.5 hours per day, the IA system sends a reminder to the patient and their caregiver. In another scenario, once the patient is near the drawer in the bedroom, the IA system can suggest to them what to wear. This approach can help the patient make everyday decisions before they get nervous.

To customize the IA system, family or caregivers can determine some image messages, including the patient's family pictures, first house, and childhood events to be sent to the patient when they are alone at home.

The mapping of the inputs to the outputs for a fuzzy model is in part characterized by a set of condition and action rules, or in modus ponens (If-Then) form,

*If premise Then consequent.*

The inputs of the fuzzy system are often linked with the hypothesis, and the outputs are connected to the consequent. These If-Then rules can be characterized in many various forms. Two standard forms are considered here; multi-input multi-output (MIMO) and multi-input single-output (MISO). The MISO style of a linguistic rule is:

*If $u_1$ is $A_j^1$ and $u_2$ is $A_k^2$ and, ..., and $u_n$ is $A_n^l$ Then $y_q$ is $B^p_q$*

It can certainly be proven that the MIMO form for a rule (for instance, one with consequents that have conditions linked with each of the fuzzy controller outputs) can be explained by several MISO rules utilizing simple rules from logic [85]. For example, the MIMO rule with n inputs and m = 2 outputs:

*If $u_1$ is $A_j^1$ and $u_2$ is $A_k^2$ and, ..., and $u_n$ is $A_n^l$ Then $y_1$ is $B^r_1$ and $y_2$ is $B^s_2$*

In our IA system, some parts of fuzzy rules in the automatic mode are as follows:

- *If (rain status is yes) and (distance from object1 is near) and (heading angle is small), then (image message is picture3) and (voice message is audio3).*
- *If (distance from object2 is near) and (heading angle is small), then (image message is picture2) and (voice message is audio2).*
- *If (plants humidity is very dry) then (text message number is text 1).*
- *If (distance from object4 is near) and (heading angle is small), then (image message is picture 6) and (voice message is audio6).*
- *If (time is late afternoon) and (distance from object5 is near) and (heading angle is small), then (image message is picture4) and (voice message is audio4).*
- *If (time is evening) then (image message is picture 5) and (voice message is audio 5).*
- *If (flame status is yes), then (relay status is yes) and (voice message is audio7).*
- *If (gas status is yes), then (relay status is yes) and (text message is text2).*
- *If (temperature is hot), then (text message is text3).*
- *If (temperature is cold), then (text message is text4).*

- *If (distance from object3 is near), then (voice message is audio8).*
- *If (distance from object1 is near) and (heading angle is small), then (image message is picture9) and (voice message is audio9).*

When an event occurs (for example, the patient stares at the specific objects or a sensor gets a new data value), input data values such as position, heading angle, and sensors' statuses are published on the cloud via MQTT protocol.

Moreover, fuzzy rules are checked, and corresponding output variables are updated and published on their Topics. Thus, according to which rules are activated, the patient receives different types of messages, or a particular actuator is activated. We describe some of them in the next Chapter.

### 4.3.4 Defuzzification

The final point is to map a fuzzy set to a crisp set. The proposed model applies the Mamdani inference system [85], which provides a fuzzy set of output membership functions. There are various methods for the defuzzification process where the center of gravity is the most prevalent one in defuzzification approach; thus, we use this method. This crisp set is an integer number. Formula 2 indicates the method's center of gravity, as follows:

$$D^* = \frac{\int D \cdot \mu_M(D) dD}{\int \mu_M(D) dD} \qquad (2)$$

### 4.4 Summary

In this Chapter, we studied IoT architecture in our system. We illustrated how data transfer and store on the cloud. The interaction with the end-user and its development were discussed. Further, we explained our cloud-based decision-making algorithm and its implementation process.

# Chapter 5 - System Evaluation and Results

In this section, we present our IA system reliability and performance according to the effective parameters and experimental results. First, we consider the minimum QR code size that is large enough to be scanned by a camera, and then we describe system response-time in different conditions.

Moreover, to assess the system performance by receiving indoor positioning data, we created an experimental setup based on AD patients' real-life scenarios presented in Chapter 3. Some of these scenarios are based on defining danger zones in the patients' homes; hence, the system acts to detect danger zone. The location of the dangerous objects can be defined by the user interface application, which is designed for the family or caregivers. Other scenarios are based on the patient's position and the sensors' data.

In this section, we present some experimental results to evaluate the reliability and accuracy of the proposed system.

## 5.1 QR Code Minimum Size

The minimum printed size of the QR code is dependent on some factors:

- Camera parameters
- Number of modules
- The distance between the code and the scanner
- Scanning environment quality

In this study, we have used version one of the QR code, which has 21*21 modules. Two minimum QR sizes are defined based on first, the environmental parameters, and the second one, the camera parameters. The final QR code size is determined according to these calculated parameters.

The first minimum QR length is calculated as:

$$Minimum\ QR\ Code\ Size\ 1: L_{min1} = \left(\frac{D_{scan}}{K_{dis}}\right) * K_{den} \quad (3)$$

Where $K_{den} = \frac{21}{25} = 0.84$ is the data density factor (the maximum module number divided by 25 to normalize it to the equivalent of a Version 2 QR); $K_{dis} = 10$ the distance factor start from a factor of 10 reduced by 1 for each of low lighting in the scanning environment, a mid-light colored QR code being used, or the scan not being done front on; and $D_{scan}$ defined 300mm as maximum scanning distance.

The second minimum QR size is calculated as following:

$$Minimum\ QR\ Code\ Size\ 2: L_{min2} = \frac{PPQ * FOV}{CCD_W} \quad (4)$$

Where FOV is the camera's field measured by experiment, in this case, 340mm; PPQ is defined as pixels in each dimension needed per QR. Considering 10 pixels needed for each module, we define:

$$Pixel\ per\ QR = PPQ = 10\ pixels * 21\ modules \quad (5)$$

And CCDW is the width of the CCD array calculated by solving the following equations:

$$Camera\ Resolution: CCD\ Area = CCD_W * CCD_H \quad (6)$$
$$CCD\ Width\ and\ Height: CCD_W = \phi * CCD_H \quad (7)$$

Where ϕ is normally the golden ration defined as:

$$Golden\ Ratio = \phi = \frac{1 + \sqrt{5}}{2} \cong 1.618 \quad (8)$$

Finally, we can find the minimum required QR size for printing according to the following logic:

$$Minimum\ QR\ Code\ Size: L_{min} = \max(L_{min1}, L_{min2}) \quad (9)$$

Based on our 12MP camera and computational result, as shown in equations, we must prepare at least 21*21mm printed QR code.

## 5.2 System Response-Time

For estimating the system performance and complexity, we have determined the application's computational response-time and analyzed the battery consumption. For example, we first published a new data value for turning on the relay actuator via MQTT protocol, and then the response-time for playing an audio message was assessed.

In another test, we scanned a QR code and measured the response-time for displaying an augmented image after publishing a new message for turning off the relay actuator via MQTT. Both experiments were then replicated under the same conditions fifty times to evaluate the system-accuracy, as shown in Fig. 5.1, and Fig. 5.2.

The system required an average 364ms to play an audio message, and for displaying a three-dimensional image based on QR code detection, the average response-time was 106ms. We found that playing an audio message instead of displaying the image, had better performance and less battery consumption. However, using QR code to display AR images could decrease the response time in comparison to playing an audio message.

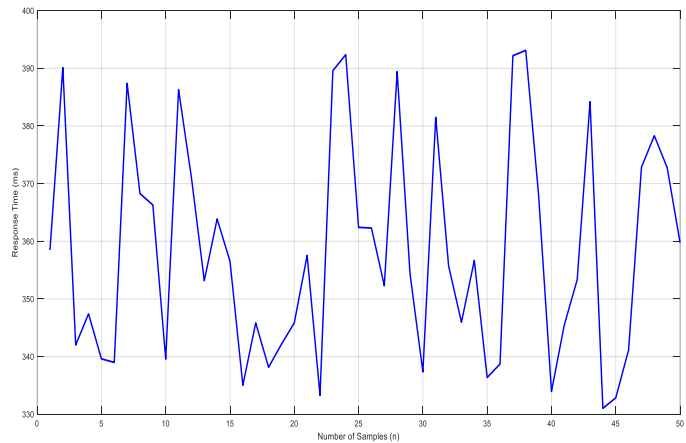

Fig. 5.1 Voice message response-time after publishing a value to the MQTT server.

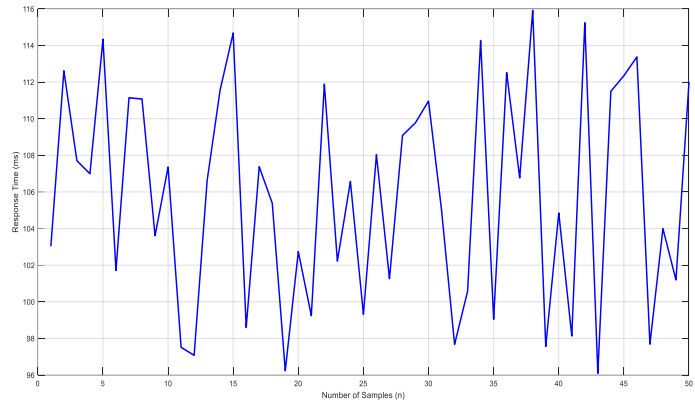

Fig. 5.2 Image message response-time after publishing a value to the MQTT server.

We are aware of the fact that the average published-subscribed latency in MQTT server is estimated to be 120ms, and the data transfer rate is limited, so our algorithm is executed in an accurate way according to the results. For real-time systems, similar to our work, this delay was adequate. Moreover, data loss did not occur after performing a series of data transfers.

## 5.3 Danger Zones and Reminders

To monitor the patient's real-time location by caregivers, we used Unity game engine because the patient's home architecture could be represented easily. Furthermore, the anchor's location and danger zones were defined by experts. All the previous algorithms we have mentioned in the previous Chapter were added to this game engine by C# libraries. In our preliminary experiments to evaluate the system's performance, we defined five objects with equal sizes of $0.52*0.7*0.23$ $m^3$ in an $8.5*4.6*2$ $m^3$ room space. To estimate the exact heading angle, the orientation data was generated using the Inertial Measurement Unit (IMU) embedded in the AR device.

In a typical scenario, once the user is near a dangerous object (for example, the fireplace), they receive an image and/or a voice alarm. This alarm reminds them to keep away from the object. As discussed in Chapter 4, the near membership function was defined from 1dm to 5dm (the danger zone).

In general, the distance between the user and the object is continuously published on its Topic to MQTT message broker. As the user walks in the room environment and the distances between the user and the objects change, each MQTT message would be updated. Therefore, the user could potentially receive different types of messages if any rules were activated by subscribing to their MQTT Topic.

A relevant experiment was performed 20 times in different conditions, as shown in Table 5.1 (by a single user). As a sample, Fig. 5.3 illustrates the user's condition in experiment number 14 (Fig. 5.3(a)) and experiment number 4 (Fig. 5.3(b)). Based on the experiments, the accuracy of the position estimation system was measured close to 1dm. Fig. 5.4 shows the danger zone areas in our experimental condition.

Table 5.1 Result of the experimental test

| Object number | Experiment number | Distance (dm) | Heading angle (degree) | Membership function of distance | Membership function of heading angle | Activated rule number |
|---|---|---|---|---|---|---|
| 1 | 1 | 3 | 12 | Near | Small | 12 |
|   | 2 | 4 | 35 | Near | Medium | - |
|   | 3 | 8 | 10 | Far | Small | - |
|   | 4 | 14 | 16 | - | small | - |
| 2 | 5 | 10 | 42 | Very Far | Medium | - |
|   | 6 | 7 | 12 | Far | Small | - |
|   | 7 | 12 | 73 | Very Far | Large | - |
|   | 8 | 2 | 11 | Near | Small | 14 |
| 3 | 9 | 5 | 122 | Near | - | 11 |
|   | 10 | 16 | 9 | - | Small | - |
|   | 11 | 6 | 38 | Far | Medium | - |
|   | 12 | 13 | 12 | Very Far | Small | - |
| 4 | 13 | 18 | 0 | - | Small | - |
|   | 14 | 4 | 10 | Near | Small | 21 |
|   | 15 | 11 | 14 | Very Far | Small | - |
|   | 16 | 7 | 72 | Far | Large | - |
| 5 | 17 | 12 | 86 | Very Far | Large | - |
|   | 18 | 2 | 12 | Near | Small | 2 |
|   | 19 | 17 | 8 | - | Small | - |
|   | 20 | 13 | 76 | Very Far | Large | - |

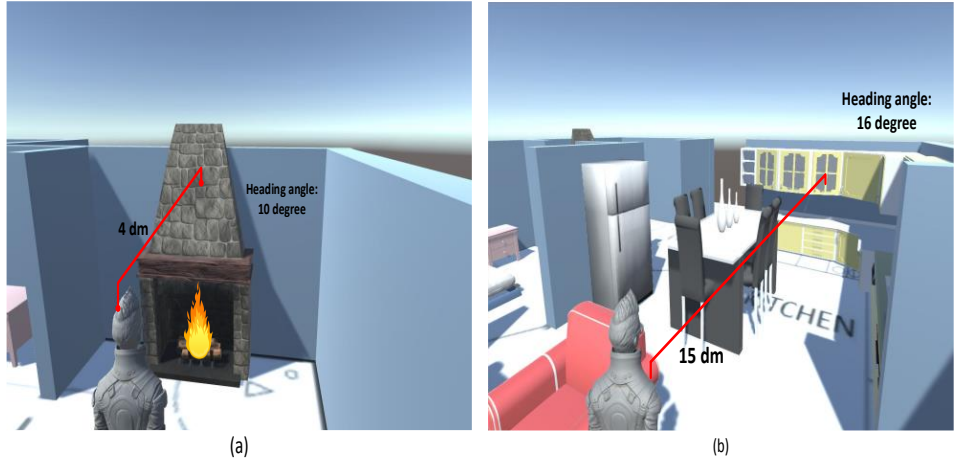

Fig. 5.3 Illustration of AD patient interaction with home objects

(a) living room, (b) kitchen.

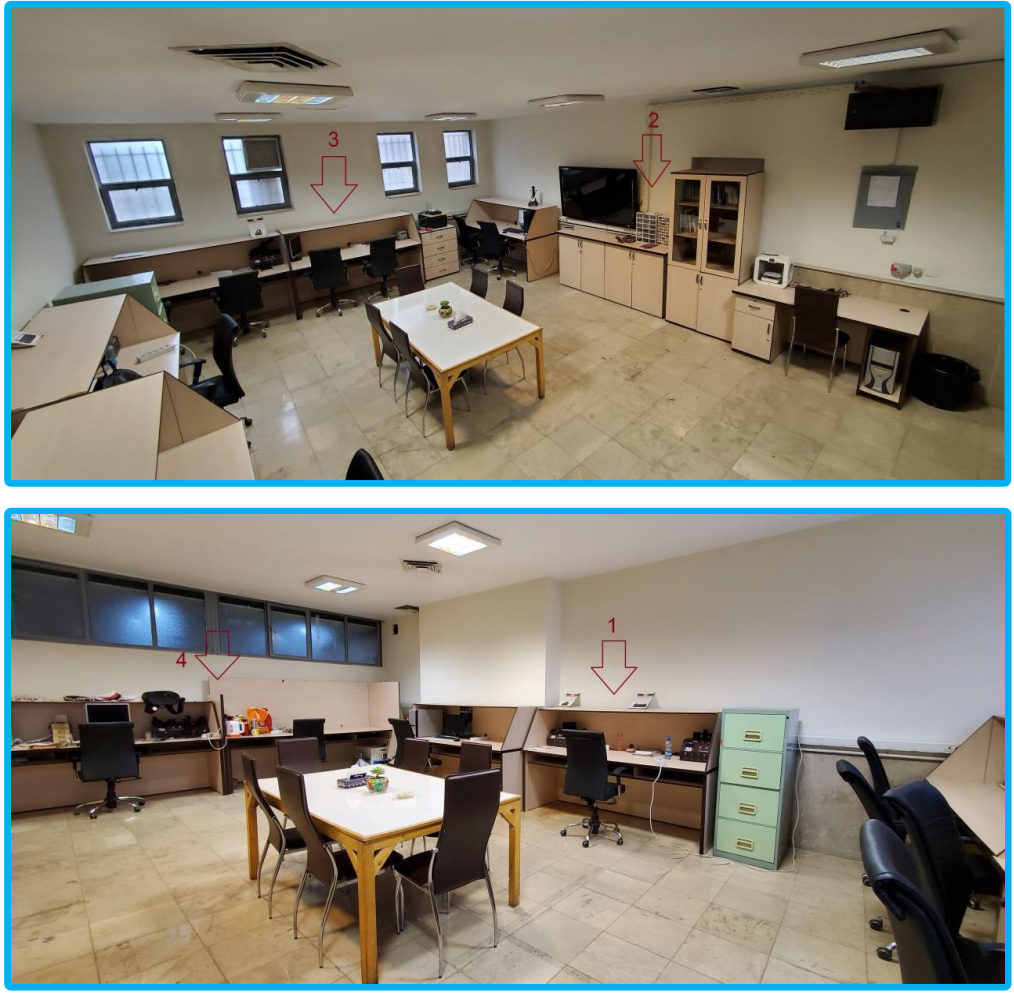

Fig. 5.4 Danger zone areas in the experimental condition.

As reported in [30], AD patients have serious difficulties about remembering their family's face and/or name; thus, in another scenario, we evaluated the system's potential as memory assistance. In the second experiment, when the user wanted to enter the room, a family member's picture was shown to them, and they received the name and age of the family member as a voice message.

Moreover, to improve AD patient's daily activity [86], we defined some other scenarios, such as reminding plant watering time based on the plant's humidity. To this intent, the humidity sensor publishes its value to the message broker, and when the humidity level drops below 40% (very dry fuzzy membership function), then fuzzy rule number 3 is activated, and the user receives a text message that contains plant care and watering reminder alarm.

To remind the patient's medication time, which could be the most critical event in AD patients' daily life (they often forget the exact time and the dosage), we developed a Windows-based application for the family members and caregivers. This application allows the caregiver to set the medication time or its period of time, so at any time of the day that fuzzy rule number 6 is activated, the system notifies the patient by sending images and audio messages to them. The messages include information about the location and dosage of the medication. Fig. 5.5 shows image samples of the result of this evaluation received by the user.

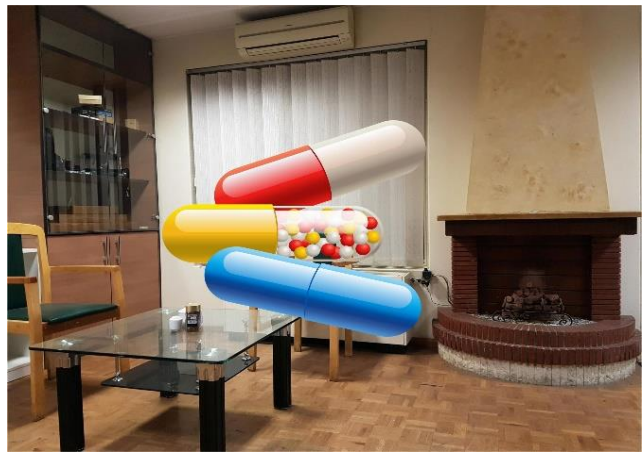

Fig. 5.5 Sample of an image message received by the user
picture message number 5 (medication).

## 5.4 Alerting Based on Data Values

In the previous Chapter, some of the fuzzy rules were presented. In most of these rules, the patient's heading angle and distance from the objects were the primary inputs to the system. Therefore, in another experiment, we evaluated the system accuracy and reliability for patient location identification.

Fig. 5.6 shows the pattern of the user's location during the first 6 seconds of this experiment. The user's location point and heading angle, anchors' (A1, A2, and A3) position, and object's location in points 1, 2, and 3, according to Fig. 5.6 are presented in Table 5.2. As shown in Fig. 5.7(a), the user changed their distance relative to a specific object (object1) while searching the environment. When the distance value lies within 1 to 5 dm, it becomes a member of the near fuzzy membership function. Fig. 5.7(b) shows the result of the pattern of the time variation of the user's heading angle during this experiment at a sampling frequency of 100 Hz. Though, as the heading angle changed between -12 to 12 degrees, it turned into a member of small fuzzy membership function.

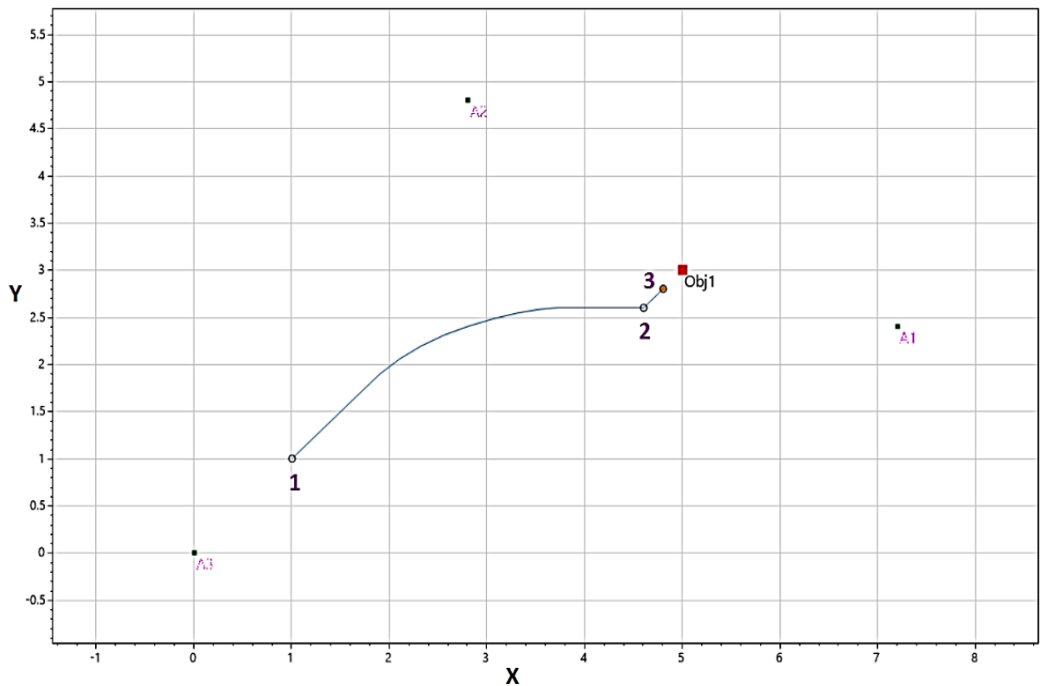

Fig. 5.6 Each anchor location and the user's position

during the first 6 second of experiment type two. Final point: (4.8, 2.8), heading angle: 3 degree.

As these two inputs become near and small, then the system checks other input values according to the fuzzy rules. After updating the user's location and orientation data values on the cloud, the system checked other inputs.

Table 5.2 Experimental condition

| Experiment number | User's location (x,y) | User's heading angle (degree) | Location of anchor1 (A1) (x,y) | Location of anchor2 (A2) (x,y) | Location of anchor3 (A3) (x,y) | Location of object (x,y) |
|---|---|---|---|---|---|---|
| 1 | (1,1) | 263 | (0,0) | (2.8,4.8) | (7.2,2.4) | (5,3) |
| 2 | (4.6,2.6) | 105 | (0,0) | (2.8,4.8) | (7.2,2.4) | (5,3) |
| 3 | (4.8,2.8) | 3 | (0,0) | (2.8,4.8) | (7.2,2.4) | (5,3) |

Measurement unit: meter, room dimensions: 8.5*4.6*2, object (Obj1) dimensions: 0.94*0.08*1.9, Rain sensor's location: (3,4.5)

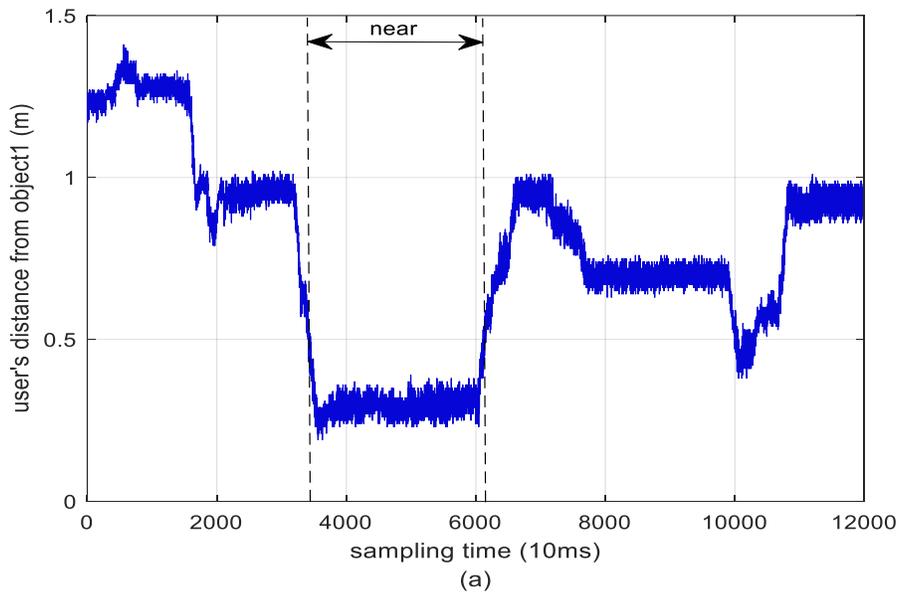

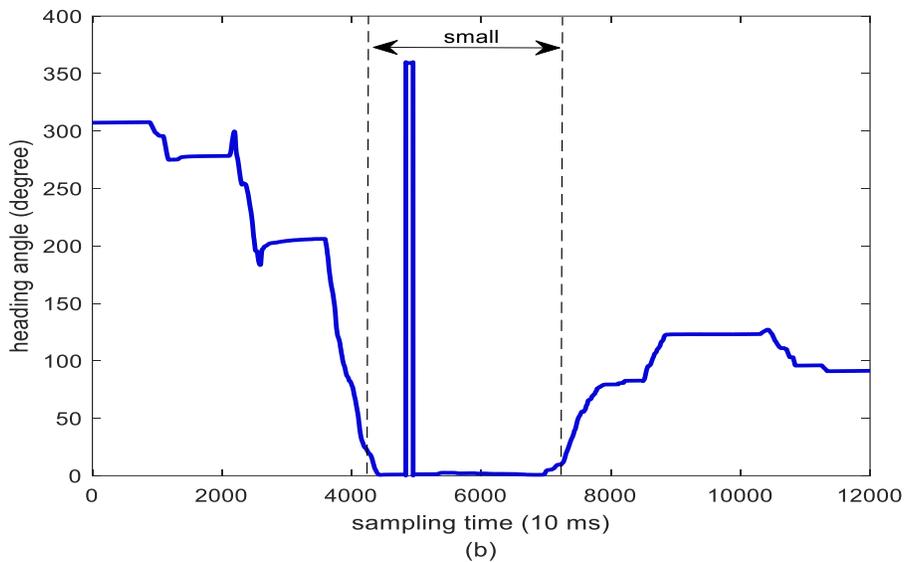

Fig. 5.7 Result of the time variation

of (a) the user's distance from a particular object during the experiment, (b) the user's heading angle during the experiment.

Moreover, during this experiment, the rain sensor value was true, thus, fuzzy rule number 1 was activated. The result of this test is shown in Fig. 5.8(a), which confirms that when rainfall status was true, and the user was in front of the object 1 and staring at it, they received the picture message number 3 and audio message number 3. Picture number 3 was an umbrella image, and audio number 3 was "the weather is raining, if you want to go outside, take your umbrella, it is in the red drawer near the door."

As shown in Fig. 5.6, the user's final location point was (4.8, 2.8), which was near to object1's location (5, 3), and the user's heading angle was 3degree, which was small. On the other hand, family or caregivers can lock or unlock the door for safety reasons. In our experiment, the door was unlocked, so relay status was off. In addition, family or caregivers can monitor these events and sensors' status by the application; thus, they are able to allow that the patient can go outside or not.

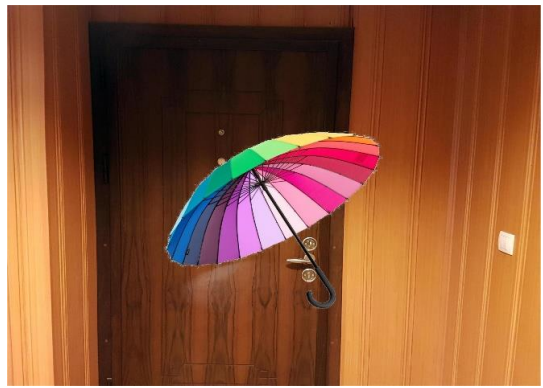

(a)

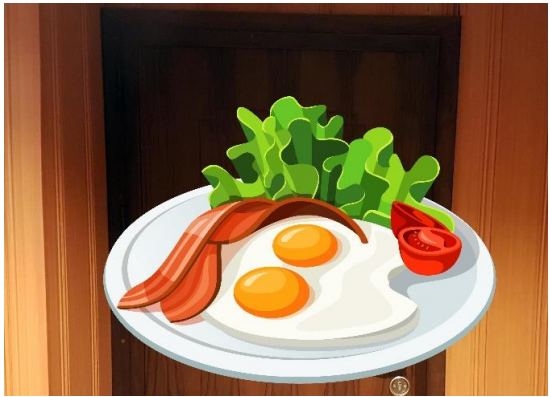

(b)

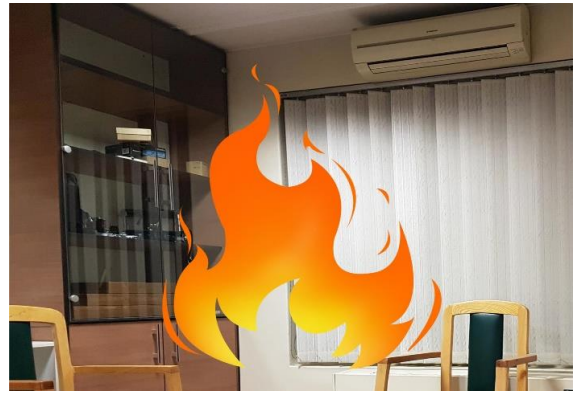

(c)

Fig. 5.8 Samples of image messages received by the user

(a) picture message number 3 (umbrella), (b) picture message number 4 (meal), and (c) picture message number 8 (fire alert).

As the heading angle changed between -12 to 12 degree, it turned into being a member of small fuzzy membership function. As these two inputs were near and small, then the system checked the other inputs' value according to fuzzy rules. In Fig. 5.9, the relation between various input (the user's distance from the cabinet, the user's heading angle) and output (audio message) of rule number 4 is shown.

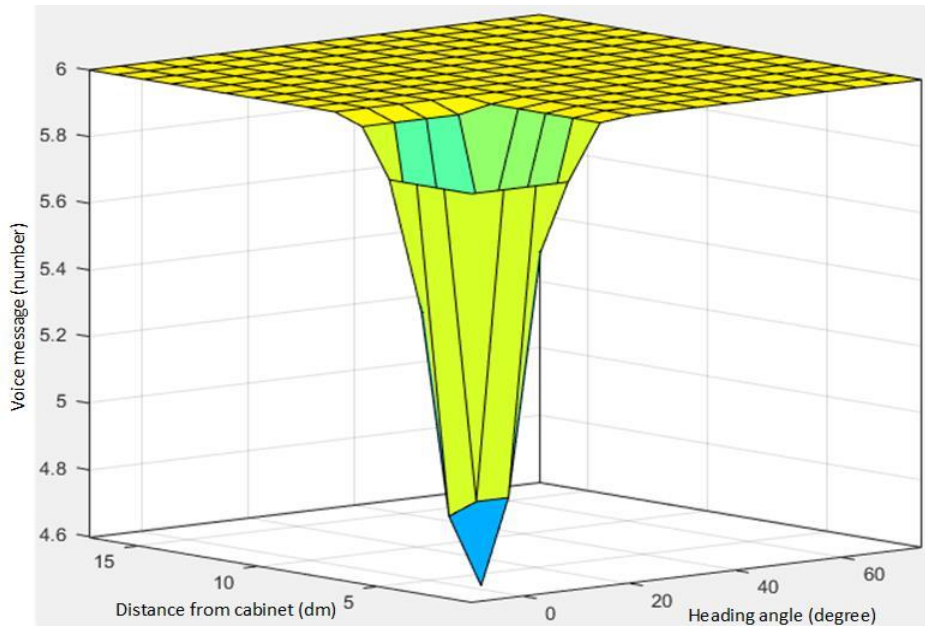

Fig. 5.9 The relation between inputs and output of fuzzy rule
inputs (the user's distance from the cabinet, and the user's heading angle) and output (voice message) of rule number4.

As we mentioned before, in the experimental test, when the user was near to the main entrance, and he was staring at the door, which meant the fuzzy membership function of the heading angle was small, then we checked the rain sensor value. If the rain sensor value was true and it was raining, then Fuzzy rule number 1 was activated.

The result of this test is shown in Fig. 5.10, which confirms that when rainfall status was true, and the user was in front of the main door and staring at it, he received the picture message number 3 and audio message number 3. Picture number 3 was an umbrella image, and audio number 3 was "the weather is raining, if you want to go outside, take your umbrella, it is in the red drawer near the door." The user's location point was (1, 2), which was near to the door, and the user's heading angle was 11 degree, which was small. On the other hand, family or caregivers can lock or unlock the door for safety reasons.

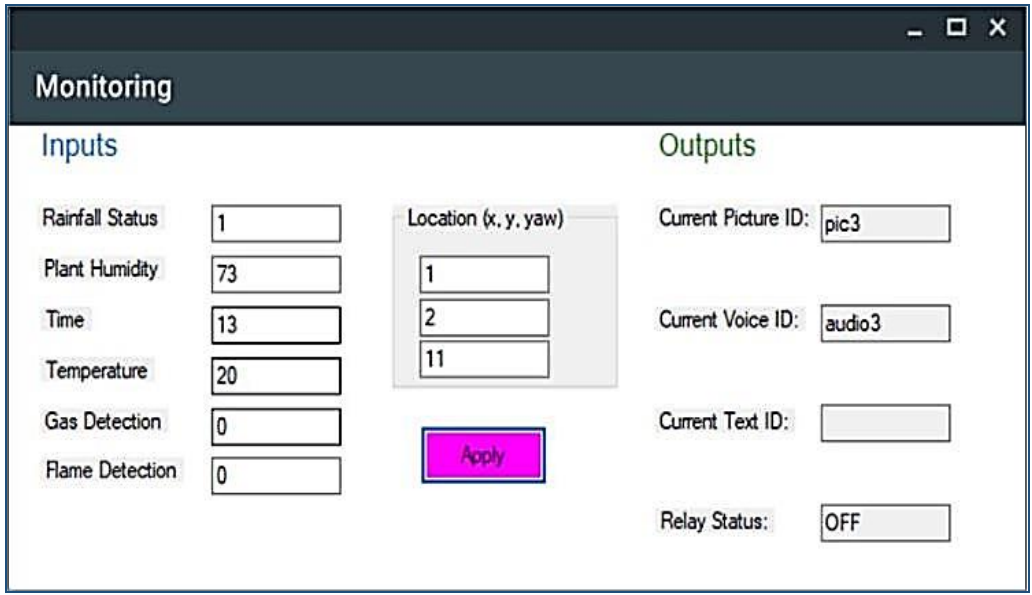

Fig. 5.10 User Interface for monitoring patient current status.

## 5.5 Network Latency and System Accuracy

In order to estimate the system's performance and complexity, we determined the system's computational response-time and analyzed the AR device's battery consumption when only fuzzy rule number 4 was activated. For instance, while the user's position data was being published to message broker via MQTT protocol, the response-time for playing an audio message was assessed.

In another test, we measured the response-time for displaying an augmented image after publishing the user's new location. Both experiments were then replicated under the same conditions thirty times to evaluate the system's accuracy. The system required an average of 247 milliseconds (msec) to play an audio message, and for displaying a three-dimensional image, the average response-time was 388msec.

We found that playing an audio message instead of displaying an image offers better performance and less battery consumption. We also realized that the average published-subscribed latency in MQTT server is estimated to be 120msec, and the data transfer rate is limited, so our algorithm is accurate enough for our application. For real-time systems, similar to our work, such time delays seem acceptable.

## 5.6 System Operation Modes

In implementing our IA system, two main rule-bases were essential to creating an adaptive decision-making process and multi-level support. The first one helped the user complete the activities independently by showing AR messages or generating automatic changes such as actuators activation. The second one allowed manual changes after real-time assessment of the user's cognitive state according to the AR-based serious game score. In this situation, some of the smart home sensors and reminders were turned off or disabled.

We considered a wide range of utilization of devices that must cover the maximum area of the user's environment. The testing environment, different objects, and anchors location, danger zones were similar to the previous experiments. The number of devices ranges from 10 to 15, including various types of sensors and actuators to realize different workloads.

Specifically, for each number of devices, we simulated the two different cases based on the system's operation mode. The complexity of the decision-making process decreases from automated mode to semi-automated mode, as the number of active devices declines from 22 to 14.

In our performance evaluation, we focus on the number of AR messages sent to the user and the number of IoT devices that were enabled during the experiment. Fig. 5.11 details the numbers of different types of AR messages included in each case and received by the user. As we can see in Fig. 5.11, the number of alarms and reminder messages decreased as the user's game score increased.

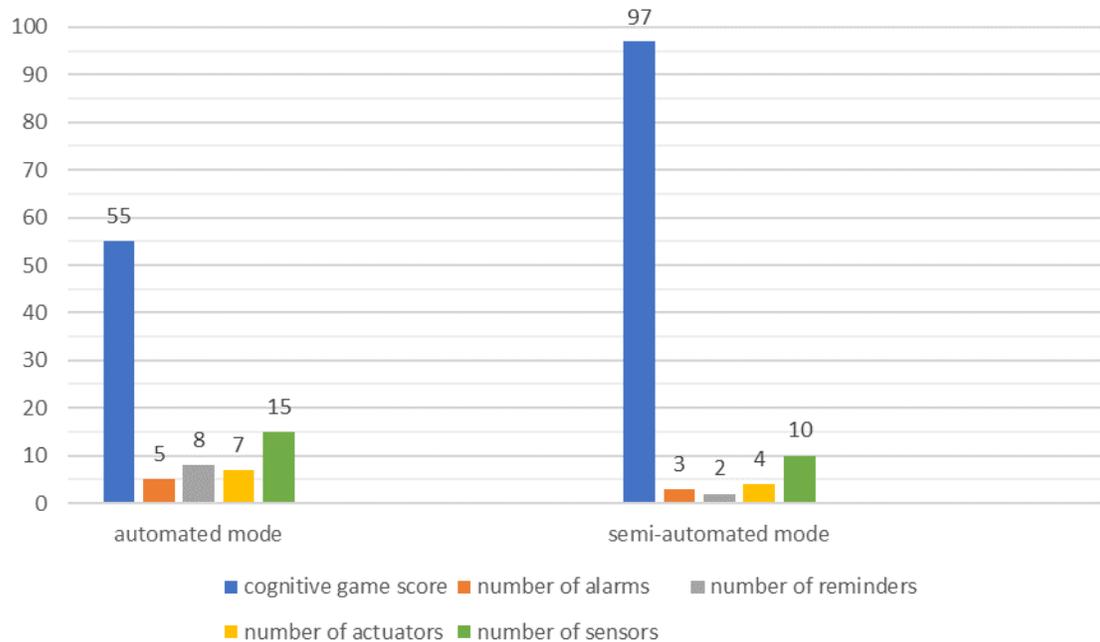

Fig. 5.11 Workload scenarios description.

As the user's cognitive game score was increased, the number of activated IoT devices, such as sensors and actuators, was reduced. This resulted in less data loss during the decision-making algorithm and the process of publishing or subscribing to the messages. It also led to less need for battery recharging or replacement.

## 5.7 Summary

In this Chapter, we presented the evaluation of our IA system and its results. The results were shown that the system response in the semi-automated mode caused less data loss than the automated mode. We also found that playing an audio message instead of displaying an image message had better performance and less battery consumption. The evolution and result of the system's performance in data rendering and providing appropriate recommendations also have been conducted.

In the next Chapter, I bring the thesis to a close and describe ideas for future work.

# Chapter 6 - Conclusion and Future Work

In this Chapter, we summarize our overall conclusions. We also propose some of the future research directions that emerge from this work.

## 6.1 Limitations

The target population and primary end-users are individuals with mild AD and older adults experiencing memory impairments, who can understand the technology they are using and its expected assistance and risk. Our proposed IA system can facilitate the daily life of AD patients in the early stages, alleviate caregivers' worries, and the health services cost.

However, this study has some limitations. One of the most important limitations is associated with personal assistance scenarios. It is difficult for the system to manage the realization of specific activities, for example, consider if the patient has finished or if she or he has appropriately performed or not a task. Moreover, it is challenging for the system to evaluate if the patient has taken their medicines correctly. Another technical limitation in the perceptual aspect of the systems is the ability to detect the anxiety of a person. Other limitations could be network disconnection and users' difficulties while making interaction with AR devices.

Results associated with user experience and performance achieved from the pilot study were only analytical, for the number of users was limited. Moreover, the space validity of this work was limited, for part of the study was presented in a laboratory environment. The simulation scenarios are particular for each patient and must be precisely designed and considered by therapists. Therefore, although the scenarios technical requirements are taken into account in our work, the particular content of the messages must be customized for each patient before the system's implementation. To achieve this goal, our IA system is capable of adding more fuzzy rules and being customized for each person.

It is essential to assess the scenarios, the AD patients' interactions, and consider if they find the system and aids engaging and helpful. We accept that the performance of the system requires more evaluation in AD patients' real life; however, we believe this framework paves the way to develop novel IA systems for further studies.

## 6.2 Conclusion

Memory loss is the main symptom experienced by patients with Alzheimer's disease. The patients forget recent incidents and have problems with remembering the information. Such situations have a considerable effect on the confidence and quality of life of both the patients and caregivers. The aim of this study was to provide an IA system for mild AD patients and improve their ability to complete everyday tasks on their own without compromising their privacy. We designed, implemented, and evaluated an intelligent assistive system based on augmented reality and Internet of Things to help people suffering from AD and make everyday decisions more independently through interaction and recall memory of events. Several ready-to-use libraries were used to facilitate system implementation on smartphones and computers.

To develop the IA system, we first designed a task prompting system based on the Augmented Reality messages without any adaptive decision-making engine to evaluate the general cognitive condition of the user or any positioning data.

The preliminary system had two main sections: the first one was the smartphone or windows application that allows caregivers/family-members to monitor a patient's status at home and be notified if the patient are at risk. The proposed configuration can be seen as a patient supervisory control engine using location-enabled fog layer services. In addition, the caregivers/family members can be able to control the actuators' states so that the patient can be reminded about different events such as medication time. The second part was capable of permitting patients to use a smartphone to recognize QR codes in the environment and receive information related to the tags in the form of audio, text, or three-dimensional image.

In the next step, we took advantage of the user's indoor positioning data. The upgraded system had four main components; a location and heading measurement system in the local fog layer, an AR device to make interactions with the AD patient, a supervisory decision-maker to handle the direct and environmental interactions with the patient, and a user interface for family or caregivers to monitor the patient's real-time situation and send reminders once required. The location information was stored in a cloud database. The collected data was then transmitted via the Internet to a fuzzy decision-making engine.

Finally, to increase the intelligence of the system and so adaptation, in this work, we also took advantage of an AR-based serious game assessment result. The collected data of the user's game result was then transmitted via the Internet to a fuzzy decision-making engine. This adaptive decision engine analyzed and compared data to make an appropriate decision based on fuzzy rules set for making interaction with the user.

The target population and primary end-users are individuals with mild AD and older adults experiencing memory impairments, who can understand the technology they are using and its expected assistance and risk. Our proposed IA system can facilitate the daily life of AD patients in the early stages, alleviate caregivers' worries, and the health services cost.

Generally, two main rule-bases were essential to creating an adaptive decision-making process and multi-level support. The first one helps the user to complete the activities in their daily life by showing AR messages or making automatic changes such as actuators activation. The second one allowed manual changes after real-time assessment of the user's cognitive state according to the AR game score. In this situation, some of the smart home sensors and reminders can be turned off or disable according to the fuzzy rules. This feature can improve the patient's self-management and self-care, and it would also slow the progression of the disease.

The system was also capable of scaling up to more sensors, actuators, and rules to enhance the AD patient's independency without compromising privacy. Furthermore, physicians can potentially use the patients' in-home collected data to identify the progress of the disease or the effectiveness of the medication.

We evaluated the prototype system to analyze its performance under several conditions. We found that playing an audio message instead of displaying an image message had better performance and less battery consumption. However, using indoor positioning data while comparing to QR codes for sending AR images could decrease the response time. The result showed that MQTT is adapted for quick and reliable messaging transport among various devices.

We also found that playing audio messages instead of displaying images offer better performance and less battery consumption. The findings of this study supported the idea that an IoT-based system can potentially help mild to moderate AD patients in the home environment and can be evaluated in AD patients' real life. The devised framework included AR images as a useful memory aid in addition to text/voice and actuators controlled by the fuzzy inference decision-maker. We also provided further evidence that the IA system's accuracy, reliability, and response-time are appropriate to be implemented in AD patients' homes.

## 6.3 Future Work

In this study, we performed an operational assessment and evaluation of the system. Some future plan could be implementing the IA system based on AR glasses for real-life scenarios of AD patient's daily life who already utilizes paper tags attached in their home by the caregivers or family members. Moreover, the data collected from the patient with AD and a coherent high-level abstraction of extracted contextual data can be considered by physicians to evaluate the patient's mental and physical health condition. Historical data, including health, records the patient's diagnoses, disease progress, current behavior and previous changes, and treatments, which can be stored on the cloud and analyzed for further studies.

In order to make the system context-aware, and more intelligence, improvements in software applications, hardware, and decision-making process can be carried out. The design of context-aware assistive systems should consider observation, interpretation, and reasoning about the patient's conditions from several perspectives, such as behavioral, physiological, and environmental. The new system can consider all the relevant contextual dimensions through which improves in the proposed IA system. For example, it can develop the current framework by considering location, time, objects, posture, frequency, and user activities history.

Monitoring and evaluating human activities, particularly the individual with dementia, is a complicated task due to the complexity of completing different types of daily activities. Behavior of the user can even change for the same person depending on mood and health conditions. This may increase the complexity of the assisting and monitoring process. In designing a new intelligent assistive tool, we can consider these factors of the user's interaction by the activity recognition process to make the system more helpful. The new system should recognize appropriate methods and techniques that can effectively realize complicated and ever-changing user behavior. Such systems must be assessed over long periods to confirm their efficiency in terms of analysis, accuracy, and adaptability of the new proposed assistive system.

In this work, we estimated the user's cognitive and mental state by importing the AR cognitive game score. Besides this assessment, the user's emotional status can also be considered. Finally, future improvements can also be made to the decision-making algorithm, and it can involve the addition of factors such as heart rate variability. This feature has shown a strong correlation with the individual's emotional status. A multi-level approach to identifying significant changes in the physiological state may also lead to better algorithm performance and make the system's recommendations more helpful for the end-user.

The prototype of the IA system can be tested with AD patients, and it includes more scenarios and functionalities as the technical challenges are addressed, and the priorities of AD patients are indicated. To move further in the interaction design process, a group of individuals with mild AD and caregivers can be invited for advance evaluation and development of the IA system. The IA system can also be implemented in the AD patients' home environment with more users to analyze their responses to the system over a more extended period of time.